\documentclass{aa}  

\usepackage{graphicx}
\usepackage{txfonts}
\usepackage[colorlinks,citecolor=blue]{hyperref}
\hypersetup{
    colorlinks = true,
    linkcolor = blue,
    anchorcolor = blue,
    citecolor = blue,
    filecolor = blue,
    urlcolor = blue
    }

\usepackage{booktabs} % for nicer tables
\usepackage{mathrsfs}
\usepackage{tikz}
\usepackage{booktabs}
\usetikzlibrary{calc}
\usepackage{mathtools}
\usepackage{siunitx}
\usepackage{svg}
\usepackage{ulem}

\newcommand{\tikzmark}[1]{\tikz[overlay,remember picture] \node (#1) {};}
\newcommand{\DrawBox}[3][]{%
    \tikz[overlay,remember picture]{
    \draw[black,#1]
      ($(#2)+(-0.2em,1.4ex)$) rectangle
      ($(#3)+(0.3em,-0.5ex)$);}
}

\begin{document} 

   \title{Quantitative spectroscopy of B-type supergiants}

   \author{D. We{\ss}mayer\inst{1}
          \and
          N. Przybilla\inst{1}
          \and
          K. Butler\inst{2}
          }

   \institute{Institut f\"ur Astro- und Teilchenphysik, Universit\"at Innsbruck, Technikerstr. 25/8, 6020 Innsbruck, Austria\\
              \email{david.wessmayer@uibk.ac.at ; norbert.przybilla@uibk.ac.at}
         \and
             LMU M\"unchen, Universit\"atssternwarte, Scheinerstr. 1, 81679 M\"unchen, Germany
             }

   \date{Received ; accepted }

% \abstract{}{}{}{}{} 
% 5 {} token are mandatory
 
  \abstract
   {B-type supergiants are versatile tools to address a number of highly-relevant astrophysical topics, ranging from 
   stellar atmospheres over stellar and galactic evolution to the characterisation of interstellar sightlines and to
   the cosmic distance scale.}
   {A hybrid non-LTE (local thermodynamic equilibrium) approach -- involving line-blanketed model atmospheres computed under the assumption of LTE in combination with line formation calculations that account for deviations from LTE 
   -- is tested for quantitative analyses of B-type supergiants of mass up to about 30\,$M_\sun$, characterising 
   a sample of 14 Galactic objects in a comprehensive way.}
   {Hydrostatic plane-parallel atmospheric structures and synthetic spectra computed with Kurucz's {\sc Atlas12} 
   code together with the non-LTE line-formation codes {\sc Detail/Surface} are compared to results from full non-LTE
   calculations with {\sc Tlusty}, and the 
   effects of turbulent pressure on the models are investigated. High-resolution spectra at signal-to-noise ratio\,$>$\,130 are analysed for atmospheric 
   parameters, using Stark-broadened hydrogen lines and multiple metal ionisation equilibria, and for elemental 
   abundances.
    Fundamental stellar parameters are derived by considering stellar evolution tracks and Gaia early data release 3 (EDR3) 
   parallaxes. Interstellar reddening and the reddening law along the sight lines towards the target stars are 
   determined by matching model spectral energy distributions to observed ones.}
   {Our hybrid non-LTE approach turns out to be equivalent to hydrostatic  
   full non-LTE modelling for the deeper photospheric layers 
   of the B-type supergiants under consideration, where most lines of the optical spectrum are formed. Turbulent pressure can
   become relevant for microturbulent velocities larger than 10\,km\,s$^{-1}$. The changes in the atmospheric density 
   structure affect many diagnostic lines, implying systematic changes in atmospheric parameters, for instance an increase in surface gravities by up to 0.05\,dex. A high precision and accuracy is achieved for all derived parameters by 
   bringing multiple indicators to agreement simultaneously. Effective temperatures are determined to 2-3\% uncertainty, 
   surface gravities to better than 0.07\,dex, masses to about 5\%, radii to about 10\%, luminosities to better than 25\%, and spectroscopic distances to 10\%
   uncertainty typically. Abundances for chemical species that are accessible from the optical spectra (He, C, N, O,
   Ne, Mg, Al, Si, S, Ar, and Fe) are derived with uncertainties of 0.05 to 0.10\,dex (1$\sigma$ standard deviations). 
   The observed spectra are reproduced 
   well by the model spectra. The derived N/C versus N/O ratios tightly follow the predictions from Geneva stellar 
   evolution models that account for rotation, and spectroscopic and Gaia EDR3 distances are closely matched.
   Finally, the methodology is tested for analyses of intermediate-resolution spectra of 
   extragalactic~B-type~supergiants.
   } 
   {}


   \maketitle
%
%-------------------------------------------------------------------

\section{Introduction}
Massive stars are drivers of the evolution of galaxies as they are crucial contributors to the 
energy and momentum budget of the interstellar medium (ISM), and they are sources of nucleosynthesis 
products \citep[e.g.][]{Matteucci08}. This is because of their ionising radiation, their strong 
stellar winds, and final fate in supernova explosions and -- under certain circumstances -- as $\gamma$-ray bursts. 
Multiple facets of the evolution of single and binary massive stars are largely understood, though many details have yet to be resolved \citep[e.g.][]{MaMe12,Langer12,Sanaetal12}, with several independent grids of evolutionary models 
being available \citep[e.g.][]{Brottetal11,Ekstroemetal12,LiCh18,Szecsietal22}. Improvements in our understanding of galactic and 
massive star evolution are driven by observational constraints, either qualitatively by consideration of new aspects,
or quantitatively by a reduction in observational uncertainties (which is of interest here).

The evolution of massive stars in the upper Hertzsprung-Rus\-sell diagram (HRD) splits overall into two domains, 
connected to the Humphreys-Davidson limit \citep{HuDa79}.
Stars more massive than $\sim$40\,$M_\sun$ remain blue objects throughout their entire life because strong stellar
winds and probably pulsational instabilities lead to the loss of their envelopes. These early and mid-O dwarfs and giants on the 
main sequence (MS) evolve into early B-type hypergiants \citep[e.g.][]{Clarketal12,Herreroetal22} and supergiants of luminosity class 
Ia, which constitute one of the more frequently populated regions of post-MS evolution in the HRD 
\citep[e.g.][]{Castroetal14}. In this evolutionary stage, they belong to the visually brightest stars in star-forming
galaxies in addition to their high energy output at UV wavelengths and they are likely to become luminous blue variables (LBVs) at more advanced evolutionary stages, and finally Wolf-Rayet (WR) stars. Realistic quantitative spectroscopy of these objects 
requires hydrodynamical stellar atmosphere models that account for deviations from local thermodynamic equilibrium 
(non-LTE) and metal-line blanketing \citep{HiMi98,Pauldrachetal01,Graefeneretal02,Pulsetal05}.

The less massive stars (i.e.~$M$\,$\lesssim$\,30\,$M_\sun$) with bolometric magnitudes larger than about $-$9.5 to $-$10\,mag evolve into red supergiants 
(RSGs) with extended hydrogen-rich envelopes at least once during their lifetime. They become B-type supergiants of luminosity classes Ib and Iab and at the lower limit of the massive star regime at $\sim$8-9\,$M_\sun$ also bright giants (luminosity class II) when
they evolve from late-O and early-B dwarfs on the MS on their way towards the RSG stage. Alternatively, some B-type 
supergiants may be post-RSG objects like Sk $-69$\degr\,202, the precursor of SN~1987A \citep{Westetal87},
with possible evolutionary channels provided by binary \citep{Podsiadlowski92} as well as single star evolution 
\citep{Hirschietal04}. Such objects should be rare. While signatures of mass-loss are still present in the spectra of 
these lower-luminosity B-type
supergiants, in the optical part of the spectrum they are restricted to a few spectral lines like H$\alpha$. 
The photospheric spectrum on the other hand is 
formed under conditions close to hydrostatic equilibrium, so that hydrostatic line-blanketed non-LTE model atmospheres
\citep{HuLa95} may be employed for their quantitative analysis. This was confirmed by a comparison of hydrostatic and 
hydrodynamic non-LTE model atmospheres for luminous early B-type supergiants by \citet{Duftonetal05}.

Galactic B-type supergiants have been investigated for a long time, starting with the early work on the B1\,Ib star $\zeta$~Per by 
\citet{Cayrel58} based on photographic plate spectra and the studies of \citet{Dufton72,Dufton79} using improved LTE 
model atmospheres. The advent of spectroscopy with CCD detectors facilitated the first spectral atlas of Galactic B-type
supergiants to be obtained at optical wavelengths and spectral line behaviours to be investigated qualitatively over 
the entire spectral type \citep{Lennonetal92,Lennonetal93}. This dataset was later employed for the first larger-scale 
investigation of atmospheric parameters and the chemical abundances of B-type supergiants, employing plane-parallel and 
hydrostatic non-LTE atmospheres composed of hydrogen and helium, plus subsequent line-formation computations for the
metals \citep{McErleanetal99}. A main focus were abundances of carbon, nitrogen, and oxygen as tracers for the presence 
of CN(O)-processed material in the atmospheres, modified from initial standard values due to evolutionary processes.
The Lennon et al. spectra were also utilised to derive stellar wind parameters for Galactic B-type supergiants by 
\citet{Kudritzkietal99} to constrain the wind momentum-luminosity relationship \citep{Pulsetal96} for distance
measurements of this kind of object, employing hydrodynamical H+He model atmospheres in non-LTE. 
About the same time some B-type supergiants were employed in the derivation of Galactic abundance gradients, 
using a differential pure LTE analysis \citep{Smarttetal01a}.

Studies with sophisticated line-blanketed non-LTE model atmospheres followed, concentrating on the derivation of 
atmospheric, stellar wind, and fundamental parameters, often employing observational data of higher resolving power and wider wavelength coverage than in the earlier work
\citep{Crowther_etal_06,Lefeveretal07,MaPu08,Searle_etal_08,Hauckeetal18}. Elemental abundances were discussed in some
of these works, focusing again on carbon, nitrogen, and oxygen as tracers for mixing of 
the atmospheric layers with nuclear-processed material from the stellar core. Models predict very tight correlations for
the surface CNO abundances independent of single or binary star evolution \citep{Przybillaetal10,Maederetal14}. 
More comprehensive chemical information (C, N, O, Mg, and Si) was derived by use of line-blanketed hydrostatic 
non-LTE model atmospheres for B-type supergiants in Galactic 
open clusters by \citet{Hunteretal09}, while \citet{Fraser_etal_10} studied atmospheric parameters, nitrogen abundances, and rotational and macroturbulent velocities based on high-resolution spectra. Similar work with an extended 
observational database was later conducted by \citet{IACOBIII}. On the cool end of the B-type supergiants and towards the early A-type supergiants a sample of objects was analysed by \citet{FiPr12}, using techniques very similar to those employed in the present work \citep{Przybillaetal06}. 

The enormous luminosities of B-type supergiants makes their spectroscopy feasible at distances beyond the Milky Way.
Objects in the Magellanic Clouds were therefore intensely studied, concentrating initially on the more metal-poor Small 
Magellanic Cloud \citep[SMC,][]{Trundleetal04,Leeetal05,Duftonetal05}, where surface enrichments with nuclear-processed 
matter due to rotational mixing were predicted to be stronger \citep[e.g.][]{MaMe01,georgy_etal_13}. Only later was 
attention turned towards the Large Magellanic Cloud \citep[LMC,][]{Hunteretal09,McEvoyetal15,Urbanejaetal17}.  

Even earlier, first studies of B-type supergiants were undertaken for more distant galaxies of the Local Group, based on
intermediate-resolution spectra and aiming at the determination of stellar parameters and elemental abundances. Work on 
M31, based on the \citet{McErleanetal99} approach or LTE techniques \citep{Smarttetal01b,Trundleetal02} and on
NGC6822 \citep{Muschieloketal99} was followed by 
studies of M33 supergiants using hydrodynamical non-LTE atmospheres \citep{Urbanejaetal05b,Uetal09}, aiming at the
derivation of abundance and metallicity gradients, and distances. Metallicities and distances 
\citep[derived via application of the Flux-weighted Gravity-Luminosity Relationship, FGLR,][]{Kudritzkietal03,Kudritzkietal08}
were also the focus of studies of the Local Group dwarf irregular galaxies IC1613
\citep{Bresolinetal07,Bergeretal18} and WLM \citep{Bresolinetal06,Urbanejaetal08}.

B-type supergiants in galaxies beyond the Local Group have also been studied, investigating not only stellar parameters,
metallicities and metallicity gradients, but also interstellar reddening in these galaxies, distances, and the galaxy 
mass-metallicity relationship \citep[e.g.][]{Lequeuxetal79,Tremontietal04,Maiolinoetal08}, which is a key to the study of 
galaxy evolution. Objects in NGC300 \citep{Bresolinetal02,Bresolinetal04,Urbanejaetal03,Urbanejaetal05a} and NGC55 
\citep{Castroetal12,Kudritzkietal16} in the Sculptor filament of galaxies were investigated, and in NGC3109
\citep{Evansetal07,Hoseketal14}, a member of the nearby Antlia-Sextans group. At even larger distances of about
3.5, 4.5, and 6.5\,Mpc, respectively, B-type supergiants were analysed in the grand design spiral galaxy M81 
\citep{Kudritzkietal12}, in the barred spiral galaxy M83 \citep{Bresolinetal16}, and in the field spiral galaxy NGC3621 \citep{Kudritzkietal14}, all based on spectra obtained 
with 8-10\,m-class telescopes.

In addition to their usefulness for stellar studies, B-type supergiants are frequently employed as background stars for studies
of diffuse interstellar bands (DIBs) because they facilitate sight lines to be covered to large distances and 
provide continuous spectra with relatively few intrinsic stellar spectral features. B-type supergiants are therefore 
not only employed to cover interstellar sight lines in the Milky Way \citep[e.g.][]{Coxetal17,Ebenbichleretal22}, but also as tracers of 
DIBs in other galaxies such as the Magellanic Clouds \citep{Coxetal06,Coxetal07} and M31 \citep{Cordineretal08,Cordineretal11}.

\begin{table*}
\caption{B-type supergiant sample.}
\label{tab:spectra_info}
\centering  
{\small
\setlength{\tabcolsep}{1mm}
\begin{tabular}{lllllcrrrccr}
\hline\hline
ID\# & Object      & Sp. T.\tablefootmark{a}    & Sp. T.\tablefootmark{b} & OB Assoc.\tablefootmark{c} &       & $V$ \tablefootmark{d}   & $B-V$\tablefootmark{d}  & $U-B$\tablefootmark{d}  & Date of Obs. & $T_{\mathrm{exp}}$         & $S/N$                \\
      &             &           & \multicolumn{1}{l}{}       &           &       & mag   & mag    & mag    & YYYY-MM-DD  & s                          &                      \\ \hline
\multicolumn{3}{l}{FOCES  $R$\,=\,40\,000}\\[1mm]
1     & \object{HD 7902}     & B6 Ib     & B6 Ia                          & NGC\,457   &       & 6.988$\pm$0.023 & 0.414$\pm$0.009  & $-$0.380$\pm$0.004  & 2001-09-27   & 896                        & 320                  \\
2     & \object{HD 14818}    & B2 Ia     & B2 Ia                          & Per OB1   &       & 6.253$\pm$0.016 & 0.301$\pm$0.007  & $-$0.613$\pm$0.009 & 2005-09-22   & 3$\times$1200              & 320                  \\
3     & \object{HD 25914}    & B5 Ia     & B6 Ia                          & Cam OB3   &       & 7.99  & 0.6    & $-$0.28  & 2005-09-25   & 2700                       & 180                  \\
4     & \object{HD 36371}    & B4 Ib     & B5 Ia                          & Aur OB1   &       & 4.766$\pm$0.014 & 0.345$\pm$0.013  & $-$0.445$\pm$0.015 & 2001-09-30   & 240                        & 480                  \\
5     & \object{HD 183143}   & B6 Ia     & B7 Ia\tablefootmark{e}       & Field     &       & 6.839$\pm$0.017  & 1.185$\pm$0.018  & 0.165$\pm$0.031  & 2001-09-25   & 900                        & 220                  \\
6     & \object{HD 184943}   & B8 Ia/Iab & B8 Iab                      & Vul OB1   &       & 8.184$\pm$0.016 & 0.725$\pm$0.009  & $-$0.073$\pm$0.011 & 2005-09-25   & 1800                       & 130                  \\
7     & \object{HD 191243}   & B6 Ib     & B5 Ib                          & Cyg OB3   &       & 6.111$\pm$0.022 & 0.151$\pm$0.014  & $-$0.447$\pm$0.034 & 2005-09-21   & 900                        & 350                  \\
8     & \object{HD 199478}   & B8 Ia     & B8 Ia                          & NGC\,6991  &       & 5.679$\pm$0.018 & 0.461$\pm$0.017  & $-$0.341$\pm$0.028 & 2001-09-26   & 1200\,+\,3$\times$600 & 240                  \\[2mm]
\multicolumn{3}{l}{FEROS $R$\,=\,48\,000}\\[1mm]
9     & \object{HD 51309}    & B3 Ib/II  & B3 Ib\tablefootmark{f}     & Field     &       & 4.380$\pm$0.014  & $-$0.064$\pm$0.008 & $-$0.704$\pm$0.018 & 2011-12-09   & 2$\times$45                & 440                  \\
10    & \object{HD 111990}   & B1/B2 Ib  & B2 Iab                          & Cen OB1   &       & 6.792$\pm$0.015 & 0.242$\pm$0.006  & $-$0.579$\pm$0.008 & 2013-08-17   & 300                        & 260                  \\
11    & \object{HD 119646}   & B1 Ib/II  & B1.5 Ib                          & Field     &       & 6.602$\pm$0.020 & 0.118$\pm$0.007  & $-$0.685$\pm$0.022 & 2005-04-23   & 400\,+\,410  & 490                  \\
12    & \object{HD 125288}   & B5 Ib/II  & B5 II                          & Field     &       & 4.336$\pm$0.013 & 0.115$\pm$0.007  & $-$0.444$\pm$0.007 & 2013-08-20   & 240                        & 410                  \\
13    & \object{HD 159110}   & B4 Ib     & B2 II                          & Field     &       & 7.578$\pm$0.009 & $-$0.022$\pm$0.010 & $-$0.685$\pm$0.012 & 2005-04-23   & 2$\times$1000              & 410                  \\
14    & \object{HD 164353}   & B5 I/Ib   & B5 Ib\tablefootmark{e}   & Coll\,359  &       & 3.961$\pm$0.019 & 0.023$\pm$0.012  & $-$0.606$\pm$0.019 & 2013-08-20   & 135\,+\,180  & 540                  \\
\hline
\end{tabular}
\tablefoot{
\tablefoottext{a}{adopted from SIMBAD} 
\tablefoottext{b}{this work}
\tablefoottext{c}{\cite{humphreys78}}
\tablefoottext{d}{\cite{Mermilliod97}}
\tablefoottext{e}{\cite{GrCo09}}
\tablefoottext{f}{Walborn's B-type standards \citep{GrCo09}}
}}
\end{table*}

Overall, B-type supergiants show enormous potential as versatile tools to address multiple astrophysical topics of 
high relevance. The present paper addresses the quantitative spectroscopy of B-type supergiants based on a 
hybrid non-LTE approach -- combining hydrostatic line-blanketed LTE atmospheres with subsequent non-LTE line formation 
--, applying state-of-the-art model atoms. The paper is organised as follows: observations and the data 
reduction are summarised in Sect.~\ref{section:observations}. The hybrid non-LTE modelling approach is introduced and 
comparisons to full non-LTE model atmospheres are made in Sect.~\ref{section:model_atmospheres}. Then, details of 
the analysis methodology are discussed in Sect.~\ref{section:spectral_analysis}.
Section~\ref{section:results} presents all results from the B-type supergiant sample analysis and the suitability of 
the method for quantitative analyses at intermediate spectral resolution is investigated in Sect.~\ref{section:intermediateR}, in preparation for
extragalactic studies applying the hybrid non-LTE approach. Conclusions are drawn in Sect.~\ref{section:conclusions}. 
An example of a detailed comparison of a tailored model with an observed spectrum is given in Appendix~\ref{section:appendixA}.

%-------------------------------------------------------------------
\section{Observations and data reduction}\label{section:observations}
High-resolution spectra of 14 Galactic B-type supergiants at high signal-to-noise ratio $S/N$ constitute the 
observational basis for the present work. The spectral range B1.5 to B8 at luminosity classes II, Ib, Iab, and Ia is
covered, extending previous work on late B and early A-type supergiants of similar masses and luminosities 
\citep{Przybillaetal06,SchPr08,FiPr12} towards higher temperatures. Basic information on the star sample 
and the observing log are summarised in Table~\ref{tab:spectra_info}. An internal ID number is given, the 
Henry-Draper catalogue designation, the spectral type, and an OB association or open cluster membership is indicated.
Spectral type information from the SIMBAD database\footnote{\url{http://simbad.u-strasbg.fr/simbad/sim-fid}} is 
summarised, as well as from a re-determination in the present work, based on anchor points of the Morgan-Keenan system 
and Walborn's B-type standards \citep{GrCo09}. The luminosity class determination was based on Balmer line appearance, in particular 
concentrating on H$\alpha$. Moreover, photometric data in the Johnson system are
given in Table~\ref{tab:spectra_info}, the $V$ magnitude and the $B-V$ and $U-B$ colours. The observing log 
provides the date of observation, exposure times, and the $S/N$ of the final spectrum, measured around 5585\,{\AA}.

The raw spectra were obtained with two instruments. Objects in the northern hemisphere were observed with the
Fibre Optics Echelle Cassegrain Spectrograph \citep[FOCES,][]{Pfeifferetal98} on the Calar Alto 2.2\,m telescope 
in two observing runs in 2001 and 2005. The spectra cover a wavelength range from 3860 to 9580\,{\AA} at a resolving
power $R$\,=\,$\lambda/\Delta\lambda$\,$\approx$\,40\,000, with 2 pixels covering a $\Delta\lambda$ resolution element. 
A median filter was applied to the raw images to remove
effects of bad pixels and cosmics in an initial step. Then, the FOCES semi-automatic pipeline \citep{Pfeifferetal98} 
was employed for the data reduction, performing subtraction of bias and dark current, flatfielding, wavelength 
calibration using Th-Ar exposures, and rectification and merging of the echelle orders. A major advantage of the FOCES 
design was that the order tilt was much more homogeneous than in similar spectrographs. This facilitated a more robust 
continuum rectification than is usually feasible, even in the case of broad features like the hydrogen Balmer lines, which 
can span more than one echelle order \citep[for a discussion see][]{Korn02}. Two objects had multiple exposures taken 
consecutively that were combined to increase the $S/N$ of the final spectrum.

Objects of the southern hemisphere were observed with the Fiberfed Extended Range Optical Spectrograph
\citep[FEROS,][]{Kauferetal99} on the European Southern Observatory (ESO)/Max-Planck Society 2.2\,m telescope in La
Silla, Chile. FEROS provides a resolving power of $R$\,$\approx$\,48\,000, with a 2.2-pixel resolution element.
The reduced Phase 3 spectra were downloaded from the ESO Science 
Portal\footnote{\url{https://archive.eso.org/scienceportal/home}}. Continuum normalisation was achieved by division 
by a spline function to carefully selected continuum windows. Only the $\sim$3800 to 9000\,{\AA} range of the full wavelength coverage of FEROS satisfied  our quality criteria for the quantitative analysis.

\begin{figure*}
\centering 
    \resizebox{0.999\textwidth}{!}{\includegraphics{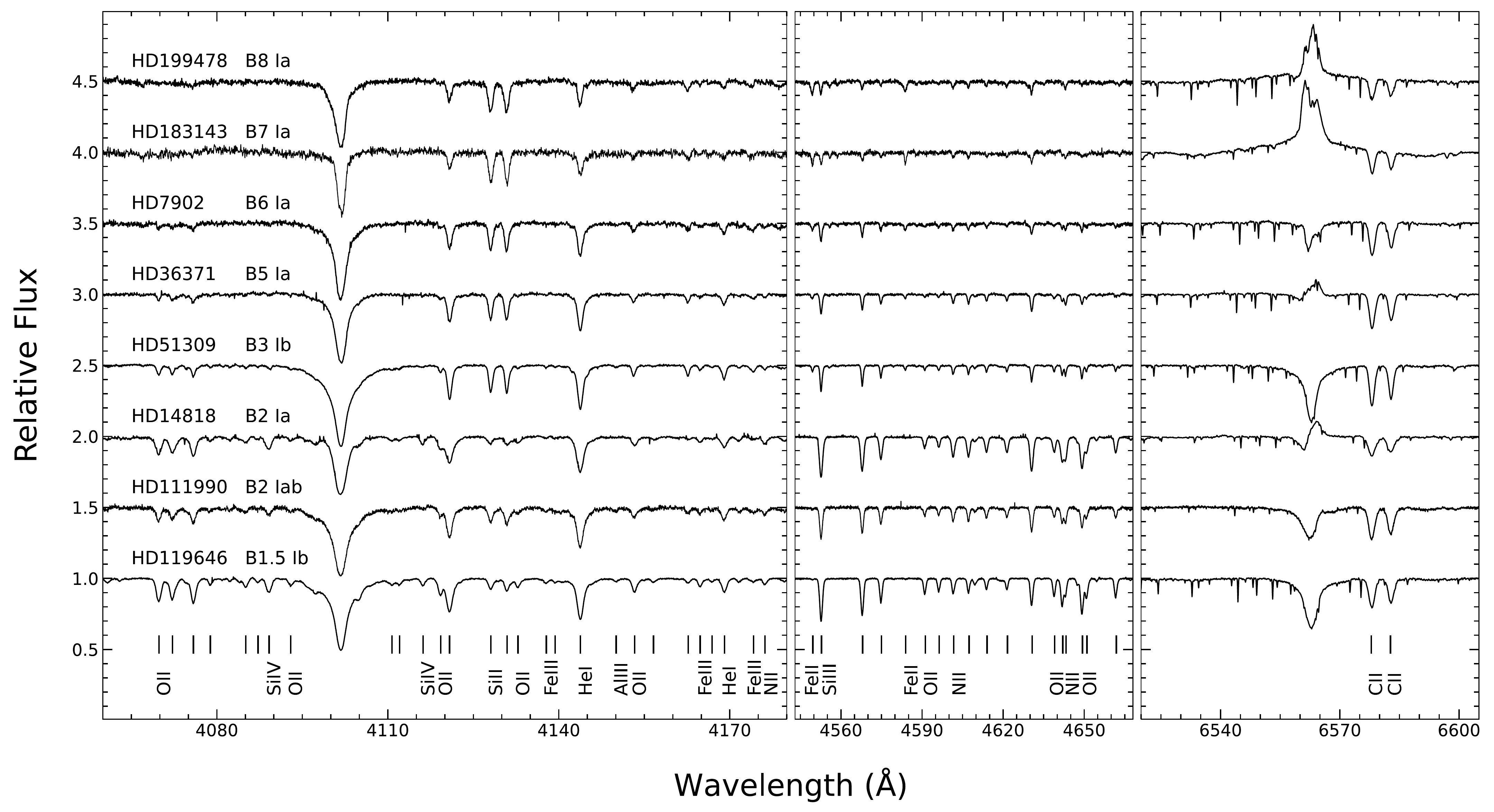}}
      \caption{Subset of the analysed spectra ordered with respect to spectral type. 
      The three panels showcase spectral windows with prominent features in the spectra of B-type supergiants, see the 
      text for a discussion.} 
         \label{fig:spect_lum_showcase}
\end{figure*}

Examples of the analysed spectra can be seen in Fig.~\ref{fig:spect_lum_showcase}, to demonstrate the data quality 
achieved for the present work. The figure focuses on three diagnostic wavelength regions: the window around
H$\delta$ with \ion{Si}{ii} and \ion{Si}{iv} lines plus several \ion{He}{i} and \ion{O}{ii} lines, among others,
the region of the \ion{Si}{iii} triplet plus numerous \ion{O}{ii} lines and on the wind-affected H$\alpha$ line with 
the adjacent strong \ion{C}{ii} doublet. We note how the density of spectral lines increases towards the earlier spectral 
types. We also note that objects at luminosity class Ib show nearly symmetric H$\alpha$ absorption, that is a 
negligible stellar wind, while the wind gives rise to pronounced H$\alpha$ emission at luminosity~class~Ia, while H$\delta$ is essentially symmetric at the luminosities covered here.

\begin{table}
\caption{IUE spectrophotometry used in the present work.}
\label{table:IUE_data}      
\centering  
    \resizebox{\linewidth}{!}{\small\begin{tabular}{llcccc} 
\hline\hline                
ID\,\# & Object    & SW     & Date       & LW     & Date       \\ \hline
2     & HD14818  & P18658 & 1982-11-26 & R14722 & 1982-11-26 \\
5     & HD183143 & P06550 & 1979-09-18 & R05637 & 1979-09-20 \\
8     & HD199478 & P07596 & 1980-01-07 & R06573 & 1980-01-07 \\
9     & HD51309  & P13936 & 1981-05-09 & R10551 & 1981-05-09 \\
10    & HD111990 &  ...   &   ...      & P13362 & 1988-06-05 \\
12    & HD125288 & P19460 & 1983-03-14 & R15489 & 1983-03-14 \\
13    & HD159110 & P45210 & 1992-07-22 & P21218 & 1991-09-11 \\
14    & HD164353 & P10172 & 1980-09-18 & R08836 & 1980-09-18\\\hline
\end{tabular}}
\end{table}

Besides the optical spectra, additional archival photometric data and UV spectrophotometry were collected to
establish the objects' spectral energy distributions (SEDs). For all analysed objects, optical photometry in the 
Johnson $U$, $B$, and $V$ bands by \cite{Mermilliod97} was adopted, $J$, $H$, and $K$ magnitudes from the Two Micron All 
Sky Survey \citep[2MASS,][]{2MASS2006} and $W1$ to $W4$ IR-photometric data from the Wide-field Infrared Survey Explorer 
(WISE) mission, from the ALLWISE data release \citep{ALLWISE_vizier}.  For a thorough comparison in the ultraviolet 
wavelength range, spectrophotometry taken by the International Ultraviolet Explorer (IUE) were preferred in our 
analysis. The designation and observation date for each IUE-spectrum used in the analysis are given in 
Table~\ref{table:IUE_data}. For both short-wavelength (SW, $\lambda\lambda$1150-1978\,{\AA}) and long-wavelength data 
(LW, $\lambda\lambda$1851-3347\,{\AA}) low-resolution spectrophotometry taken with the large aperture was favoured; in 
cases where only high-resolution data were available, the spectra were artificially degraded in resolution for the 
analysis. Whenever possible, SW and LW data observed close in time were employed. 

For several of the stars of our sample, IUE data was either unavailable or inconsistent with the optical and infrared 
photometry (possibly because of a misalignment of the aperture). In these cases photometric measurements by the 
Astronomical Netherlands Satellite \citep[ANS,][]{wesseliusetal82} or from the Belgian/UK Ultraviolet Sky Survey 
Telescope \citep[S2/68,][]{Thompson95} on board the European Space Research Organisation (ESRO) TD1 satellite were 
used.

%-------------------------------------------------------------------
\section{Model atmospheres and spectrum synthesis}\label{section:model_atmospheres}
Our methodology for the analysis of B-type supergiants is based on a hybrid non-LTE approach of calculating 
static, plane-parallel line-blanketed LTE model atmospheres, which serve as the basis of non-LTE line formation 
computations. The basic approach was outlined by \citet{Przybillaetal06} where its potential to accurately 
reproduce all relevant spectral features of late B- and early A-type (BA-type) supergiants was shown. This methodology
was validated in a direct comparison with full non-LTE hydrodynamic line-blanketed model atmospheres 
\citep{NiSi11} and was used to derive high-precision atmospheric and fundamental stellar parameters and abundances for 
many chemical species in early B-type MS stars \citep{NiPr12,NiPr14,Irrgangetal14}. Moreover, the hybrid 
approach is applicable to analyses of a wide range of other B-type stars, such as subdwarf B-stars 
\citep{Przybillaetal06b,Schaffenrothetal21}, MS Bp \citep{Przybillaetal08a}, He-strong stars 
\citep{Przybillaetal16,Przybillaetal21}, and supergiant extreme helium stars \citep{Kupferetal17}.
In the following, we therefore briefly recap the basic principles and the model codes and will 
concentrate on new aspects relevant for the present work.

\begin{table}
\caption{Model atoms for non-LTE calculations with {\sc Detail}.}             
\label{table:modelatoms}      
\centering                        
{\small
\begin{tabular}{llll}        
\hline\hline
Ion                         & Terms           & Transitions    & Reference \\ \hline
H                           & 20              & 190            & {[}1{]}   \\
He\,{\sc i}                 & 29+6            & 162            & {[}2{]}   \\
C\,{\sc ii/iii}             & 68/70           & 425/373        & {[}3{]}   \\
N\,{\sc i/ii}               & 89/77           & 668/462        & {[}4{]}   \\
O\,{\sc i/ii}               & 51/176+2        & 243/2559       & {[}5{]}   \\
Ne\,{\sc i/ii}              & 153/78          & 952/992        & {[}6{]}   \\
Mg\,{\sc ii}                & 37              & 236            & {[}7{]}   \\
Al\,{\sc ii/iii}            & 54+6/46+1       & 378/272        & {[}8{]}   \\
Si\,{\sc ii/iii/iv}         & 52+3/68+4/33+2  & 357/572/242    & {[}9{]}   \\
S\,{\sc ii/iii}             & 78/21           & 302/34         & {[}10{]}  \\
\ion{Ar}{ii}                & 56              & 596            & {[}11{]}  \\
Fe\,{\sc ii/iii/iv}         & 265/60+46/65+70 & 2887/2446/2094 & {[}12{]}\\\hline
\end{tabular}
\tablebib{[1] \cite{PrBu04}; [2] \cite{przybilla05}; [3]~\cite{NiPr06}, \cite{NiPr08}; [4] \cite{PrBu01}; [5] \cite{Przybillaetal00}, Przybilla \& Butler (in prep.); [6]~\cite{MoBu08}; [7] \cite{Przybillaetal01a}; [8] Przybilla (in prep.); [9] Przybilla \& Butler (in prep.); [10] \citet{Vranckenetal96}, updated; [11] Butler (in prep.); [12] \cite{Becker98}, \cite{Moreletal06}.
}}
\end{table}

\subsection{Models and programmes}
 The LTE line-blanketed model atmospheres used in this work were calculated with the \,{\sc Atlas12} code 
 \citep{kurucz05}, which assumes plane-parallel geometry, a stationary and hydrostatic stratification, and chemical
 homogeneity. In contrast to the previous version {\sc Atlas9} \citep[which is still required to provide converged starting models]{kurucz93} it does not rely on pretabulated opacity
 distribution functions (ODFs), but evaluates the opacities via opacity sampling (OS). 
 The code thus facilitates model atmospheres to be calculated for freely specified input abundances and microturbulent 
 velocities, including the turbulent pressure in the hydrostatic equation for a self-consistent solution. 
 
 The LTE model atmospheres were then used to compute non-LTE level population densities via an updated and extended 
 version of \,{\sc Detail} 
 \citep{giddings81} by solving the coupled radiative transfer and statistical equillibrium equations adopting an
 accelerated lambda iteration scheme by \cite{RyHu91}, and considering line blocking based on the Kurucz' OS scheme. 
 State-of-the-art model atoms were employed, as summarised in Table \ref{table:modelatoms}. There, for each chemical 
 species the considered ions are listed, the number of explicit terms (+ superlevels) and radiative bound-bound 
 transitions, and references. All model atoms are completed by the ground term of the next higher ionisation stage not 
 indicated here.
 
 While most of the model atoms were used previously in other studies, a new model atom for \ion{O}{ii} was employed in the 
 present work for the first time. We describe it briefly. Level energies were adopted from 
 \citet{Martinetal93} and combined into 176 LS-coupled terms up to principal quantum number $n$\,=\,8 and the levels for
 $n$\,=\,9 combined into two superlevels, one each for the doublet and quartet spin systems. Oscillator strengths and 
 photoionisation cross-sections were for the most part adopted from the Opacity Project \citep[OP, e.g.][]{Seatonetal94}, with several 
 improved data taken from \citet{Wieseetal96}, and supplemented by Kurucz' recently computed oscillator 
 strengths\footnote{\url{http://kurucz.harvard.edu/atoms.html}} for missing transitions. Electron impact-excitation data
 for a large number of transitions were available from the ab-initio calculations of \citet{Tayal07} and 
 \citet{Maoetal20}. Missing data were provided by use of Van Regemorter's formula \citep{vanRegemorter62} for 
 radiatively permitted or Allen's formula \citep{Allen73} for forbidden transitions. All collisional ionisation data 
 were provided via the Seaton formula \citep{Seaton62} using OP photoionisation threshold cross-sections or hydrogenic 
 values.
 
 \begin{figure}
\centering
    \resizebox{0.96\linewidth}{!}{\includegraphics[width=\hsize]{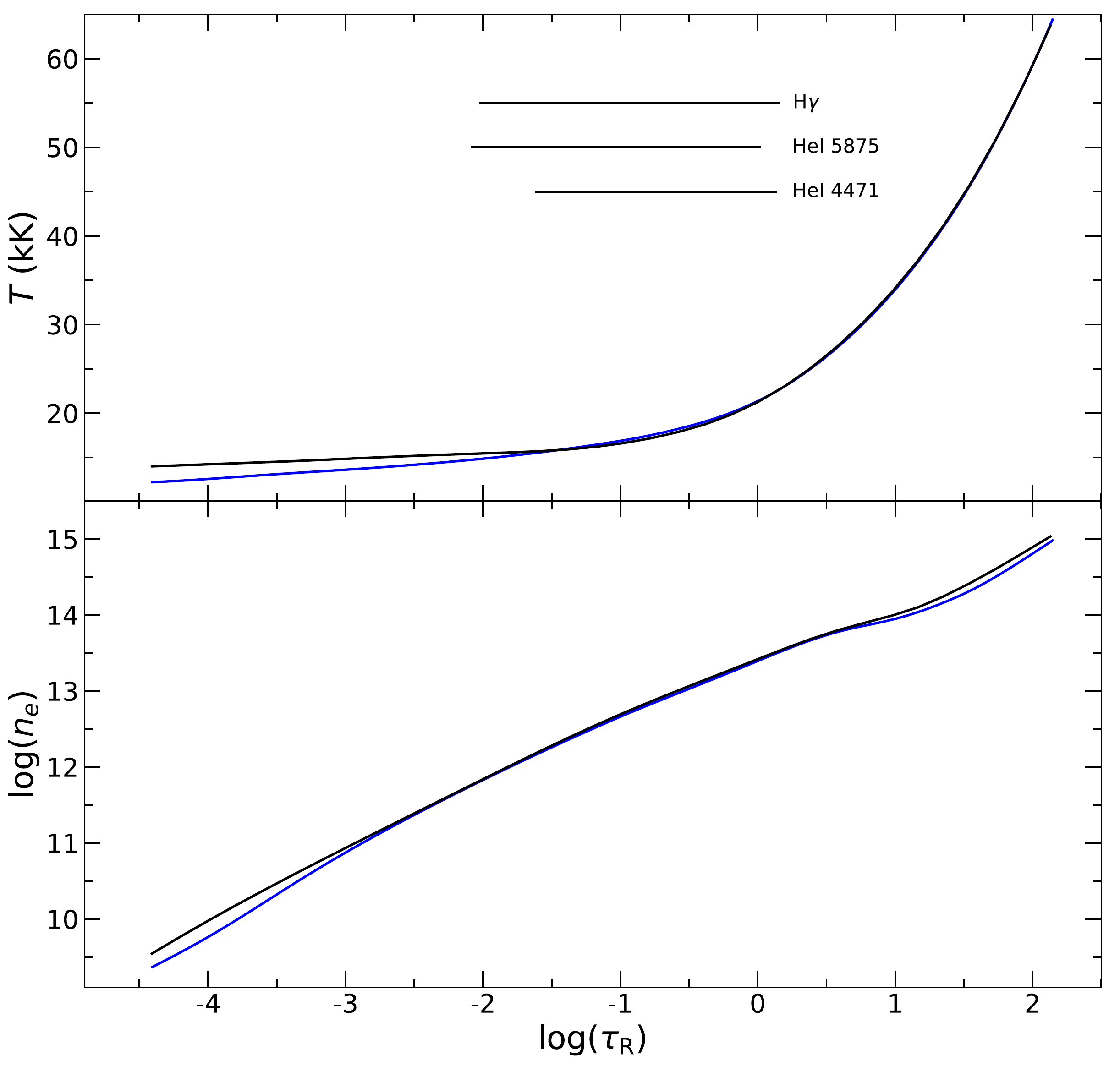}}
    \caption{Comparison of an {\sc Atlas12} (blue) and {\sc Tlusty} (black lines) atmospheric structures for a model with $T_\mathrm{eff}$\,=\,20\,000\,K, $\log g$\,=\,2.50, and $\xi$\,=\,10\,km\,s$^{-1}$. The upper and lower panel depict the run of temperature and the electron density, respectively, as a function of the Rosseland optical depth $\tau_{\mathrm{R}}$. The line formation regions of H$\gamma$ and two lines of \ion{He}{i} are indicated in the upper panel. See the text for a discussion.}
        \label{fig:tlusty_atlas12_structure_20k250v10}
\end{figure}
 
 Finally, synthetic spectra based on the non-LTE population numbers were calculated using an updated and extended 
 version of {\sc Surface} \citep{BuGi85}, employing refined fine-structure transition data and line-broadening theories. 
 Oscillator strengths from \citet{Wieseetal96} and Kurucz were replaced by data computed based on the multi-configuration Hartree-Fock 
 method by \citet{FFT04} for \ion{O}{ii} \citep[as for other elements and ions, also accounting for data from][]{FFTI06}, which was 
 decisive in achieving the close match with observations.
 For both {\sc Detail} and {\sc Surface} an occupation probability formalism \citep{HuMi88} -- as realised by 
 \citet{Hubenyetal94} -- was considered for hydrogen, in order to facilitate a better modelling of the series limits.
 
 Grids of synthetic spectra were calculated with {\sc Atlas12}, {\sc Detail}, and {\sc Surface} -- abbreviated as {\sc Ads} in
 the following -- for the entire parameter space of B-type supergiants. For the primary
 analysis of Balmer-lines and ionisation equilibria of all metals, effective temperatures $T_\mathrm{eff}$ were varied 
 from 11\,000 to 23\,000\,K in steps of 500 to 700\,K, logarithmic surface gravities $\log g$ from 1.70 to 3.70 (in
 cgs-units) in steps of 0.2\,dex, and elemental abundances in steps of 0.2\,dex, centred on cosmic abundance standard
 values \citep{NiPr12,Przybillaetal08b,Przybillaetal13}. For nitrogen, much higher values up to 1\,dex above standard
 were covered because of the expected enrichment. Microturbulent velocities $\xi$ were varied with increments of 
 4\,km\,s$^{-1}$ initially and refined later to as low as 1~km\,s$^{-1}$. The analysis was carried out -- depending on 
 the convergence of the model atmospheres -- on grids ranging from 0 up to 16\,km\,s$^{-1}$ in microturbulence, that is subsonic velocities.
 
 \begin{figure*}
\centering 
    \resizebox{0.78\textwidth}{!}{\includegraphics{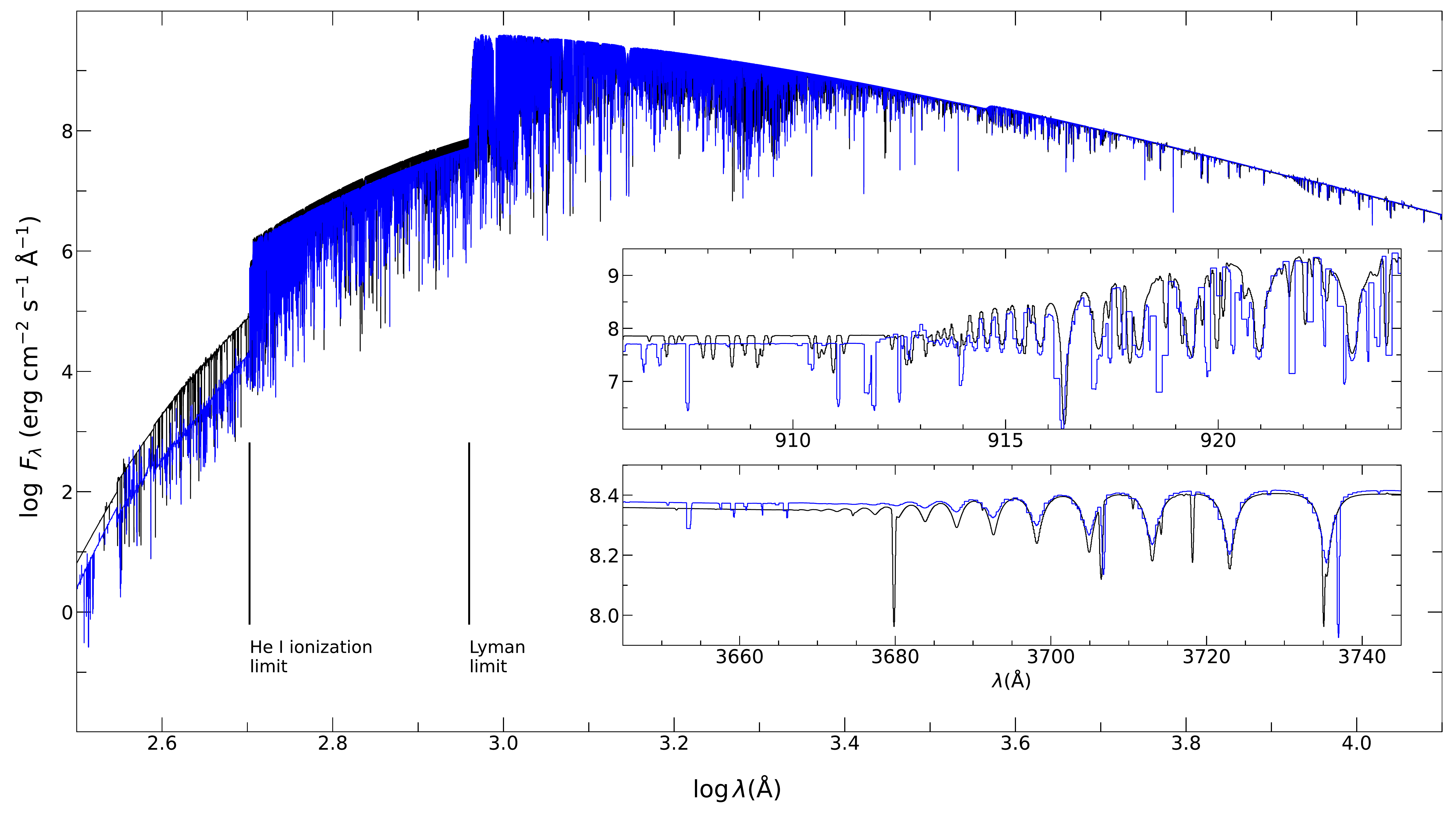}}
      \caption{Comparison of {\sc Detail} (blue) and {\sc Tlusty} (black lines) SEDs for a model with $T_\mathrm{eff}$\,=\,20\,000\,K, $\log g$\,=\,2.50, and $\xi$\,=\,10\,km\,s$^{-1}$. The locations of the Lyman- and He\,{\sc i} ionisation limits are indicated towards the lower left. The insets focus on the Lyman and Balmer jumps, respectively.}
         \label{fig:tlusty_atlas12_sed_20k250v10}
\end{figure*}
 
 We employed the Spectral Plotting and Analysis Suite ({\sc Spas}, \citealt{Hirsch09}) to compare the synthetic 
 and observed spectra. The programme allows instrumental, (radial-tangential) macroturbulent and 
 rotational broadening to be applied flexibly to the models and can be used to interpolate to the actual parameters with bi-cubic splines and fit up to three different parameters on the pre-calculated grid.
 To achieve this task, the programme employs the downhill simplex algorithm \citep{NeMe65} to find minima in the 
 $\chi^2$-landscape.

\subsection{Comparison with full non-LTE models}
Non-LTE effects gain in importance for more intense radiation fields (i.e. with increasing $T_\mathrm{eff}$) and reduced
collision rates (i.e. with decreasing particle density). The atmospheric structures of B-type supergiants are therefore
likely to be subject to non-LTE effects. Our hybrid non-LTE approach will only be successful if solutions from 
full non-LTE modelling can be closely recovered. 

A comparison of a full non-LTE model atmosphere for solar metallicity \citep{GrSa98} as adopted from the BSTAR grid \citep{LaHu07} that
was computed using the  {\sc Tlusty} code \citep{HuLa95} with an {\sc Atlas12} structure is shown in
Fig.~\ref{fig:tlusty_atlas12_structure_20k250v10}, for model parameters $T_\mathrm{eff}$\,=\,20\,000\,K, 
$\log g$\,=\,2.50, and $\xi$\,=\,10\,km\,s$^{-1}$. The temperature $T$ and electron density $n_\mathrm{e}$ stratification
as  a function of Rosseland optical depth $\tau_\mathrm{R}$ is shown. Line-formation depths for some
of the strongest diagnostic spectral features in the optical spectrum are also indicated, with the bulk of the metal lines being formed towards the inner boundary of this region. We want to note that the metallicity of the
{\sc Atlas12} model \citep[computed explicitly for abundances according to][]{GrSa98} was reduced
by 0.2\,dex in order to account empirically for non-LTE effects on the line blanketing. At the same metallicity, supergiant atmospheres in LTE and non-LTE show different temperature gradients because of the different amount of backwarming because of line blanketing and blocking. Empirically, a reduction of metallicity of LTE atmospheres by 0.2\,dex can compensate this differences (see Fig.~\ref{fig:tlusty_atlas12_structure_20k250v10}). The necessity for such an adjustment also follows on observational grounds. In order to reproduce the observed spectral lines, the real temperature gradient in the stellar atmosphere has to be matched by the model, as the different formation depths from line cores to the wings near the continuum-forming layers of the entire ensemble of the lines map the temperature 
(and density) structure in the atmosphere in detail. Achieving a match between observation and model as shown in the figures in Appendix~\ref{section:appendixA} requires the reproduction of the actual atmospheric structure by the model. To reproduce the observed SEDs, in particular for the cases where IUE spectrophotometry is available, also requires the reduction of the overall {\sc Atlas} atmosphere's metallicity.
Otherwise the line absorption in the UV is stronger than observed. Thus, the empirical metallicity adjustment mimics non-LTE effects on the line opacity. 
The effect of a reduction of metalli\-city by 0.2\,dex in the LTE model can be applied globally to supergiant models covering the range of effective temperatures investigated here, and only  diminishes for models towards the main sequence.
It can even be extended to early B-type supergiants as tested for a model with $T_\mathrm{eff}$\,=\,27\,000\,K, 
$\log g$\,=\,3.00, and $\xi$\,=\,10\,km\,s$^{-1}$. In all cases the agreement of the adapted {\sc Atlas12} 
stratifications with the {\sc Tlusty} structures is good throughout the photospheric line-formation depths, 
with differences less than 2\% in $T$ and 8\% in $n_\mathrm{e}$ and only starts to deviate more for $\log \tau_\mathrm{R}$\,$<$\,$-$2, where effects of the mass outflow would start to lead to departures from hydrostatic 
equilibrium in a real B-type supergiant atmosphere anyway. 

\begin{figure*}[!htb]
\centering 
    \includegraphics[width=0.81\textwidth]{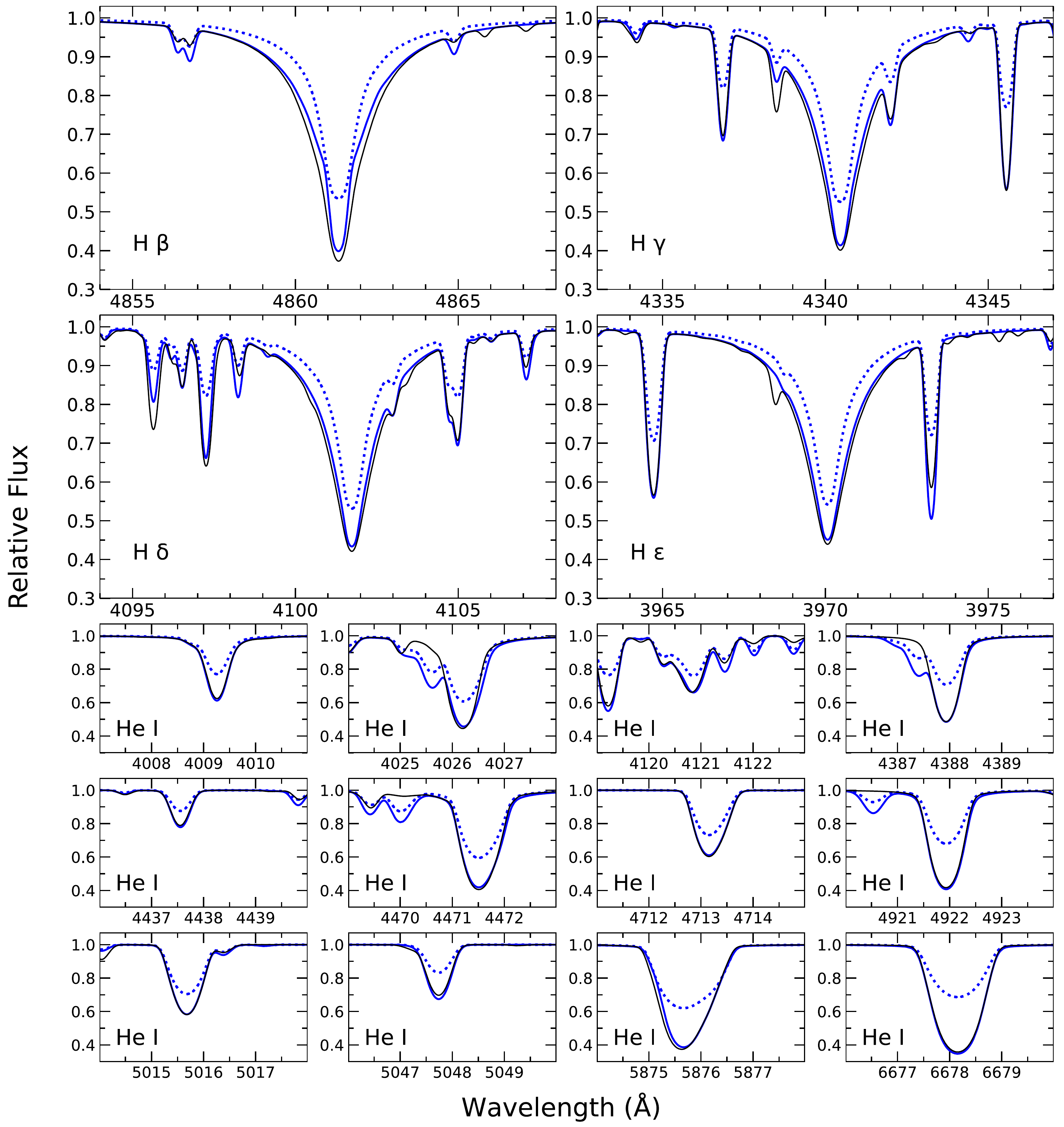}
        \caption{Comparison of Balmer lines H$\beta$, H$\gamma$, H$\delta$, H$\varepsilon$, and selected lines of He\,{\sc i} in synthetic spectra ($T_\mathrm{eff}$\,=\,20\,000\,K, $\log g$\,=\,2.50, and $\xi$\,=\,10\,km\,s$^{-1}$) calculated with {\sc Ads} (blue lines) and {\sc Tlusty/Synspec} (black lines). The dotted blue lines correspond to a model computed with {\sc Surface} only on an {\sc Atlas12} atmospheric structure, i.e. representing an
        LTE solution using  otherwise the same input data as in the {\sc Ads} non-LTE model.}
    \label{fig:tlusty_atlas12_HHE_20k250v10}
\end{figure*}

The comparison of the {\sc Tlusty} and {\sc Detail} non-LTE SEDs for the above parameters is shown in 
Fig.~\ref{fig:tlusty_atlas12_sed_20k250v10}. The agreement longward of the Lyman limit is excellent overall, with the 
differences amounting to only a few percent. Also the hydrogen series limits (see the insets in
Fig.~\ref{fig:tlusty_atlas12_sed_20k250v10}) resemble each other closely because the same occupation probability formalism
is employed in both codes. Larger differences occur at wavelengths below the Lyman limit, and in particular below the 
\ion{He}{i} ionisation limit (locations indicated in Fig.~\ref{fig:tlusty_atlas12_sed_20k250v10}), where {\sc Tlusty} predicts (significantly) higher ionising fluxes. The atmospheric layers
that emit this extreme-ultraviolet radiation are located in the outermost regions of the model atmosphere. 
Consequently the differences are not relevant for the photospheric lines investigated here. Moreover, as these layers 
are not in hydrostatic equilibrium in real B-type supergiant atmospheres, the predictive power of both models presented 
here is limited and would be better investigated with hydrodynamical stellar atmosphere models.

A further comparison of profiles for a selection of diagnostic hydrogen Balmer and \ion{He}{i} lines as calculated by 
{\sc Tlusty/Synspec} and {\sc Ads} for the above atmospheric parameters is shown in 
Fig.~ \ref{fig:tlusty_atlas12_HHE_20k250v10}. The match between the two non-LTE synthetic spectra for these two chemical
species is excellent except for some fine details. These concern the line cores of the Balmer lines, with the
differences diminishing towards the higher series members, and some of the forbidden components of the \ion{He}{i} 
lines, which are explained by the use of different broadening tables. However, the corresponding LTE model shows much 
weaker lines throughout, with the equivalent widths differing by factors of up to two to three. Overall, the 
differences increase towards the red. Pure LTE modelling is inapplicable for quantitative analyses of B-type
supergiants.

In particular, the panels in Fig.~\ref{fig:tlusty_atlas12_HHE_20k250v10} that show the Balmer lines cover a
wider wavelength range and also depict spectral lines from other chemical species. While most of these cases show only moderate differences between the two non-LTE models, some lines are noticeably discrepant. However, a straightforward comparison
of these should not be made, because -- in contrast to H and He, with their rather well-established atomic input data --
most of the differences stem from the different atomic data used and the different assumptions made in the 
construction of the model atoms that were used in the two approaches. A detailed discussion of these aspects for 
the case of OB-type main-sequence stars was presented by \citet{Przybillaetal11}, which we do not repeat here.
The basic conclusion is that the ability of different models to reproduce observations in a consistent way is decisive. 

   \begin{figure}
   \centering
       \resizebox{0.9\linewidth}{!}{\includegraphics[width=\hsize]{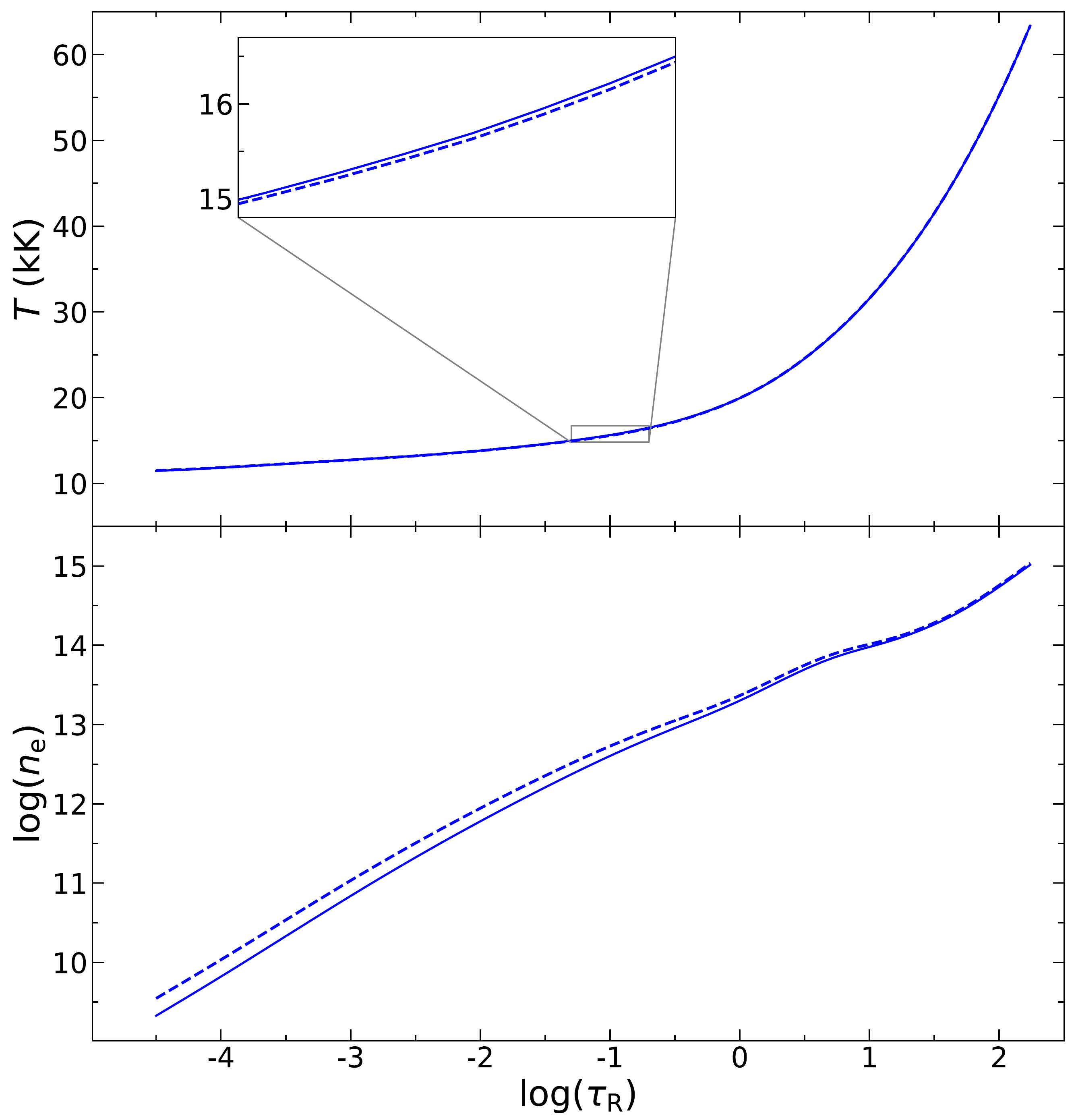}}
      \caption{Effects of turbulent pressure on the atmospheric stratification. Upper panel: temperature stratification,
      lower panel: electron density as a function of the logarithmic Rosseland optical depth. Displayed are 
      {\sc Atlas12} stratifications computed with (full line) and without considering turbulent pressure (dashed line)
      for $T_\mathrm{eff}$\,=\,18\,600\,K and $\log g$\,=\,2.45, i.e. the atmospheric parameters for HD~14818.}
         \label{fig:structure_turbulence}
   \end{figure}

\subsection{Turbulent pressure}\label{subsection:turb_pressure}
The {\sc Atlas12} code allows the effects of turbulent motions with velocity $\varv_{\mathrm{turb}}$ (i.e. the microturbulent velocity) on the model atmosphere computations to be taken into account. An additional turbulent 
pressure $P_{\mathrm{turb}}$ term is considered in the hydrostatic equilibrium equation in the form of
\begin{equation}\label{eq:turb_press}
P_{\mathrm{turb}} = \frac{\rho \varv_{\mathrm{turb}}^2}{2},
\end{equation}
where $\rho$ is the atmospheric density. This additional term increases in importance for stars approaching the Eddington 
limit because of the diminishing r\^ole of the gas pressure, and for increasing $\varv_\mathrm{turb}$.
Since it is possible to enable and disable turbulent pressure in the model specification of \,{\sc Atlas12}, we can directly
compare the effects of this term on the atmospheric structure and the synthetic spectra, while keeping all other 
parameters fixed. As a test, we chose the sample star HD~14818, at $T_\mathrm{eff}$\,=\,18\,600\,K, 
$\log g$\,=\,2.45, and a derived high luminosity, $\log L/L_\sun$\,=\,5.41. We expected to find a maximised impact on the model atmospheric structure because of its large $\xi$\,=\,14\,km\,s$^{-1}$.

Figure~\ref{fig:structure_turbulence} visualises the run of temperature $T$ (upper panel) and the logarithmic electron 
density $n_\mathrm{e}$ (lower panel) as a function of log-scale Rosseland optical depth $\tau_\mathrm{R}$ in the model
atmosphere of HD~14818 for the two cases of turbulent pressure switched on and off, respectively.
While the temperature hardly changes, with a maximum difference of about 50\,K (being higher in the model with 
turbulent pressure), the electron density is noticeably lower for $\log \tau_\mathrm{R}$\,$<$\,0 when turbulent pressure
is considered because of the more extended atmosphere.
Here, the absolute difference is about 0.12\,dex in $\log n_\mathrm{e}$ at $\log \tau_\mathrm{R}$\,=\,$-1$.

\begin{figure*}
\centering 
    \resizebox{0.7\textwidth}{!}{\includegraphics{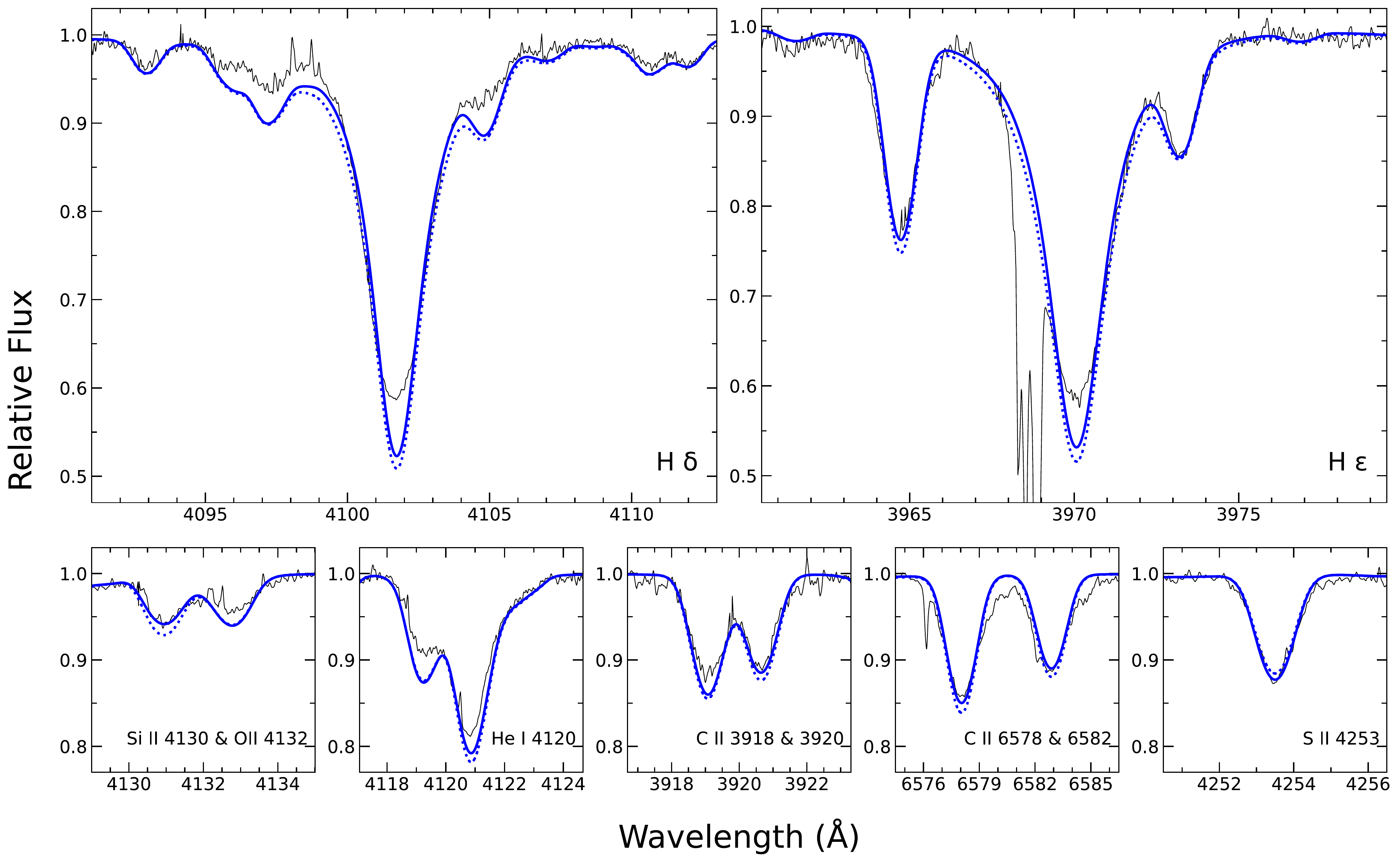}}
      \caption{Effects of turbulent pressure on line profiles. The solid blue line depicts the best fitting synthetic 
      spectrum for several diagnostic lines in the observed spectrum of HD~14818 (black) derived from model atmospheres 
      that account for turbulent pressure. The dotted blue line depicts the same solution without assuming turbulent pressure. The strong line in the blue wing of H\,$\varepsilon$ that is absent in the model is the interstellar~Ca~H line.}
         \label{fig:turbulent_pressure_lines}
\end{figure*}

Figure \ref{fig:turbulent_pressure_lines} shows the effects of the models with and without turbulent pressure for otherwise
identical parameters on various spectral line profiles.
It can be seen that for the fitted lines of hydrogen (H$\delta$ and H$\varepsilon$) the decreased density in the 
atmospheres with turbulent pressure corresponds to reduced pressure broadening of the Balmer line wings. Conversely, the model without turbulent pressure appears like a model with increased pressure broadening corresponding to the effect of an increase in surface gravity of about $\Delta \log g \approx 0.05$ dex. A 
systematic effect can also be detected in lines of helium and some metallic lines (\ion{Si}{ii}, \ion{C}{ii}, and \ion{S}{ii}) which
mostly show enhanced line strength for models without turbulent pressure (the exception being the sulphur line at 
4253\,{\AA})\footnote{We note that the panels depicting H$\delta$, \ion{Si}{ii} $\lambda$4130 and \ion{He}{i} $\lambda$4120 show lines of (and blends with) \ion{O}{ii} that systematically suggest a lower oxygen abundance. Originating from two multiplets sharing the same lower energy term, these lines appear too strong throughout the sample stars and were therefore excluded from the further quantitative analysis.}. The effect stems from a shift in the ionisation balance, yielding a higher degree of ionisation
in the model that accounts for microturbulent pressure. This amounts to a reduction of equivalent widths of individual \ion{Si}{ii} lines by $\sim$15 to 25\% for the example of HD~14818 (the equivalent width of \ion{Si}{ii} $\lambda$4130\,{\AA} in Fig.~\ref{fig:turbulent_pressure_lines} is e.g. reduced by 16\%), while the lines of the main ionisation stage \ion{Si}{iii} remain essentially unchanged, and \ion{Si}{iv} lines experience a slight strengthening by $<$5\% in equivalent width.
This impacts the atmospheric parameter and abundance determination to some small, but systematic, degree. Turbulent pressure is therefore considered in all analyses in the present work.

\section{Spectral analysis}\label{section:spectral_analysis}
\subsection{Atmospheric parameter and abundance determination}
The basic atmospheric parameters were determined via an analysis of the spectral features of multiple ionisation stages
of seven different chemical species (C, N, O, Ne, Al, Si, and Fe) as well as the analysis of the neutral helium lines 
and the Balmer lines of hydrogen. These parameters, effective temperature $T_{\mathrm{eff}}$, surface gravity $\log g$,
helium number fraction $y$, microturbulent velocity $\xi$, projected rotational velocity $\varv \sin i$, macroturbulence
$\zeta$ as well as the elemental abundances $\varepsilon\left(X\right)$\,=\,$\log\left(X/\mathrm{H}\right)$\,+\,12, 
were derived on the basis of spectrum synthesis, aiming at the reproduction of the detailed line profiles of features
spanning the entire observed visual to near-infrared spectra. An iterative approach was employed to overcome 
ambiguities because of strong correlations, until all parameters were constrained in a consistent way and a single global
solution for the synthetic spectrum was found that matches closely the entire observed spectrum.

\subsubsection{Effective temperature and surface gravity}
In order to begin the analysis, an initial guess on the basis of spectral type and the shape and strength of the 
Balmer lines suffices for an estimation to within $\Delta T_{\mathrm{eff}}$\,$<$\,1500\,K and 
$\Delta \log g$\,<\,0.4\,dex. Ambiguities in these two parameters arise due to their counteracting nature: in the regime
of B-type supergiants, the Balmer lines grow weaker with $T_{\mathrm{eff}}$ as hydrogen is increasingly ionised, while 
increasing in strength with surface gravity due to the pressure broadening. Hence, multiple combinations of 
$T_{\mathrm{eff}}$ and $\log g$ fit the observations. This means that Balmer line fitting alone is insufficient 
for a thorough analysis. The problem is solved by independently constraining $T_\mathrm{eff}$ and $\log g$ 
using multiple ionisation equilibria of the studied elements, that is requiring that lines from the different ionisation stages of a chemical element are reproduced at the same elemental abundance value (within the mutual uncertainties). 
Table~\ref{tab:ionisation_equillibria} summarises which ionisation balances were employed for the analysis of the 
sample stars, sorted from highest to lowest $T_\mathrm{eff}$. Dots indicate that lines from the respective ionisation 
stage were analysed, the blue boxes then frame the achieved ionisation balance. Some combinations were useful throughout
the entire $T_\mathrm{eff}$-range, for example \ion{O}{i/ii} or \ion{Si}{ii/iii}, while other ionisation stages appear only 
towards the highest $T_\mathrm{eff}$-values, like \ion{C}{iii}, \ion{Ne}{ii} or \ion{Si}{iv} and others such as 
\ion{N}{i}, \ion{Al}{ii} or \ion{Fe}{ii} are no longer visible. Four to seven ionisation balances were matched simultaneously per 
star, with the tightest constraints occurring if three consecutive ionisation stages could be employed, as in the case
of \ion{Si}{ii/iii/iv}. Overall, ionisation balances are more sensitive to $T_\mathrm{eff}$-variations, while the 
Balmer lines are more sensitive to $\log g$-variations. The finally adopted values of effective temperature and surface
gravity, and their uncertainties, were then calculated as the arithmetic mean and standard deviation of the values 
implied by the individual indicators. In most of the sample objects, H$\alpha$ (see Fig.~\ref{fig:spect_lum_showcase} 
for examples) had to be excluded from the analysis because of line asymmetries or the occurrence of emission due to the 
stellar wind. In the most luminous stars, H$\beta$, H$\gamma$, and even H$\delta$ may show signs of influence from
the stellar wind and they were also omitted from the fitting process.

\begin{table}
\setlength{\tabcolsep}{2pt}
\caption{Ionisation balances used for the atmospheric parameter determination.}
\label{tab:ionisation_equillibria}
\centering   
\resizebox{20.5cm}{!}{\begin{tabular}{lclllllll}
\hline\hline
ID\,\# & $T_{\mathrm{eff}}$ & \ion{C}{ii/iii}                                                   & \ion{N}{i/ii}                                                     & \ion{O}{i/ii}                                                     & \ion{Ne}{i/ii}                                                     & \ion{Al}{ii/iii}                                                  & \ion{Si}{ii/iii/iv}                                                   & \ion{Fe}{ii/iii}                                                   \\
& kK\\
\hline
11   & 19.7               & \tikz\draw[black,fill=black] (0,0) circle (.5ex);~~~~                                                                                              & \tikzmark{top left 3}\tikz\draw[black,fill=black] (0,0) circle (.5ex);~~\tikz\draw[black,fill=black] (0,0) circle (.5ex);\tikzmark{bottom right 3}   & \tikzmark{top left 16}\tikz\draw[black,fill=black] (0,0) circle (.5ex);~~\tikz\draw[black,fill=black] (0,0) circle (.5ex);\tikzmark{bottom right 16} & \tikzmark{top left 27}\tikz\draw[black,fill=black] (0,0) circle (.5ex);~~\tikz\draw[black,fill=black] (0,0) circle (.5ex);\tikzmark{bottom right 27} & ~~~~\tikz\draw[black,fill=black] (0,0) circle (.5ex);                                                                                                & \tikzmark{top left 41}\tikz\draw[black,fill=black] (0,0) circle (.5ex);~~\tikz\draw[black,fill=black] (0,0) circle (.5ex);~~\tikz\draw[black,fill=black] (0,0) circle (.5ex);\tikzmark{bottom right 41} & ~~~~\tikz\draw[black,fill=black] (0,0) circle (.5ex);                                                                                                \\
13   & 19.5               & \tikzmark{top left 1}\tikz\draw[black,fill=black] (0,0) circle (.5ex);~~\tikz\draw[black,fill=black] (0,0) circle (.5ex);\tikzmark{bottom right 1} & \tikzmark{top left 2}\tikz\draw[black,fill=black] (0,0) circle (.5ex);~~\tikz\draw[black,fill=black] (0,0) circle (.5ex);\tikzmark{bottom right 2}   & \tikzmark{top left 15}\tikz\draw[black,fill=black] (0,0) circle (.5ex);~~\tikz\draw[black,fill=black] (0,0) circle (.5ex);\tikzmark{bottom right 15} & \tikz\draw[black,fill=black] (0,0) circle (.5ex);~~~~                                                                                                & \tikzmark{top left 29}\tikz\draw[black,fill=black] (0,0) circle (.5ex);~~\tikz\draw[black,fill=black] (0,0) circle (.5ex);\tikzmark{bottom right 29} & \tikzmark{top left 40}\tikz\draw[black,fill=black] (0,0) circle (.5ex);~~\tikz\draw[black,fill=black] (0,0) circle (.5ex);~~\tikz\draw[black,fill=black] (0,0) circle (.5ex);\tikzmark{bottom right 40} & \tikzmark{top left 54}\tikz\draw[black,fill=black] (0,0) circle (.5ex);~~\tikz\draw[black,fill=black] (0,0) circle (.5ex);\tikzmark{bottom right 54} \\
2    & 18.6               & \tikz\draw[black,fill=black] (0,0) circle (.5ex);~~~~                                                                                              & ~~~~\tikz\draw[black,fill=black] (0,0) circle (.5ex);                                                                                                & \tikzmark{top left 3}\tikz\draw[black,fill=black] (0,0) circle (.5ex);~~\tikz\draw[black,fill=black] (0,0) circle (.5ex);\tikzmark{bottom right 3}                                                                                                & \tikzmark{top left 28}\tikz\draw[black,fill=black] (0,0) circle (.5ex);~~\tikz\draw[black,fill=black] (0,0) circle (.5ex);\tikzmark{bottom right 28} & ~~~~\tikz\draw[black,fill=black] (0,0) circle (.5ex);                                                                                                & \tikzmark{top left 42}\tikz\draw[black,fill=black] (0,0) circle (.5ex);~~\tikz\draw[black,fill=black] (0,0) circle (.5ex);~~\tikz\draw[black,fill=black] (0,0) circle (.5ex);\tikzmark{bottom right 42} & ~~~~\tikz\draw[black,fill=black] (0,0) circle (.5ex);                                                                                                \\
10   & 17.2               & \tikz\draw[black,fill=black] (0,0) circle (.5ex);~~~~                                                                                              & ~~~~\tikz\draw[black,fill=black] (0,0) circle (.5ex);    & \tikzmark{top left 17}\tikz\draw[black,fill=black] (0,0) circle (.5ex);~~\tikz\draw[black,fill=black] (0,0) circle (.5ex);\tikzmark{bottom right 17} & \tikz\draw[black,fill=black] (0,0) circle (.5ex);~~~~                                                                                                & ~~~~\tikz\draw[black,fill=black] (0,0) circle (.5ex);  & \tikzmark{top left 43}\tikz\draw[black,fill=black] (0,0) circle (.5ex);~~\tikz\draw[black,fill=black] (0,0) circle (.5ex);~~\tikz\draw[black,fill=black] (0,0) circle (.5ex);\tikzmark{bottom right 43} & \tikzmark{top left 55}\tikz\draw[black,fill=black] (0,0) circle (.5ex);~~\tikz\draw[black,fill=black] (0,0) circle (.5ex);\tikzmark{bottom right 55} \\
9    & 15.6               & \tikz\draw[black,fill=black] (0,0) circle (.5ex);~~~~                                                                                              & \tikzmark{top left 5}\tikz\draw[black,fill=black] (0,0) circle (.5ex);~~\tikz\draw[black,fill=black] (0,0) circle (.5ex);\tikzmark{bottom right 5}   & \tikzmark{top left 18}\tikz\draw[black,fill=black] (0,0) circle (.5ex);~~\tikz\draw[black,fill=black] (0,0) circle (.5ex);\tikzmark{bottom right 18} & \tikz\draw[black,fill=black] (0,0) circle (.5ex);~~~~                                                                                                & \tikzmark{top left 30}\tikz\draw[black,fill=black] (0,0) circle (.5ex);~~\tikz\draw[black,fill=black] (0,0) circle (.5ex);\tikzmark{bottom right 30} & \tikzmark{top left 44}\tikz\draw[black,fill=black] (0,0) circle (.5ex);~~\tikz\draw[black,fill=black] (0,0) circle (.5ex);~~\tikz\draw[black,fill=black] (0,0) circle (.5ex);\tikzmark{bottom right 44} & \tikzmark{top left 56}\tikz\draw[black,fill=black] (0,0) circle (.5ex);~~\tikz\draw[black,fill=black] (0,0) circle (.5ex);\tikzmark{bottom right 56} \\
14   & 14.7               & \tikz\draw[black,fill=black] (0,0) circle (.5ex);~~~~                                                                                              & \tikzmark{top left 6}\tikz\draw[black,fill=black] (0,0) circle (.5ex);~~\tikz\draw[black,fill=black] (0,0) circle (.5ex);\tikzmark{bottom right 6}   & \tikzmark{top left 19}\tikz\draw[black,fill=black] (0,0) circle (.5ex);~~\tikz\draw[black,fill=black] (0,0) circle (.5ex);\tikzmark{bottom right 19} & \tikz\draw[black,fill=black] (0,0) circle (.5ex);~~~~                                                                                                &  \tikzmark{top left 4}\tikz\draw[black,fill=black] (0,0) circle (.5ex);~~\tikz\draw[black,fill=black] (0,0) circle (.5ex);\tikzmark{bottom right 4}   & \tikzmark{top left 45}\tikz\draw[black,fill=black] (0,0) circle (.5ex);~~\tikz\draw[black,fill=black] (0,0) circle (.5ex);\tikzmark{bottom right 45}                                                    & \tikzmark{top left 57}\tikz\draw[black,fill=black] (0,0) circle (.5ex);~~\tikz\draw[black,fill=black] (0,0) circle (.5ex);\tikzmark{bottom right 57} \\
4    & 14.6               & \tikz\draw[black,fill=black] (0,0) circle (.5ex);~~~~                                                                                              & \tikzmark{top left 7}\tikz\draw[black,fill=black] (0,0) circle (.5ex);~~\tikz\draw[black,fill=black] (0,0) circle (.5ex);\tikzmark{bottom right 7}   & \tikzmark{top left 20}\tikz\draw[black,fill=black] (0,0) circle (.5ex);~~\tikz\draw[black,fill=black] (0,0) circle (.5ex);\tikzmark{bottom right 20} & \tikz\draw[black,fill=black] (0,0) circle (.5ex);~~~~                                                                                                & \tikzmark{top left 31}\tikz\draw[black,fill=black] (0,0) circle (.5ex);~~\tikz\draw[black,fill=black] (0,0) circle (.5ex);\tikzmark{bottom right 31} & \tikzmark{top left 46}\tikz\draw[black,fill=black] (0,0) circle (.5ex);~~\tikz\draw[black,fill=black] (0,0) circle (.5ex);~~\tikz\draw[black,fill=black] (0,0) circle (.5ex);\tikzmark{bottom right 46} & \tikzmark{top left 58}\tikz\draw[black,fill=black] (0,0) circle (.5ex);~~\tikz\draw[black,fill=black] (0,0) circle (.5ex);\tikzmark{bottom right 58} \\
1    & 14.1               & \tikz\draw[black,fill=black] (0,0) circle (.5ex);~~~~                                                                                              & \tikzmark{top left 8}\tikz\draw[black,fill=black] (0,0) circle (.5ex);~~\tikz\draw[black,fill=black] (0,0) circle (.5ex);\tikzmark{bottom right 8}   & \tikzmark{top left 21}\tikz\draw[black,fill=black] (0,0) circle (.5ex);~~\tikz\draw[black,fill=black] (0,0) circle (.5ex);\tikzmark{bottom right 21} & \tikz\draw[black,fill=black] (0,0) circle (.5ex);~~~~                                                                                                & \tikzmark{top left 32}\tikz\draw[black,fill=black] (0,0) circle (.5ex);~~\tikz\draw[black,fill=black] (0,0) circle (.5ex);\tikzmark{bottom right 32} & \tikzmark{top left 47}\tikz\draw[black,fill=black] (0,0) circle (.5ex);~~\tikz\draw[black,fill=black] (0,0) circle (.5ex);\tikzmark{bottom right 47}                                                    & \tikzmark{top left 59}\tikz\draw[black,fill=black] (0,0) circle (.5ex);~~\tikz\draw[black,fill=black] (0,0) circle (.5ex);\tikzmark{bottom right 59} \\
7    & 14.0               & \tikz\draw[black,fill=black] (0,0) circle (.5ex);~~~~                                                                                              & \tikzmark{top left 9}\tikz\draw[black,fill=black] (0,0) circle (.5ex);~~\tikz\draw[black,fill=black] (0,0) circle (.5ex);\tikzmark{bottom right 9}   & \tikzmark{top left 22}\tikz\draw[black,fill=black] (0,0) circle (.5ex);~~\tikz\draw[black,fill=black] (0,0) circle (.5ex);\tikzmark{bottom right 22} & \tikz\draw[black,fill=black] (0,0) circle (.5ex);~~~~                                                                                                & \tikzmark{top left 33}\tikz\draw[black,fill=black] (0,0) circle (.5ex);~~\tikz\draw[black,fill=black] (0,0) circle (.5ex);\tikzmark{bottom right 33} & \tikzmark{top left 48}\tikz\draw[black,fill=black] (0,0) circle (.5ex);~~\tikz\draw[black,fill=black] (0,0) circle (.5ex);\tikzmark{bottom right 48}                                                    & \tikzmark{top left 60}\tikz\draw[black,fill=black] (0,0) circle (.5ex);~~\tikz\draw[black,fill=black] (0,0) circle (.5ex);\tikzmark{bottom right 60} \\
12   & 13.7               & \tikz\draw[black,fill=black] (0,0) circle (.5ex);~~~~                                                                                              & \tikzmark{top left 10}\tikz\draw[black,fill=black] (0,0) circle (.5ex);~~\tikz\draw[black,fill=black] (0,0) circle (.5ex);\tikzmark{bottom right 10} & \tikzmark{top left 23}\tikz\draw[black,fill=black] (0,0) circle (.5ex);~~\tikz\draw[black,fill=black] (0,0) circle (.5ex);\tikzmark{bottom right 23} & \tikz\draw[black,fill=black] (0,0) circle (.5ex);~~~~                                                                                                & \tikzmark{top left 34}\tikz\draw[black,fill=black] (0,0) circle (.5ex);~~\tikz\draw[black,fill=black] (0,0) circle (.5ex);\tikzmark{bottom right 34} & \tikzmark{top left 49}\tikz\draw[black,fill=black] (0,0) circle (.5ex);~~\tikz\draw[black,fill=black] (0,0) circle (.5ex);\tikzmark{bottom right 49}                                                    & \tikzmark{top left 61}\tikz\draw[black,fill=black] (0,0) circle (.5ex);~~\tikz\draw[black,fill=black] (0,0) circle (.5ex);\tikzmark{bottom right 61} \\
3    & 13.6               & \tikz\draw[black,fill=black] (0,0) circle (.5ex);~~~~                                                                                              & \tikzmark{top left 11}\tikz\draw[black,fill=black] (0,0) circle (.5ex);~~\tikz\draw[black,fill=black] (0,0) circle (.5ex);\tikzmark{bottom right 11} & \tikzmark{top left 24}\tikz\draw[black,fill=black] (0,0) circle (.5ex);~~\tikz\draw[black,fill=black] (0,0) circle (.5ex);\tikzmark{bottom right 24} & \tikz\draw[black,fill=black] (0,0) circle (.5ex);~~~~                                                                                                & \tikzmark{top left 35}\tikz\draw[black,fill=black] (0,0) circle (.5ex);~~\tikz\draw[black,fill=black] (0,0) circle (.5ex);\tikzmark{bottom right 35} & \tikzmark{top left 50}\tikz\draw[black,fill=black] (0,0) circle (.5ex);~~\tikz\draw[black,fill=black] (0,0) circle (.5ex);\tikzmark{bottom right 50}                                                    & \tikzmark{top left 62}\tikz\draw[black,fill=black] (0,0) circle (.5ex);~~\tikz\draw[black,fill=black] (0,0) circle (.5ex);\tikzmark{bottom right 62} \\
5    & 12.8               & \tikz\draw[black,fill=black] (0,0) circle (.5ex);~~~~                                                                                              & \tikzmark{top left 12}\tikz\draw[black,fill=black] (0,0) circle (.5ex);~~\tikz\draw[black,fill=black] (0,0) circle (.5ex);\tikzmark{bottom right 12} &\tikzmark{top left 66}\tikz\draw[black,fill=black] (0,0) circle (.5ex);~~\tikz\draw[black,fill=black] (0,0) circle (.5ex);\tikzmark{bottom right 66}& \tikz\draw[black,fill=black] (0,0) circle (.5ex);~~~~                                                                                                & \tikzmark{top left 36}\tikz\draw[black,fill=black] (0,0) circle (.5ex);~~\tikz\draw[black,fill=black] (0,0) circle (.5ex);\tikzmark{bottom right 36} & \tikzmark{top left 51}\tikz\draw[black,fill=black] (0,0) circle (.5ex);~~\tikz\draw[black,fill=black] (0,0) circle (.5ex);\tikzmark{bottom right 51}                                                    & \tikzmark{top left 63}\tikz\draw[black,fill=black] (0,0) circle (.5ex);~~\tikz\draw[black,fill=black] (0,0) circle (.5ex);\tikzmark{bottom right 63} \\
8    & 12.7               & \tikz\draw[black,fill=black] (0,0) circle (.5ex);~~~~                                                                                              & \tikzmark{top left 13}\tikz\draw[black,fill=black] (0,0) circle (.5ex);~~\tikz\draw[black,fill=black] (0,0) circle (.5ex);\tikzmark{bottom right 13} & \tikzmark{top left 25}\tikz\draw[black,fill=black] (0,0) circle (.5ex);~~\tikz\draw[black,fill=black] (0,0) circle (.5ex);\tikzmark{bottom right 25} & \tikz\draw[black,fill=black] (0,0) circle (.5ex);~~~~                                                                                                & \tikzmark{top left 37}\tikz\draw[black,fill=black] (0,0) circle (.5ex);~~\tikz\draw[black,fill=black] (0,0) circle (.5ex);\tikzmark{bottom right 37} & \tikzmark{top left 52}\tikz\draw[black,fill=black] (0,0) circle (.5ex);~~\tikz\draw[black,fill=black] (0,0) circle (.5ex);\tikzmark{bottom right 52}                                                    & \tikzmark{top left 64}\tikz\draw[black,fill=black] (0,0) circle (.5ex);~~\tikz\draw[black,fill=black] (0,0) circle (.5ex);\tikzmark{bottom right 64} \\
6    & 11.9               & \tikz\draw[black,fill=black] (0,0) circle (.5ex);~~~~                                                                                              & \tikzmark{top left 14}\tikz\draw[black,fill=black] (0,0) circle (.5ex);~~\tikz\draw[black,fill=black] (0,0) circle (.5ex);\tikzmark{bottom right 14} & \tikzmark{top left 67}\tikz\draw[black,fill=black] (0,0) circle (.5ex);~~\tikz\draw[black,fill=black] (0,0) circle (.5ex);\tikzmark{bottom right 67}& \tikz\draw[black,fill=black] (0,0) circle (.5ex);~~~~                                                                                                & \tikzmark{top left 38}\tikz\draw[black,fill=black] (0,0) circle (.5ex);~~\tikz\draw[black,fill=black] (0,0) circle (.5ex);\tikzmark{bottom right 38} & \tikzmark{top left 53}\tikz\draw[black,fill=black] (0,0) circle (.5ex);~~\tikz\draw[black,fill=black] (0,0) circle (.5ex);\tikzmark{bottom right 53}                                                    & \tikzmark{top left 65}\tikz\draw[black,fill=black] (0,0) circle (.5ex);~~\tikz\draw[black,fill=black] (0,0) circle (.5ex);\tikzmark{bottom right 65}\\ \hline
\end{tabular}
\DrawBox[thick,blue]{top left 1}{bottom right 1}
\DrawBox[thick,blue]{top left 2}{bottom right 2}
\DrawBox[thick,blue]{top left 3}{bottom right 3}
\DrawBox[thick,blue]{top left 4}{bottom right 4}
\DrawBox[thick,blue]{top left 5}{bottom right 5}
\DrawBox[thick,blue]{top left 6}{bottom right 6}
\DrawBox[thick,blue]{top left 7}{bottom right 7}
\DrawBox[thick,blue]{top left 8}{bottom right 8}
\DrawBox[thick,blue]{top left 9}{bottom right 9}
\DrawBox[thick,blue]{top left 10}{bottom right 10}
\DrawBox[thick,blue]{top left 11}{bottom right 11}
\DrawBox[thick,blue]{top left 12}{bottom right 12}
\DrawBox[thick,blue]{top left 13}{bottom right 13}
\DrawBox[thick,blue]{top left 14}{bottom right 14}
\DrawBox[thick,blue]{top left 15}{bottom right 15}
\DrawBox[thick,blue]{top left 16}{bottom right 16}
\DrawBox[thick,blue]{top left 17}{bottom right 17}
\DrawBox[thick,blue]{top left 18}{bottom right 18}
\DrawBox[thick,blue]{top left 19}{bottom right 19}
\DrawBox[thick,blue]{top left 20}{bottom right 20}
\DrawBox[thick,blue]{top left 21}{bottom right 21}
\DrawBox[thick,blue]{top left 22}{bottom right 22}
\DrawBox[thick,blue]{top left 23}{bottom right 23}
\DrawBox[thick,blue]{top left 24}{bottom right 24}
\DrawBox[thick,blue]{top left 25}{bottom right 25}
\DrawBox[thick,blue]{top left 27}{bottom right 27}
\DrawBox[thick,blue]{top left 28}{bottom right 28}
\DrawBox[thick,blue]{top left 29}{bottom right 29}
\DrawBox[thick,blue]{top left 30}{bottom right 30}
\DrawBox[thick,blue]{top left 31}{bottom right 31}
\DrawBox[thick,blue]{top left 32}{bottom right 32}
\DrawBox[thick,blue]{top left 33}{bottom right 33}
\DrawBox[thick,blue]{top left 34}{bottom right 34}
\DrawBox[thick,blue]{top left 35}{bottom right 35}
\DrawBox[thick,blue]{top left 36}{bottom right 36}
\DrawBox[thick,blue]{top left 37}{bottom right 37}
\DrawBox[thick,blue]{top left 38}{bottom right 38}
\DrawBox[thick,blue]{top left 40}{bottom right 40}
\DrawBox[thick,blue]{top left 41}{bottom right 41}
\DrawBox[thick,blue]{top left 42}{bottom right 42}
\DrawBox[thick,blue]{top left 43}{bottom right 43}
\DrawBox[thick,blue]{top left 44}{bottom right 44}
\DrawBox[thick,blue]{top left 45}{bottom right 45}
\DrawBox[thick,blue]{top left 46}{bottom right 46}
\DrawBox[thick,blue]{top left 47}{bottom right 47}
\DrawBox[thick,blue]{top left 48}{bottom right 48}
\DrawBox[thick,blue]{top left 49}{bottom right 49}
\DrawBox[thick,blue]{top left 50}{bottom right 50}
\DrawBox[thick,blue]{top left 51}{bottom right 51}
\DrawBox[thick,blue]{top left 52}{bottom right 52}
\DrawBox[thick,blue]{top left 53}{bottom right 53}
\DrawBox[thick,blue]{top left 54}{bottom right 54}
\DrawBox[thick,blue]{top left 55}{bottom right 55}
\DrawBox[thick,blue]{top left 56}{bottom right 56}
\DrawBox[thick,blue]{top left 57}{bottom right 57}
\DrawBox[thick,blue]{top left 58}{bottom right 58}
\DrawBox[thick,blue]{top left 59}{bottom right 59}
\DrawBox[thick,blue]{top left 60}{bottom right 60}
\DrawBox[thick,blue]{top left 61}{bottom right 61}
\DrawBox[thick,blue]{top left 62}{bottom right 62}
\DrawBox[thick,blue]{top left 63}{bottom right 63}
\DrawBox[thick,blue]{top left 64}{bottom right 64}
\DrawBox[thick,blue]{top left 65}{bottom right 65}
\DrawBox[thick,blue]{top left 66}{bottom right 66}
\DrawBox[thick,blue]{top left 67}{bottom right 67}}
\end{table}

\subsubsection{Helium abundance}\label{section:helium_fitting}
In the first step of the iterative procedure, the helium number density was set to the cosmic value of $y$\,=\,0.089 
\citep{NiPr12} in order to derive a satisfactory estimate for effective temperature and surface gravity. With these 
values, fitting of the weakest helium lines permitted a refined abundance estimate for helium with relative
uncertainties between $\delta y$\,$\coloneqq$\,$\Delta y/y$\,$\approx$\,5--15\% to be derived. Stronger lines were 
generally excluded from the analysis, as they are less sensitive to abundance changes.
Constraining the helium abundance is of importance not only per se. It strongly influences the molecular weight of the 
atmospheric elemental mixture and thereby changes the density and pressure stratification of the atmosphere, leading to 
changes in the derived surface gravity of up to $\Delta\log g$\,=\,0.05\,dex. Further changes of the helium abundance in
the  following steps of the iteration scheme (i.e. after correcting $T_{\mathrm{eff}}$, $\log g$, and $\xi$) were 
considered, but were found to lie within the uncertainties of the first determination.

\begin{figure*}
\centering 
    \resizebox{0.76\textwidth}{!}{\includegraphics{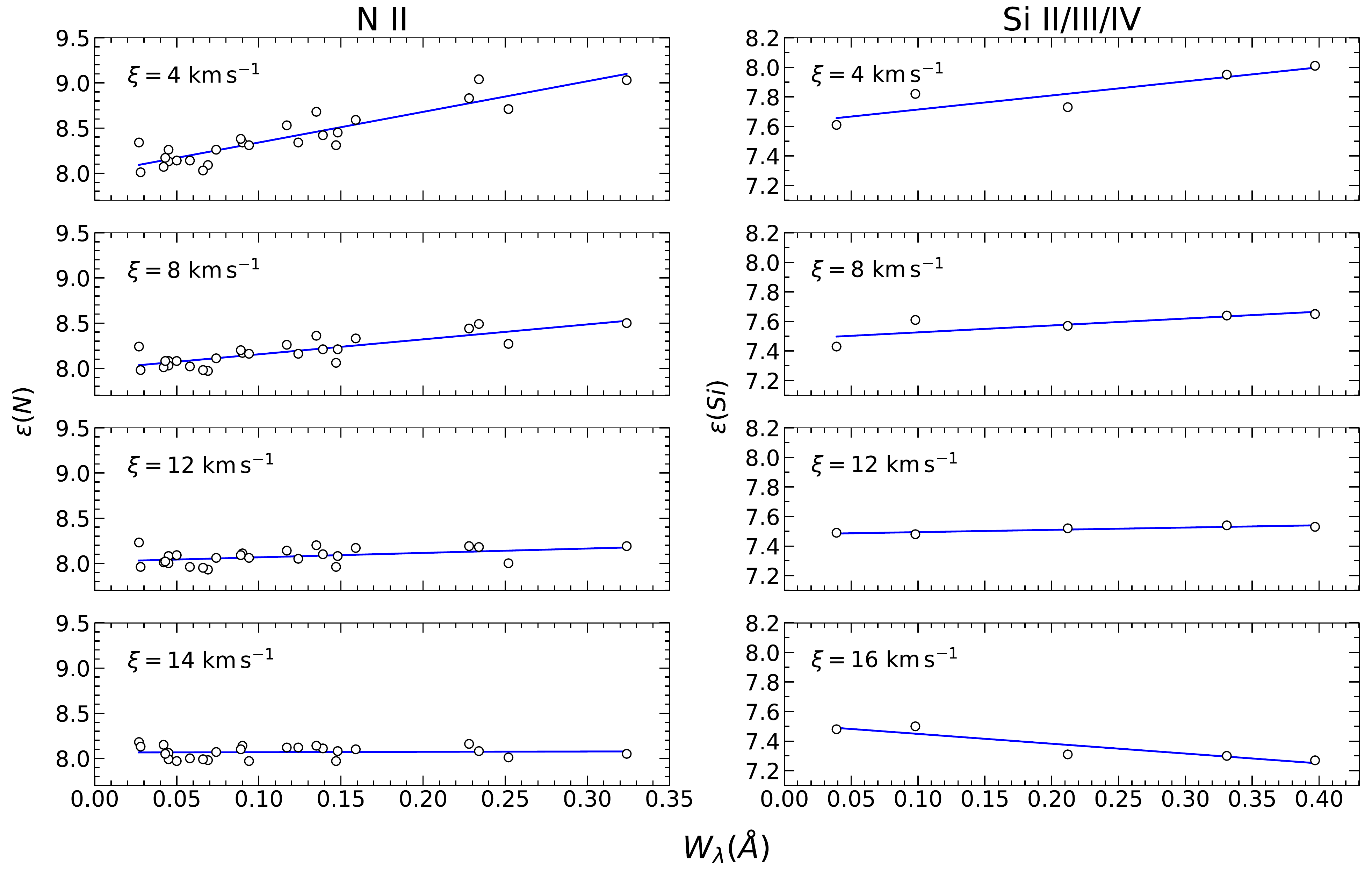}}
      \caption{Constraining the microturbulent velocity for HD~119646. The nitrogen (left panels) and 
      silicon abundances (right panels) are displayed as a function of equivalent width
      for varying values of microturbulence, as indicated. 
      The blue lines depict the best linear fit to the data.} 
         \label{fig:microturbulence_hd93840}
\end{figure*}

\subsubsection{Microturbulence}
Turbulent flows of matter on scales smaller than unit optical-depth can influence the shape and strength of spectral 
lines. They are parameterised as an ad-hoc microturbulence broadening parameter $\xi$ (measured in km\,s$^{-1}$) in addition to thermal 
broadening. Since this microturbulent velocity directly influences the broadening and therefore also the strength of the fitted metal
lines, an incorrect value will lead to offsets in ionisation balances and consequently to inaccurate estimates of effective 
temperature. In fact, because of the pressure of the turbulent matter flows, microturbulence can also change the 
density structure of the atmosphere noticeably (see Sect.~\ref{subsection:turb_pressure}), affecting the surface gravity determination. 

The appropriate value for $\xi$ can be found by enforcing the criterion that the abundances derived from various 
spectral lines of a given element are independent of the strength of the spectral lines. For the analysis of our sample,
we measured the equivalent widths $W_{\lambda}$ of several trustworthy lines of \ion{N}{i/ii} and \ion{Si}{ii/iii/iv} as auxiliary quantities by
direct integration of the observed spectral lines and compared their fitted
abundances (from spectrum synthesis) for multiple values of $\xi$. This procedure is shown in Fig.~\ref{fig:microturbulence_hd93840}: as 
$\xi$ increases, the equivalent widths of the strong lines are affected more markedly by microturbulent broadening such that 
the abundance values necessary to fit them are reduced. The correct value for $\xi$ is found when the fitted 
individual line abundances of a given element no longer  correlate with their $W_{\lambda}$ and the 
line-to-line abundance scatter is minimised. Uncertainties of the equivalent widths are of the order of the symbol size. In the given sample plot, the determinations are consistent with a 
microturbulent velocity of $\xi$\,$\approx$\,14\,km\,s$^{-1}$. The value so derived was then checked for consistency 
with multiple lines of \ion{C}{ii} and \ion{Mg}{ii} in later steps of the atmospheric parameter iteration, and with
\ion{Fe}{ii/iii} lines in a further inspection. Corrections of $\Delta \xi$\,$\approx$\,1--2\,km\,s$^{-1}$ to the initial 
value were implied with respect to the initial value in some cases, such that a final value was obtained,
fully consistent with the available indicators from several chemical species and ions.

\begin{figure*}
\centering 
    \includegraphics[width=0.74\textwidth]{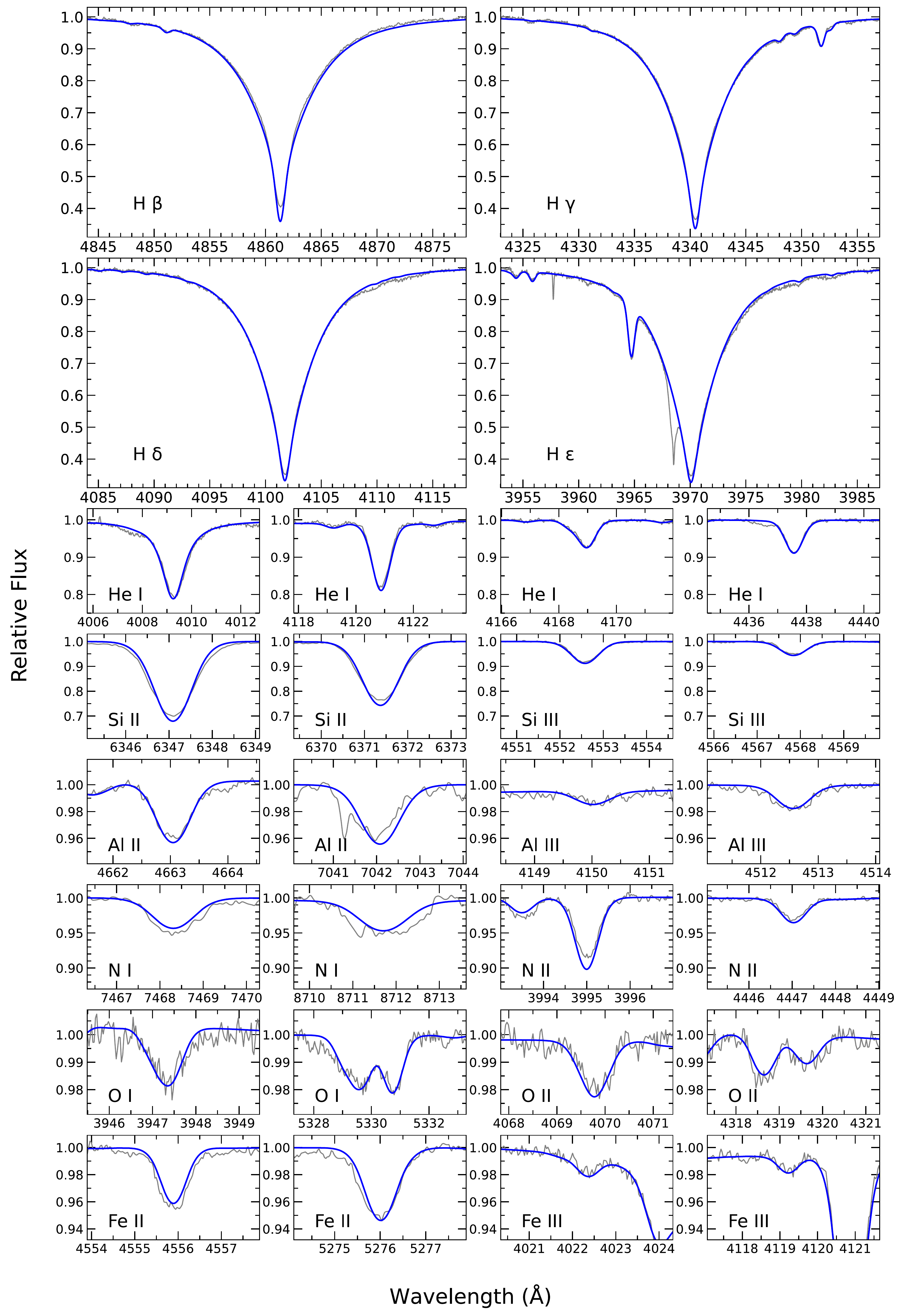}
        \caption{Comparison of the observed spectrum for HD~125288 (black) with the model spectrum for some diagnostic
        lines at the best fitting parameters (blue lines). All lines of different atomic species and ionisation stages 
        fit equally well, regardless of the strength of the individual lines.}
    \label{fig:sample_plot}
\end{figure*}

\subsubsection{Projected rotational velocity and macroturbulence}
Since the parameters of projected rotational velocity $\varv \sin i$ and macroturbulent velocity $\zeta$ do not affect the
model atmosphere structure nor influence the line formation, their derivation is not related to time-expensive grid 
calculations. By convolution of the synthetic spectrum with the corresponding rotational and macroturbulent broadening functions
\citep[realised here by a radial-tangential model,][]{Gray75} and by fitting weak metal lines of 
the observed spectra, we can find well-fitting values to within about 10\% uncertainty for both $\varv \sin i$ and $\zeta$. 
As has been pointed out in previous studies, multiple values of a pair of these parameters can lead to a similarly  
satisfactory fit to individual lines \citep[e.g.][]{Ryansetal02,FiPr12,SiHe14}. However, this ambiguity of solutions 
may be minimised using suited line blends, see for example Fig.~11 of \citet{Przybillaetal06} or Fig.~5 of \citet{FiPr12}.

The existence of non-rotational broadening in B-type supergiants is well established. Physically, surface motions 
due to a sub-surface convection zone and stellar pulsations, among others, are the phenomena that are subsumed 
by the macroturbulence parameter \citep{IACOBIII}.

\subsubsection{Elemental abundances}
Having derived a consistent solution for all primary atmospheric parameters, abundances of all investigated elements and
ions were once more determined in a last step in order to allow a consistent fit of a single synthetic spectrum to 
the observed spectrum to be made. The final abundances of the individual elements were then computed as the
arithmetic mean of the entire sample of fitted lines and the respective uncertainty as the $1\sigma$ standard deviation.
For the specification of the abundance of an element $X$, the customary logarithmic scale normalised to 12 was chosen, 
such that $\varepsilon(X)$\,=\,$\log(X/H)$\,+\,12. 

A selected sub-sample of lines in one of the analysed spectra is shown in Fig.~\ref{fig:sample_plot} compared 
to the best-fitting model. The simultaneous reproduction of the Balmer lines, the helium and metal lines of different ionisation stages 
regardless of individual strength demonstrates the consistency of the derived solution.

\subsection{Stellar mass and age}\label{section:masses_radii_method}
\begin{figure}
   \centering
      \includegraphics[width=.95\linewidth]{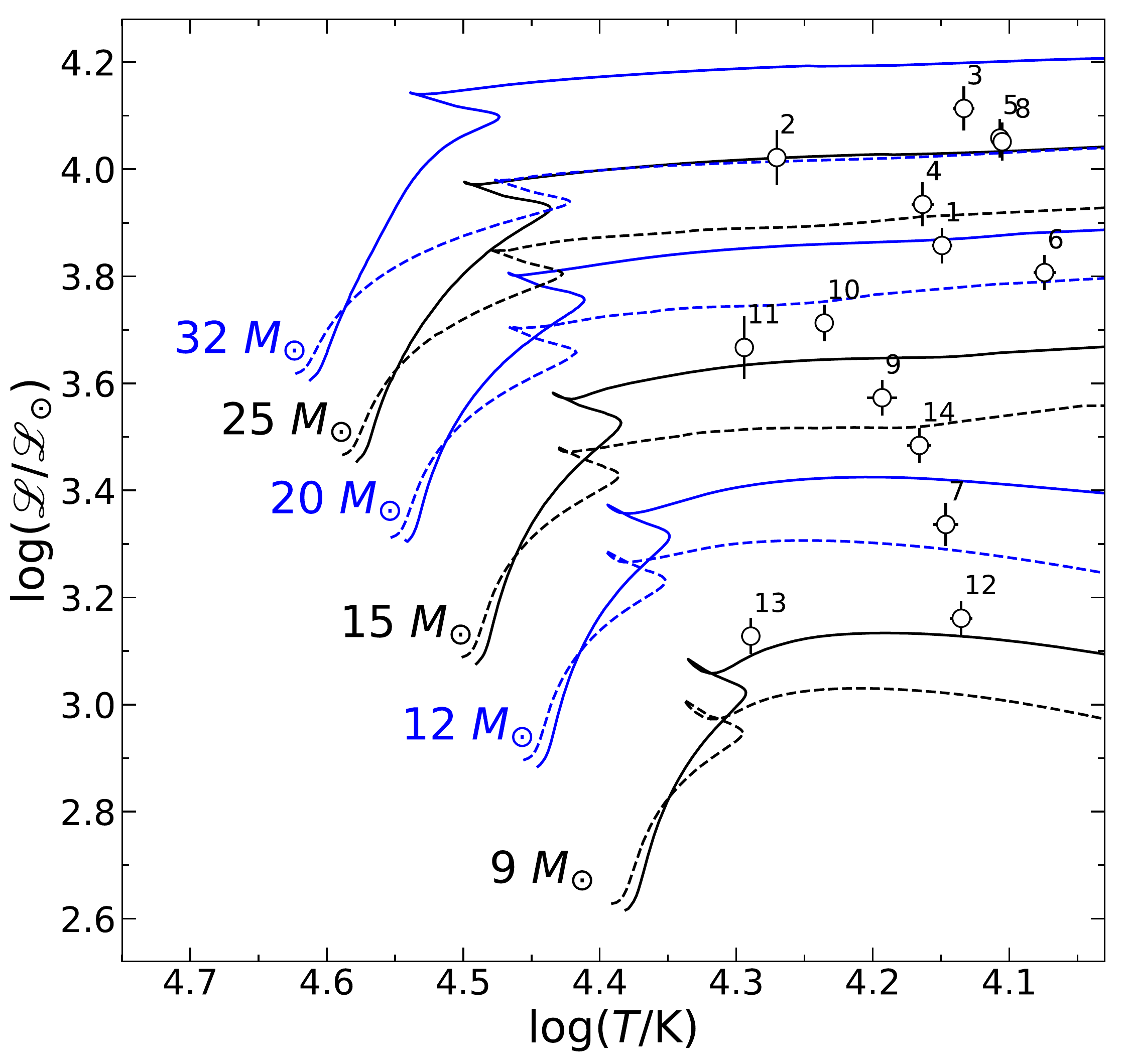}
      \caption{Sample objects located in the spectroscopic HRD, spectroscopic luminosity  
      versus log-scale effective temperature, with 1$\sigma$ error bars. 
      Stellar evolution tracks \citep[][for $\varv_{\mathrm{rot}}$\,=\,$0.4 \varv_{\mathrm{crit}}$ and $Z$\,=\,0.014]{Ekstroemetal12} ranging in ZAMS-masses from $9$ to $32$\,$M_\sun$ are indicated in alternating colours of blue and black. The dashed lines depict the corresponding tracks for non-rotating models. Numbers denote the internal sample star IDs.}
         \label{fig:stellar_evolution}
   \end{figure}

The derivation of the spectroscopically accessible atmospheric parameters (in conjunction with the photometric data)
allowed the determination of stellar masses. Effective temperature and surface gravity alone suffice to derive
the initial, or zero-age main-sequence stellar mass $M_{\mathrm{ZAMS}}$. For this, we define the 
spectroscopic luminosity $\mathscr{L}$ as
\begin{equation}\label{eq:spectroscopic_luminosity}
\mathscr{L}/\mathscr{L}_{\odot} = \frac{\left(T_{\mathrm{eff}}/T_{\mathrm{eff},\odot}\right)^4}{\left(g/g_{\odot}\right)}\,,
\end{equation}
\citep{LaKu14}, where the values for the solar effective temperature and surface gravity are  
$T_{\mathrm{eff},\odot}$\,=\,5777\,K and $\log  g_{\odot}$\,=\,4.44. Figure~\ref{fig:stellar_evolution} 
shows the sample stars in a 'spectroscopic' 
Hertzsprung-Russell diagram (sHRD) with tracks of stellar evolution models according to \cite{Ekstroemetal12} 
tracing the
spectroscopic luminosity as a function of log-scale effective temperature. Interpolating on this grid of rotating models, we 
derived $M_{\mathrm{ZAMS}}$. Under the assumption of a normal, single-star, red-ward evolution in the HRD we then interpolated the model tracks in effective 
temperature to estimate the objects' current masses $M_{\mathrm{evol}}$ in their respective evolved state. Depending on 
the initial mass of the objects, the evolved stars lost more than 2\,$M_{\odot}$ through their stellar wind at the high 
mass limit of the sample and negligibly little ($<$\,0.05\,$M_{\odot}$) for the least massive objects. From a comparison of rotating versus non-rotating models, it is apparent that the derived masses depend on the initial rotational velocity. Specifically, we find that values of $M_{\mathrm{evol}}$ -- as derived from non-rotating models -- are larger by up to 7\,$M_{\odot}$ at the high mass limit and about 1\,$M_\odot$ larger at the lower mass limit of our sample. The initial rotational velocities of the sample stars on the ZAMS are unknown. However, mass loss (and therefore angular momentum loss) and the expansion to supergiant dimensions lead to a strong reduction of the rotational velocities, for example from about 300\,km\,s$^{-1}$ on the ZAMS to about 50\,km\,s$^{-1}$ for the rotating models of \citet{Ekstroemetal12} at the boundary spanned 
by the stars ID\#2 to \#10 to \#12 in Fig.~\ref{fig:stellar_evolution}. As our sample stars show $\varv \sin i$-values between about 20 to 50\,km\,s$^{-1}$ one would expect them to stem from stars with about 
average rotation on the ZAMS, and we can exclude initially very slow rotators with confidence,
as they would be seen near zero $\varv \sin i$~at the supergiant stage. We note, however, that the predictive power of the stellar evolution models needs to be treated with caution because of remaining uncertainties of the models. We therefore would like to stress that our results were obtained under the stated assumptions, and some small-scale systematic errors are likely to be present in the fundamental stellar parameters, but these cannot be quantified at the current time.

Ages of the stars were derived from interpolation in the isochrones for the (rotating) stellar evolution models of \citet{Ekstroemetal12}. Again, a normal red-ward evolution as single stars was assumed for the sample objects.

\subsection{SED fitting}\label{section:sed_fitting}
To fit the multi-band photometry and UV-data, synthetic \,{\sc Atlas9}\footnote{The {\sc Atlas9} starting
models are equivalent to {\sc Atlas12} models for the purpose of SED fitting, as the temperature structures
are practically identical for nearly scaled-solar abundances (as realised here). However, they can readily 
be employed for the comparison without requiring adjustment to the low-resolution observations.} model SEDs of all sample objects were reddened 
according to the mean extinction law of \cite{fitzpatrick99}. The flux of the observed magnitudes was calibrated with 
zero-points and fluxes according to the SVO Filter Profile 
Sevice\footnote{\url{http://svo2.cab.inta-csic.es/theory/fps/}} \citep{svoI,svoII}. Models were then fitted for 
the two-parameter solution by \cite{fitzpatrick99} -- total-to-selective extinction $R_V$\,=\,$A_V/E\left(B-V\right)$ 
and colour excess $E\left(B-V\right)$ -- in order to match the observations. The visual extinction $A_V$ is 
then simply the product of the two parameters. A different approach had to be employed for HD~183143 because of a highly 
anomalous reddening law, see \citet{Ebenbichleretal22} for details.

This method of examining the SED of our final solution generally worked very well and produced a high-precision 
characterisation of the interstellar medium along the sight lines towards the sample objects. In addition to being 
consistent with small uncertainties (see Sect.~\ref{section:sightlines}) the method can detect unusual features in the
extinction curve, such as excess radiation in the WISE pass bands, hinting at black body radiation contributions to the 
stellar SED. Specifically, this process can detect anormalies in the composition of the interstellar medium along these sight lines, 
as in the case of HD~183143 mentioned above.

\subsection{Spectroscopic distance}\label{sec:spec_dist_method}
Having derived spectroscopic and fundamental parameters, as well as a precise reddening law for all sample objects, spectroscopic distances 
$d_{\mathrm{spec}}$ were calculated using an expression by \cite{Ramspecketal2001}
\begin{equation}\label{eq:spec_dist}
d_{\mathrm{spec}} = 7.11 \times 10^4 \sqrt{H_{\nu}~M_{\mathrm{evol}}~10^{0.4 m_{V_0}-\log g}},
\end{equation}
where $H_\nu$ denotes the Eddington flux, given in units of erg\,cm$^{-2}$\,s$^{-1}$\,Hz$^{-1}$ at 550 nm, 
$M_{\mathrm{evol}}$ the evolutionary mass in units of $M_{\odot}$, $m_{V_0}$\,=\,$m_V-A_V$ the dereddened Johnson $V$ 
magnitude in mag, and $\log g$ the logarithmic surface gravity in cgs units. Equation \ref{eq:spec_dist} utilises the 
Vega flux calibration according to \cite{Heberetal84} and provides distances in units of pc.

We have to stress once more that our $M_\mathrm{evol}$-values were derived under the assumption of the overall applicability of the evolution tracks for rotating stars by \citet{Ekstroemetal12}. As we have argued in Sect.~\ref{section:masses_radii_method}, the true initial rotational velocities of the sample stars are unknown, therefore some additional systematic uncertainty applies to the spectroscopic distances according to Eqn.~\ref{eq:spec_dist} that may either increase or decrease the derived value.

These spectroscopic distances $d_{\mathrm{spec}}$ may be compared to distances $d_{\mathrm{Gaia}}$ derived from Gaia early data release 3
(EDR3) parallaxes \citep{Gaia2016,Gaia2020}. One potential issue is a mismatch of the Gaia distance with the spectroscopic distance because of a biased evolutionary mass, however the effects are only of order $\propto$\,$M_\mathrm{evol}^{1/2}$. Alternatively, this can uncover potentially undetected problems with the
spectroscopic analysis. For instance, widely diverging estimations of distances can hint at an incorrect value for the 
surface gravity as this parameter contributes most of the uncertainty to the equation. It can, however, also 
uncover an unusual evolutionary development of an object in question (see Sect.~\ref{section:spectroscopic_distances}). Gaia EDR3 parallaxes may also be affected by bias, such as increased uncertainties for the five brightest supergiants of the sample with Gaia $G$ magnitude smaller than 6, or for objects with a large renormalised unit weight error (RUWE), like HD~51309 (ID\#9), HD~125288 (ID\#12), and HD~164353 (ID\#14), which have a RUWE of about 2 -- all other objects have RUWE-values around 1.

%RUWE values:
%7902   0.921
%14818  0.950
%25914  0.925
%36371  1.162
%183143 0.929
%184943 0.920
%191243 0.877
%199478 0.904
%51309  2.267
%111990 1.178
%119646 0.755
%125288 1.878
%159110 0.789
%164353 2.506

\subsection{Bolometric correction, luminosity and radius}\label{section:bolometric_correction}
For the calculation of the bolometric correction $B.C.$, we defined the bolometric magnitude $m_{\mathrm{bol}}$ for each
star as the direct integration of its \,{\sc Atlas9} model SED over all wavelengths, with the integration constant 
chosen such that a solar \,{\sc Atlas9} model satisfies $M_{\mathrm{bol},\odot}$\,=\,4.74 \citep[see][]{Besseletal98}.
The $B.C.$ was then calculated as the difference between $m_{\mathrm{bol}}$ and the synthetic $m_V$.

In order to determine the stellar luminosity $L$, the absolute $V$-band magnitude $M_V$ was calculated from the observed
apparent magnitude $m_V$ \citep{Mermilliod97} utilising the derived spectroscopic distances $d_{\mathrm{spec}}$ 
(Sect.~\ref{sec:spec_dist_method}), as well as the total-to-selective extinction $R_V$, and colour excess $E(B-V)$ 
(Sect.~\ref{section:sed_fitting}) in the distance modulus. Correction of $M_V$ by $B.C.$ yielded the absolute bolometric
magnitude $M_{\mathrm{bol}}$, from which $L$ was derived using the above value for $M_{\mathrm{bol},\odot}$. The effective 
temperature and luminosity were finally utilised to determine the stellar radius $R$ by application of the 
Stefan-Boltzmann law.

\begin{table*}
\caption{Stellar parameters of the sample objects.}
\label{tab:stellar_parameters}
\centering   
{\small
\setlength{\tabcolsep}{1.6mm}
\begin{tabular}{rlr@{\hspace{0.1mm}}rrcrrrlccrrrrrr}
\hline\hline
ID\#  & Object     &       & $T_{\mathrm{eff}}$ & $\log g$ & $y$ & $\xi$                & $\varv  \sin i$          & $\zeta$              & $R_V$ & $E\left(B-V\right)$ & $B.C.$     & $M_{\mathrm{evol}}$ & $R$         & $\log L/L_\sun$ & $\log \tau_\mathrm{evol}$ & $d_{\mathrm{spec}}$ & $d_{\mathrm{Gaia}}$\tablefootmark{a} \\ \cline{7-9}
     &        &       & kK                 & (cgs)      & & \multicolumn{3}{c}{$\mathrm{km\,s}^{-1}$} &       & mag                 & mag      & $M_{\odot}$         & $R_{\odot}$ &  & yr & pc                  & pc                  \\ \hline
1                    & HD 7902                 &                      & 14.1                 & 2.13                 & 0.089                   & 9                    & 36                   & 35                   & 3.16                 & 0.58                 & $-0.999$             & 19.2                 & 64                   & 5.16    & 6.98             & 2900                 & 2487                 \\
                     &                      & $\pm$                & 0.2                  & 0.05                 & 0.006                   & 2                    & 5                    & 5                    & 0.1                  & 0.03                 &                      & 0.8                  & 7                    & 0.09       & 0.3          & 280                  & $^{110}_{80}$        \\
2                    & HD 14818                &                      & 18.6                 & 2.45                 & 0.095                   & 14                   & 48                   & 40                   & 3.09                 & 0.56                 & $-1.645$             & 23.6                 & 49                   & 5.41    & 6.89             & 2150                 & 2121                 \\
                     &                      & $\pm$                & 0.3                  & 0.07                 & 0.008                   & 2                    & 6                    & 5                    & 0.1                  & 0.03                 &                      & 1.9                  & 6                    & 0.11       & 0.03          & 250                  & $^{150}_{110}$       \\
3                    & HD 25914                &                      & 13.6                 & 1.81                 & 0.092                   & 11                   & 35                   & 40                   & 2.97                 & 0.76                 & $-0.943$             & 25.7                 & 106                  & 5.54    & 6.86             & 6030                 & 5431                 \\
                     &                      & $\pm$                & 0.2                  & 0.06                 & 0.004                   & 2                    & 5                    & 5                    & 0.1                  & 0.03                 &                      & 1.6                  & 13                   & 0.10       & 0.03          & 640                  & $^{790}_{440}$       \\
4                    & HD 36371                &                      & 14.6                 & 2.11                 & 0.086                   & 11                   & 36                   & 35                   & 3.35                 & 0.52                 & $-1.081$             & 21.1                 & 68                   & 5.28    & 6.94             & 1200                 & 1214                 \\
                     &                      & $\pm$                & 0.3                  & 0.06                 & 0.005                   & 2                    & 5                    & 5                    & 0.1                  & 0.03                 &                      & 1.2                  & 8                    & 0.10       & 0.03          & 130                  & $^{380}_{220}$       \\
5                    & HD 183143               &                      & 12.8                 & 1.76                 & 0.099                   & 7                    & 37                   & 27                   & 3.3                  & 1.22                 & $-0.793$             & 24.2                 & 109                  & 5.46    & 6.88             & 1530                 & 2168                 \\
                     &                      & $\pm$                & 0.2                  & 0.05                 & 0.005                   & 2                    & 5                    & 5                    & 0.1                  & 0.03                 &                      & 1.4                  & 15                   & 0.11       & 0.03          & 170                  & $^{120}_{120}$       \\
6                    & HD 184943               &                      & 11.9                 & 1.88                 & 0.099                   & 9                    & 35                   & 25                   & 2.97                 & 0.84                 & $-0.600$             & 17.7                 & 82                   & 5.07    & 7.02             & 4040                 & 4090                 \\
                     &                      & $\pm$                & 0.2                  & 0.05                 & 0.002                   & 2                    & 6                    & 5                    & 0.1                  & 0.03                 &                      & 0.8                  & 10                   & 0.10       & 0.04          & 400                  & $^{240}_{240}$       \\
7                    & HD 191243               &                      & 14.0                 & 2.64                 & 0.087                   & 8                    & 27                   & 25                   & 2.88                 & 0.33                 & $-0.972$             & 11.0                 & 27                   & 4.39    & 7.32             & 1220                 & 1205                 \\
                     &                      & $\pm$                & 0.3                  & 0.06                 & 0.011                   & 2                    & 6                    & 5                    & 0.1                  & 0.03                 &                      & 0.5                  & 3                    & 0.09       & 0.05          & 120                  & $^{30}_{30}$         \\
8                    & HD 199478               &                      & 12.7                 & 1.76                 & 0.107                   & 8                    & 40                   & 40                   & 3.03                 & 0.62                 & $-0.783$             & 24.0                 & 111                  & 5.46    & 6.88             & 2440                 & 2423                 \\
                     &                      & $\pm$                & 0.2                  & 0.05                 & 0.004                   & 2                    & 6                    & 5                    & 0.1                  & 0.03                 &                      & 1.3                  & 12                   & 0.09       & 0.03          & 230                  & $^{230}_{220}$       \\
9                    & HD 51309                &                      & 15.6                 & 2.59                 & 0.081                   & 10                   & 30                   & 35                   & 3.08                 & 0.11                 & $-1.236$             & 13.7                 & 32                   & 4.72    & 7.17             & 950                  & 1108                 \\
                     &                      & $\pm$                & 0.4                  & 0.05                 & 0.003                   & 2                    & 6                    & 5                    & 0.1                  & 0.03                 &                      & 0.5                  & 4                    & 0.09       & 0.04          & 90                   & $^{410}_{210}$       \\
10                   & HD 111990               &                      & 17.2                 & 2.62                 & 0.089                   & 12                   & 40                   & 40                   & 3.3                  & 0.45                 & $-1.464$             & 16.1                 & 33                   & 4.94    & 7.07             & 1940                 & 2418                 \\
                     &                      & $\pm$                & 0.3                  & 0.05                 & 0.003                   & 2                    & 6                    & 5                    & 0.1                  & 0.03                 &                      & 0.7                  & 4                    & 0.09       & 0.04          & 180                  & $^{160}_{180}$       \\
11                   & HD 119646               &                      & 19.7                 & 2.9                  & 0.100                   & 14                   & 37                   & 40                   & 3.53                 & 0.34                 & $-1.781$             & 15.4                 & 23                   & 4.87    & 7.10             & 1620                 & 1721                 \\
                     &                      & $\pm$                & 0.2                  & 0.07                 & 0.009                   & 2                    & 6                    & 5                    & 0.1                  & 0.03                 &                      & 1.2                  & 3                    & 0.11       & 0.05          & 190                  & $^{80}_{70}$         \\
12                   & HD 125288               &                      & 13.7                 & 2.77                 & 0.094                   & 6                    & 23                   & 30                   & 3.65                 & 0.30                 & $-0.934$             & 9.3                  & 21                   & 4.14    & 7.46             & 390                  & 438                  \\
                     &                      & $\pm$                & 0.3                  & 0.05                 & 0.007                   & 2                    & 4                    & 5                    & 0.1                  & 0.03                 &                      & 0.3                  & 2                    & 0.09       & 0.05          & 40                   & $^{50}_{30}$         \\
13                   & HD 159110               &                      & 19.5                 & 3.42                 & 0.095                   & 3                    & 17                   & 15                   & 3.3                  & 0.22                 & $-1.826$             & 9.3                  & 10                   & 4.11    & 7.45             & 1290                 & 1362                 \\
                     &                      & $\pm$                & 0.3                  & 0.05                 & 0.001                   & 2                    & 4                    & 5                    & 0.1                  & 0.03                 &                      & 0.3                  & 1                    & 0.09       & 0.05          & 120                  & $^{80}_{70}$         \\
14                   & HD 164353               &                      & 14.7                 & 2.57                 & 0.091                   & 8                    & 20                   & 32                   & 3.61                 & 0.19                 & $-1.081$             & 12.6                 & 31                   & 4.60    & 7.22             & 620                  & 797                  \\
                     &                      & $\pm$                & 0.3                  & 0.05                 & 0.004                   & 2                    & 4                    & 5                    & 0.1                  & 0.03                 &                      & 0.4                  & 4                    & 0.09       & 0.04          & 60                   & $^{200}_{130}$       \\  \hline
\end{tabular}
\tablefoot{Uncertainties are 1$\sigma$-values, except where noted otherwise. 
\tablefoottext{a}{\cite{Gaia2016,Gaia2020} - distances and uncertainties correspond to 'photogeometric 
distances' and associated $14^{\mathrm{th}}$ and $86^{\mathrm{th}}$ confidence percentiles \citep{Bailer-Jones_etal_2021}.}
}}
\end{table*}

\section{Results}\label{section:results}
The results of the analysis of the sample stars are summarised in Table~\ref{tab:stellar_parameters}. The
parameters listed are: internal identification number, HD-designation, effective temperature, surface gravity,
surface helium abundance, microturbulent, projected rotational and macroturbulent velocities, total-to-selective 
extinction parameter, colour excess, bolometric correction, evolutionary mass, radius, luminosity, evolutionary age, spectroscopic and Gaia EDR3 distances 
\citep[probabilistic estimations of 'photogeometric' distances,][]{Bailer-Jones_etal_2021}. The respective 
uncertainties, given in the line below the observed values, denote $1\sigma$-intervals. 

\subsection{Atmospheric and fundamental stellar parameters}\label{section:stellar_parameters}
For the effective temperature and surface gravity, the uncertainties roughly match the values derived in 
previous work analysing BA-type supergiants with a similar analysis approach as employed here \citep{FiPr12}, with
$\delta T_{\mathrm{eff}}$\,$\approx$\,1--3\% and $\Delta \log g$\,$\approx$\,0.05\,dex. 
Abundances of helium were fitted with uncertainties of $\delta y$\,$\approx$\,5--15\% owing to the line-to-line scatter 
from the weakest \ion{He}{i} features analysed, that is those most sensitive to abundance variations.

The uncertainty of the microturbulent velocity is generally limited by the size of the grid used during the fitting 
process. Though observed values of $\xi$ were in some cases inspected on scales of 1\,km\,s$^{-1}$, a conservative 
estimate of $\Delta \xi$\,$\approx$\,2\,km\,s$^{-1}$ is adopted throughout. Projected rotational velocities show 
relative uncertainties amounting to typically $\delta \varv \sin(i)$\,$\approx$\,10--15\% owing to the degeneracy in the
joint derivation with macroturbulent velocities. For the macroturbulence $\zeta$ the uncertainties were estimated at a
value of 5\,km\,s$^{-1}$. 

A similar but weaker ambiguity in deduced values is generally present in the estimation of total-to-selective 
extinction parameter and colour excess. The fitting procedure described in Sect.~\ref{section:sed_fitting} produces error
margins of the order of $\Delta R_V$\,$\approx$\,0.1 and $\Delta E(B-V)$\,$\approx$\,0.03\,mag. Although the exact 
margins in this derivation depend on the available data (in particular in the UV-range) these values generally 
represent the typical uncertainties for the entire sample. For the derivation of the $B.C.$~no detailed analysis of 
uncertainties was conducted, though variation of parameters in input models hinted at an uncertainty range
of $\Delta B.C.$\,$\approx$\,0.04--0.06\,mag for both hotter and cooler sample stars. 

The parameter of evolved mass $M_{\mathrm{evol}}$ was derived using the ZAMS mass estimates, tracing their mass loss in
evolution tracks by \cite{Ekstroemetal12}, see Sect.~\ref{section:mass_estimates_and_discrepancy}, such that the
uncertainties of the evolved masses are assumed to be identical to the ZAMS mass uncertainty of about 
$\delta M_{\mathrm{ZAMS}}$\,=\,$\delta M_{\mathrm{evol}}$\,$\approx$\,5\%. Radii of sample objects show relative 
errors of about $\delta R$\,$\approx$\,10\%, stemming largely from the associated uncertainty in luminosity, which
amounts to typically $\Delta \log L/L_{\odot}$\,$\approx$\,0.1\,dex. The distances derived in this work show 
consistent relative uncertainties of $\delta d_{\mathrm{spec}}$\,$\approx$\,10\%, matching the sample mean relative 
difference between the deduced values and those derived from parallactic distances by the Gaia mission (see 
Sect.~\ref{section:spectroscopic_distances}). The fundamental parameters can be expected to be
subject to a small amount of additional systematic error because a set of stellar evolution models for a particular rotational velocity were adopted, see the discussion in Sect.~\ref{section:masses_radii_method}.

\begin{figure*}[ht]
\centering 
    \resizebox{0.87\textwidth}{!}{\includegraphics[width=\hsize]{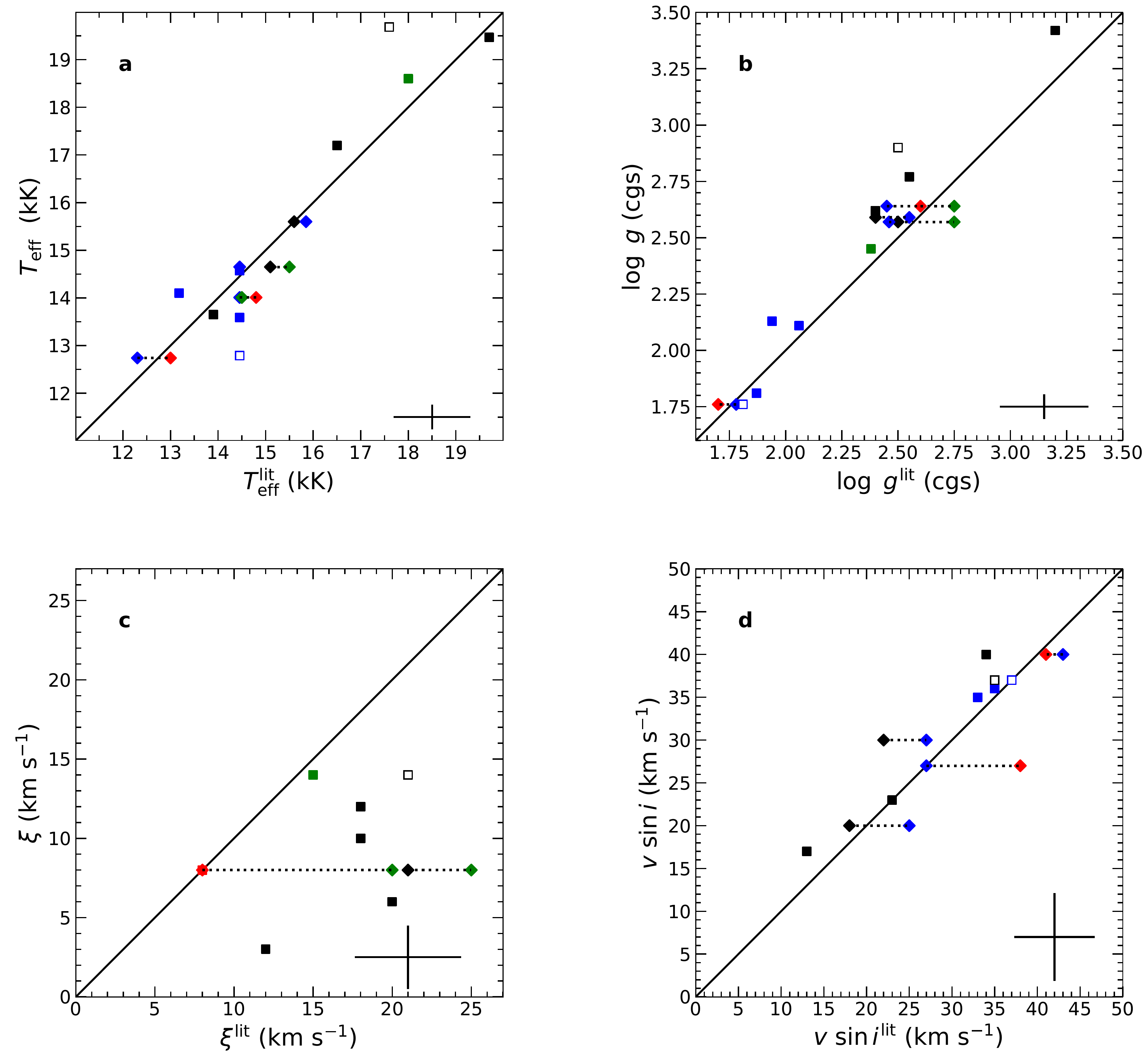}}
      \caption{Comparison of values for effective temperature $T_{\mathrm{eff}}$ (panel a), surface gravity $\log g$\, (panel b), microturbulence $\xi$\, (panel c), and projected rotational velocity $\varv 
\sin i$ (panel d) as derived in the present work with previous studies: \citet[][black symbols]{Fraser_etal_10}, \citet[][blue]{IACOBIII}, \citet[][red]{MaPu08}, and \citet[][green]{Searle_etal_08}. In cases in which an object is present in two or multiple studies the values are depicted by diamonds and connected with dotted lines. The symbols for HD~183143 and HD~119646 are marked by open symbols. For better visibility, the mean uncertainties are indicated in the lower right corner.}  
         \label{fig:comparison_combined}
\end{figure*}

\begin{table*}
\caption{Metal abundances $\varepsilon (X)$\,=\,$\log (X/\mathrm{H})+12$ and metallicity $Z$ (by mass) of the sample objects.}
\label{tab:abundances}
\centering   
{\small
\setlength{\tabcolsep}{1.5mm}
\begin{tabular}{lll@{\hspace{0.1mm}}llllllllllc}
\hline\hline
ID\#  & Object&          & C         & N         & O         & Ne        & Mg        & Al       & Si        & S         & Ar        & Fe  & $Z$      \\ \hline
1     & HD 7902   &       & 8.25 (9)  & 8.27 (27) & 8.75 (17) & 7.96 (12) & 7.51 (8)  & 6.44 (5) & 7.54 (10) & 6.96 (13) & 6.41 (4)  & 7.59 (24) & 0.014 \\
      &           & $\pm$ & 0.04      & 0.06      & 0.05      & 0.05      & 0.07      & 0.09     & 0.09      & 0.07      & 0.07      & 0.11      & 0.002 \\
2     & HD 14818  &       & 8.00 (9)  & 8.33 (24) & 8.50 (22) & 8.08 (6)  & 7.49 (4)  & 6.17 (5) & 7.62 (8)  & 6.94 (5)  & 6.54 (4)  & 7.37 (17) & 0.011 \\
      &           & $\pm$ & 0.05      & 0.09      & 0.06      & 0.05      & 0.11      & 0.05     & 0.07      & 0.03      & 0.05      & 0.07      & 0.002 \\
3     & HD 25914  &       & 8.09 (8)  & 8.22 (25) & 8.58 (18) & 7.96 (11) & 7.28 (3)  & 6.19 (4) & 7.29 (9)  & 6.72 (10) & 6.38 (3)  & 7.50 (22) & 0.010 \\
      &           & $\pm$ & 0.09      & 0.09      & 0.05      & 0.06      & 0.11      & 0.06     & 0.08      & 0.09      & 0.06      & 0.07      & 0.002 \\
4     & HD 36371  &       & 8.10 (11) & 8.33 (27) & 8.57 (26) & 8.06 (14) & 7.40 (5)  & 6.34 (3) & 7.44 (10) & 6.96 (14) & 6.34 (7)  & 7.48 (26) & 0.012 \\
      &           & $\pm$ & 0.08      & 0.06      & 0.07      & 0.05      & 0.07      & 0.02     & 0.08      & 0.09      & 0.05      & 0.11      & 0.002 \\
5     & HD 183143 &       & 8.31 (9)  & 8.69 (26) & 8.78 (12) & 8.09 (12) & 7.69 (6)  & 6.44 (6) & 7.58 (7)  & 7.10 (10) & 6.56 (2)  & 7.66 (26) & 0.018 \\
      &           & $\pm$ & 0.07      & 0.06      & 0.05      & 0.07      & 0.09      & 0.06     & 0.06      & 0.08      & 0.08      & 0.10      & 0.002 \\
6     & HD 184943 &       & 8.43 (6)  & 8.63 (19) & 8.84 (8)  & 8.02 (12) & 7.63 (4)  & 6.47 (7) & 7.66 (7)  & 7.05 (12) & 6.63 (2)  & 7.75 (16) & 0.019 \\
      &           & $\pm$ & 0.04      & 0.06      & 0.06      & 0.05      & 0.01      & 0.08     & 0.08      & 0.07      & 0.12      & 0.10      & 0.002 \\
7     & HD 191243 &       & 8.28 (9)  & 8.24 (35) & 8.72 (27) & 7.94 (14) & 7.57 (9)  & 6.34 (4) & 7.52 (10) & 6.98 (13) & 6.39 (7)  & 7.62 (31) & 0.014 \\
      &           & $\pm$ & 0.08      & 0.08      & 0.08      & 0.05      & 0.05      & 0.13     & 0.06      & 0.07      & 0.07      & 0.09      & 0.002 \\
8     & HD 199478 &       & 8.20 (6)  & 8.63 (26) & 8.74 (15) & 7.99 (12) & 7.64 (5)  & 6.40 (4) & 7.58 (7)  & 7.11 (13) & 6.54 (5)  & 7.76 (23) & 0.016 \\
      &           & $\pm$ & 0.05      & 0.06      & 0.08      & 0.05      & 0.08      & 0.09     & 0.06      & 0.08      & 0.09      & 0.09      & 0.002 \\
9     & HD 51309  &       & 8.29 (12) & 8.23 (27) & 8.71 (24) & 8.02 (11) & 7.52 (8)  & 6.52 (3) & 7.56 (9)  & 7.03 (13) & 6.41 (9)  & 7.55 (28) & 0.014 \\
      &           & $\pm$ & 0.06      & 0.05      & 0.07      & 0.05      & 0.09      & 0.10     & 0.06      & 0.08      & 0.09      & 0.16      & 0.002 \\
10    & HD 111990 &       & 8.13 (13) & 8.21 (31) & 8.65 (17) & 8.06 (12) & 7.50 (3)  & 6.17 (6) & 7.50 (7)  & 7.07 (11) & 6.44 (10) & 7.49 (22) & 0.012 \\
      &           & $\pm$ & 0.06      & 0.06      & 0.08      & 0.07      & 0.06      & 0.11     & 0.05      & 0.09      & 0.08      & 0.07      & 0.002 \\
11    & HD 119646 &       & 8.24 (15) & 8.02 (24) & 8.65 (16) & 8.15 (11) & 7.51 (5)  & 6.22 (5) & 7.54 (7)  & 7.01 (4)  & 6.47 (7)  & 7.41 (19) & 0.012 \\
      &           & $\pm$ & 0.08      & 0.06      & 0.05      & 0.04      & 0.06      & 0.08     & 0.03      & 0.09      & 0.05      & 0.06      & 0.002 \\
12    & HD 125288 &       & 8.35 (8)  & 8.50 (29) & 8.80 (20) & 8.06 (12) & 7.54 (10) & 6.26 (6) & 7.62 (11) & 7.11 (13) & 6.52 (10) & 7.60 (34) & 0.017 \\
      &           & $\pm$ & 0.07      & 0.07      & 0.06      & 0.06      & 0.09      & 0.07     & 0.09      & 0.05      & 0.08      & 0.08      & 0.002 \\
13    & HD 159110 &       & 8.53 (19) & 7.92 (35) & 8.85 (22) & 8.10 (21) & 7.49 (10) & 6.34 (5) & 7.54 (7)  & 7.20 (16) & 6.54 (16) & 7.55 (21) & 0.016 \\
      &           & $\pm$ & 0.06      & 0.05      & 0.07      & 0.07      & 0.06      & 0.04     & 0.10      & 0.09      & 0.06      & 0.08      & 0.002 \\
14    & HD 164353 &       & 8.31 (10) & 8.37 (33) & 8.81 (20) & 8.05 (19) & 7.51 (7)  & 6.32 (4) & 7.65 (10) & 7.11 (12) & 6.47 (12) & 7.63 (32) & 0.016 \\
      &           & $\pm$ & 0.04      & 0.06      & 0.06      & 0.06      & 0.05      & 0.04     & 0.10      & 0.07      & 0.08      & 0.11      & 0.002 \\ \hline
& CAS~$^{a,b}$       &       & 8.35      & 7.79      & 8.76      & 8.09      & 7.56      & 6.30     & 7.50      & 7.14      & 6.50 & 7.52 & 0.014\\
&           & $\pm$ & 0.04      & 0.04      & 0.05      & 0.05      & 0.05      & 0.07     & 0.06   & 0.06      & 0.08 & 0.03 & 0.002\\ \hline
\end{tabular}
\tablefoot{Uncertainties are 1$\sigma$-values from the line-to-line scatter. Numbers in parentheses quantify the
analysed lines.\\ $^{(a)}$~\citet{NiPr12}~~~$^{(b)}$~\citet{Przybillaetal13}}}
\end{table*}  

\subsection{Comparison with previous analyses}\label{section:comparison_previous_analyses}
Many of our sample stars were analysed in previous studies that employed full non-LTE model atmospheres. For comparison, data from the following studies were considered:\\
{\sc i)} \cite{MaPu08} employed {\sc Fastwind} for their analyses. They utilised hydrogen, helium, and \ion{Si}{ii/iii/iv} 
lines to derive temperature, surface gravity, and microturbulence iteratively. The derivation of projected rotational velocities was based on the analysis of the shape of the Fourier transform (FT) of absorption lines \citep{Gray75, SiHe07}. Two objects are in common.\\
{\sc ii)} \cite{Searle_etal_08} used the stellar atmosphere codes {\sc Tlusty} and {\sc Cmfgen} \citep{HiMi98}. To estimate the temperature, 
the diagnostic silicon lines of \ion{Si}{iv} 4089 and \ion{Si}{iii} 4552--4574\,{\AA} were used in supergiants of 
spectral types B0 to B2 and \ion{Si}{ii} 4128--4130 and \ion{Si}{iii} 4552--4574\,{\AA} for B2.5 to B5 supergiants. The 
luminosity was then constrained by inferred values of the absolute visual magnitude $M_V$ and corrected if necessary. Surface gravity 
$\log g$ was determined by fitting H$\gamma$ and H$\delta$. The microturbulent velocity was determined by analysing the 
\ion{Si}{iii} triplet lines. Three objects are in common.\\
{\sc iii)} \cite{Fraser_etal_10} used the hydrostatic line-blanketed non-LTE codes {\sc Tlusty} and {\sc Synspec} 
\citep{hubeny_88,HuLa95} that consider plane-parallel geometry. Effective temperatures were estimated on the basis of silicon 
ionisation equilibria and surface gravities from a fit of the H$\gamma$ and H$\delta$ lines. For the determination of microturbulence they relied solely on the analysis of the \ion{Si}{iii} triplet at 4552--4574\,{\AA} and projected rotational velocities were derived using the FT method. Six objects are shared.\\ 
{\sc iv)} \cite{IACOBIII} used the hydrodynamic line-blanketed non-LTE code {\sc Fastwind} \citep{Santolaya-Rey97,Pulsetal05} 
that accounts for spherical geometry, following the spectroscopic analysis strategy described by \cite{Castroetal12}. They analysed H$\beta$, H$\gamma$, H$\delta$, multiple lines of \ion{He}{i},
as well as the silicon multiplets \ion{Si}{ii} 4128-4130\,{\AA}, \ion{Si}{iii} 4552--4574\,{\AA}, and \ion{Si}{iv}\,4116\,{\AA} 
to derive $T_\mathrm{eff}$ and $\log g$. For the derivation of projected rotational velocities they used the \,{\sc iacob-broad} tool \citep{SiHe14} on lines of O, Si, Mg, and C, depending on the spectral type of the star. Eight objects are common to the present work.

Furthermore, a sample of 25 O9.5--B3 Galactic supergiants were analysed by \citet{Crowther_etal_06} based on {\sc Cmfgen} 
models. We do not compare with this paper since it has only one object (HD~14818) in common with the present work.

Figure~\ref{fig:comparison_combined}, panel a, shows a comparison of the effective temperatures from the literature 
$T_{\mathrm{eff}}^{\mathrm{lit}}$ with those derived here. An overall good correlation is found with a mean 
relative difference $\delta T_{\mathrm{eff}}^{\mathrm{comp}}$\,=\,$\frac{1}{N-1}\sum^{N}_{i} (T_{\mathrm{eff},i} -  
T_{\mathrm{eff},i}^{\mathrm{lit}})/T_{\mathrm{eff},i}^{\mathrm{lit}}$ of less than 1\%, and a standard deviation of 5\%. 
Two objects, HD~119646 and HD~183143 (marked with open symbols), differ by $\sim$12\% in literature versus present-work $T_\mathrm{eff}$ with
both higher and lower values realised. Some objects are present in two or more of the cited studies, in which case 
the objects are depicted as diamonds. For such objects the values never scatter by more than about 1000\,K.

The comparison of surface gravities is shown graphically in Fig.~\ref{fig:comparison_combined}, panel b. Again, values occurring in more than one study are depicted as diamonds, but in this comparison, some objects show a larger scatter among the 
various studies. For HD~164353 estimates range from $\log g$\,=\,2.46 in \cite{IACOBIII}, $\log g$\,=\,2.50 in 
\cite{Fraser_etal_10} to $\log g$\,=\,2.75 in \cite{Searle_etal_08}; for HD~191243 \cite{IACOBIII} derive $\log g$\,=\,2.45, 
while \cite{MaPu08} finds $\log g$\,=\,2.61 and \cite{Searle_etal_08} again give the largest value of $\log g$\,=\,2.75,
that is differences up to a factor of 2. Treating the literature values as one complete comparison set, a systematic offset 
towards higher $\log g$ values in the present work may be noticed. One may speculate that this trend may be due to the inclusion of $P_{\mathrm{turb}}$ in our analysis, resulting in a systematic increase of $\log g$. However, we have demonstrated that the effect of this term in the model is limited to $\sim$0.05 dex even in objects with large microturbulent velocities close to the Eddington limit (see Sect.~\ref{subsection:turb_pressure}). In light of the large uncertainties of the literature values, small number statistics, and general differences between the trends in the different comparison studies, we 
cannot draw definite conclusions on the origin of these differences.

Figure~\ref{fig:comparison_combined}, panel c shows the comparison of microturbulent velocities. Both the correlation of literature and
present values, as well as the agreement between literature values of different studies is poor. In the case of HD~191243, a 
maximum difference of $\Delta \xi$\,=\,13\,km\,s$^{-1}$ is found between \cite{MaPu08} and our work on the one hand and 
\cite{Searle_etal_08} on the other. Overall, our assessments of microturbulent velocity are systematically lower by 
$\sim$10\,km\,s$^{-1}$ than the literature values, with the exception of \citet{MaPu08}. While our microturbulent velocities
remain subsonic, literature values are often found to be supersonic. Such systematically lower microturbulent velocities were 
also found in previous work on B-type main-sequence stars \citep{NiPr12}, where a broad variety of microturbulence indicators 
were employed versus the usual reliance on the \ion{Si}{iii} 4552--4574\,{\AA} triplet alone. We note that microturbulence velocities were not provided by \citet{IACOBIII}.

Finally, a comparison of projected rotational velocities is shown in Figure~\ref{fig:comparison_combined}, panel d. Good agreement is achieved overall, though there are some small-scale differences between the compared works. Values by \citet{IACOBIII} agree very well with ours, showing little to no offset and small scatter, while values derived in this work are systematically larger by $\sim$4\,km\,s$^{-1}$ in comparison with data of \cite{Fraser_etal_10}. The only significant outlier
is the $\varv \sin i$-value of HD~191243 in the work by \cite{MaPu08}, which is likely a statistical outlier, given the good accordance of the corresponding value in \cite{IACOBIII}.

\subsection{Elemental abundances and stellar metallicity}\label{section:abundances_and_metallicity}
The mean abundances of all the metal species studied here (which constitute the ten most abundant elements besides hydrogen and helium) 
along with their uncertainties and the number of analysed lines are summarised in Table \ref{tab:abundances}. In addition, 
the resulting metallicities of the sample stars are shown. For a conservative estimate of the error margins the $1\sigma$ 
sample standard deviation of individual line abundances was chosen, as tests on single line statistical uncertainty resulted in
unreasonably low margins. In general, these statistical uncertainties range from $\sim$0.05--0.10\,dex, and rarely exceed the 
latter value. The number of lines analysed per species and object is at least two in very few cases and usually much larger. 
Standard errors of the mean therefore amount to typically 0.02--0.03\,dex for the elemental abundances in each star. 
Metal mass fractions $Z$ ('metallicities') of the sample stars were calculated from the available metal abundances and are indicated
in the last column of Table~\ref{tab:abundances}. As these cover the ten most abundant metal species these should be representative for the 
sum of all metals. 

The systematic uncertainties depend primarily on the quality of the respective model atoms and on the uncertainties in
effective temperatures, surface gravities, and microturbulent velocities, see for example the discussions by
\citet{Przybillaetal00,Przybillaetal01a,Przybillaetal01b} and \citet{PrBu01}. Given the experience gained in these works, we expect the 
systematic uncertainties of the elemental abundances to amount to $\sim$0.1\,dex.

The derivation of abundances for all chemical species that show spectral lines in the optical allowed global 
synthetic spectra to be calculated, that is one model spectrum based on the derived atmospheric parameters and abundances per 
star. This also includes the blended features that were excluded from the chemical analysis.
As can be expected from the small abundance uncertainties, the reproduction of the observed spectra by the global 
synthetic spectrum is excellent overall, as shown for the exemplary case of HD~164353 in Appendix~\ref{section:appendixA},
Figs.~\ref{fig:HD164353_3900_4500} to \ref{fig:HD164353_8100_8700}. Apart from some occasional very weak features, for instance of \ion{S}{ii} where
the model atom would need to be extended to include more energy levels, all important
stellar spectral lines are included in the spectrum synthesis. Noticeable omissions are several interstellar ('IS') atomic 
features, such as the Ca H and K lines, the Na D lines, a \ion{K}{i} resonance line\footnote{Only the \ion{K}{i} $\lambda$7698.9\,{\AA} 
line is clearly visible in this case, while the other fine-structure component \ion{K}{i} $\lambda$7664.9\,{\AA} overlaps with a
saturated telluric O$_2$ line \citep[see e.g.][]{Kimeswengeretal21}, which depends on the radial velocity of the target star.}, the diffuse
interstellar bands (DIBs), and the telluric absorption 
features typically due to O$_2$ and H$_2$O bands that occur with increasing frequency towards the near-IR. Some residual problems 
remain for a few stellar lines, for example the mismatch of the H$\alpha$ Doppler core, which is likely caused by the 
(weak) stellar wind in this object and not accounted for by the present modelling approach. The widths of the two strongest 
\ion{He}{i} lines $\lambda$5875 and 6678\,{\AA} are not perfectly matched. It would certainly be worthwhile to investigate this 
further as the widths of all other helium lines are reproduced well, but this is beyond the scope of the present paper.
A few metal lines also show somewhat larger deviations, such as the \ion{C}{ii} $\lambda$6578/82\,{\AA} doublet, which may hint 
at the possibility that the model atom may need to be improved with respect to these lines. However, in view of the overall 
solution these are minor details, the few discrepant features were not considered for the analysis.

Previous work on abundances of early B-type stars in the solar neighbourhood (distances out to $\sim$400\,pc from the Sun) 
has found chemical homogeneity, establishing a present-day cosmic abundance standard 
\citep[CAS,][]{NiPr12,Przybillaetal13}, see Table~\ref{tab:abundances}. Such a comparison of abundances between the present
sample stars and the CAS is inappropriate here because of the widely different distances of the sample objects from the
Galactic centre (see Sect.~\ref{section:spectroscopic_distances}), for example $\sim$7\,kpc for HD~184943 versus $\sim$13\,kpc for 
HD~25914. For the same reason, an important test to verify the independence of abundances of atmospheric parameters such as 
$T_\mathrm{eff}$ and $\log g$ that could be made by \citet{NiPr12}, cannot be repeated here.
We note, however, that the supergiants closest to the Sun in the sample, HD~125288 and HD~164353, are consistent with the
CAS values within the mutual uncertainties, but they show overall larger abundances.
In particular the surface abundances of nitrogen show clear indication of mixing of the atmospheres with
CN-processed material from the stellar cores.

   \begin{figure}
   \centering
   \includegraphics[width=\hsize]{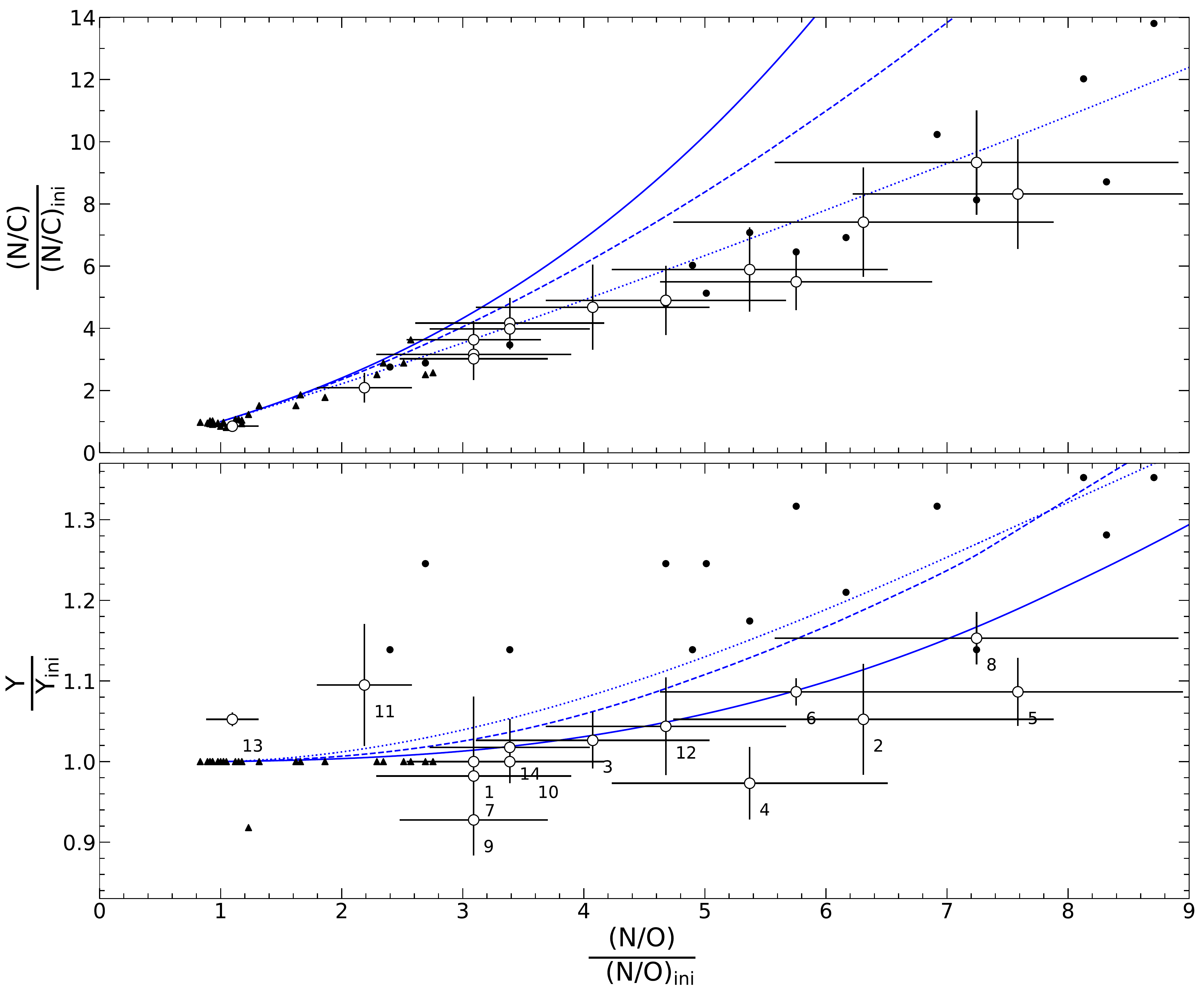}
      \caption{Nitrogen-to-carbon ratio (upper panel) and helium abundance (lower panel) versus nitrogen-to-oxygen ratio,
      normalised to cosmic abundance standard values \citep{NiPr12}. Objects from the present work (open
      symbols), B-type main-sequence stars \citep[][black triangles]{NiPr12}, and BA-type supergiants \citep[][black dots]{Przybillaetal10} 
      are compared to predictions from stellar evolution models. Solid lines: 15\,$M_\sun$, 
      $\Omega_{\mathrm{rot}}$\,=\,0.95\,$\Omega_{\mathrm{crit}}$ model by \cite{georgy_etal_13}; 
      dashed lines: 15\,$M_\sun$, $\Omega_{\mathrm{rot}}$\,=\,0.568\,$\Omega_{\mathrm{crit}}$  model by 
      \citet{Ekstroemetal12}; dotted lines: 25\,$M_\sun$, $\Omega_{\mathrm{rot}}$\,=\,0.568\,$\Omega_{\mathrm{crit}}$ model by 
      \citet{Ekstroemetal12}. For all tracks, a metallicity of $Z$\,=\,0.014 was assumed. Abundances in the models were 
      normalised with respect to their initial model values.}
         \label{fig:cnoy}
   \end{figure}

\subsection{Signatures of mixing with CNO-processed material}\label{section:CNO}
Different physical mechanisms can lead to mixing of CNO-cycled matter from the stellar core to the surface of rotating stars.
Examples are meridional circulation or shear mixing due to differential rotation \citep[e.g][]{MaMe12,Langer12} further modified by 
the presence of magnetic fields. As a consequence, ratios of the surface carbon, nitrogen, and oxygen mass fractions, and the 
helium mass fractions are expected to appear in relatively narrow regions in diagnostic diagrams \citep{Przybillaetal10,Maederetal14}
as shown in Fig.~\ref{fig:cnoy}. 
All ratios
were normalised to the initial values so as to make the comparison to the evolution tracks easier -- the observations were normalised relative 
to CAS abundances (see Table~\ref{tab:abundances}, $Y_{\mathrm{ini}}$\,=\,0.276), the models to their respective (solar) initial values.

   \begin{figure}
   \centering
   \includegraphics[width=.95\linewidth]{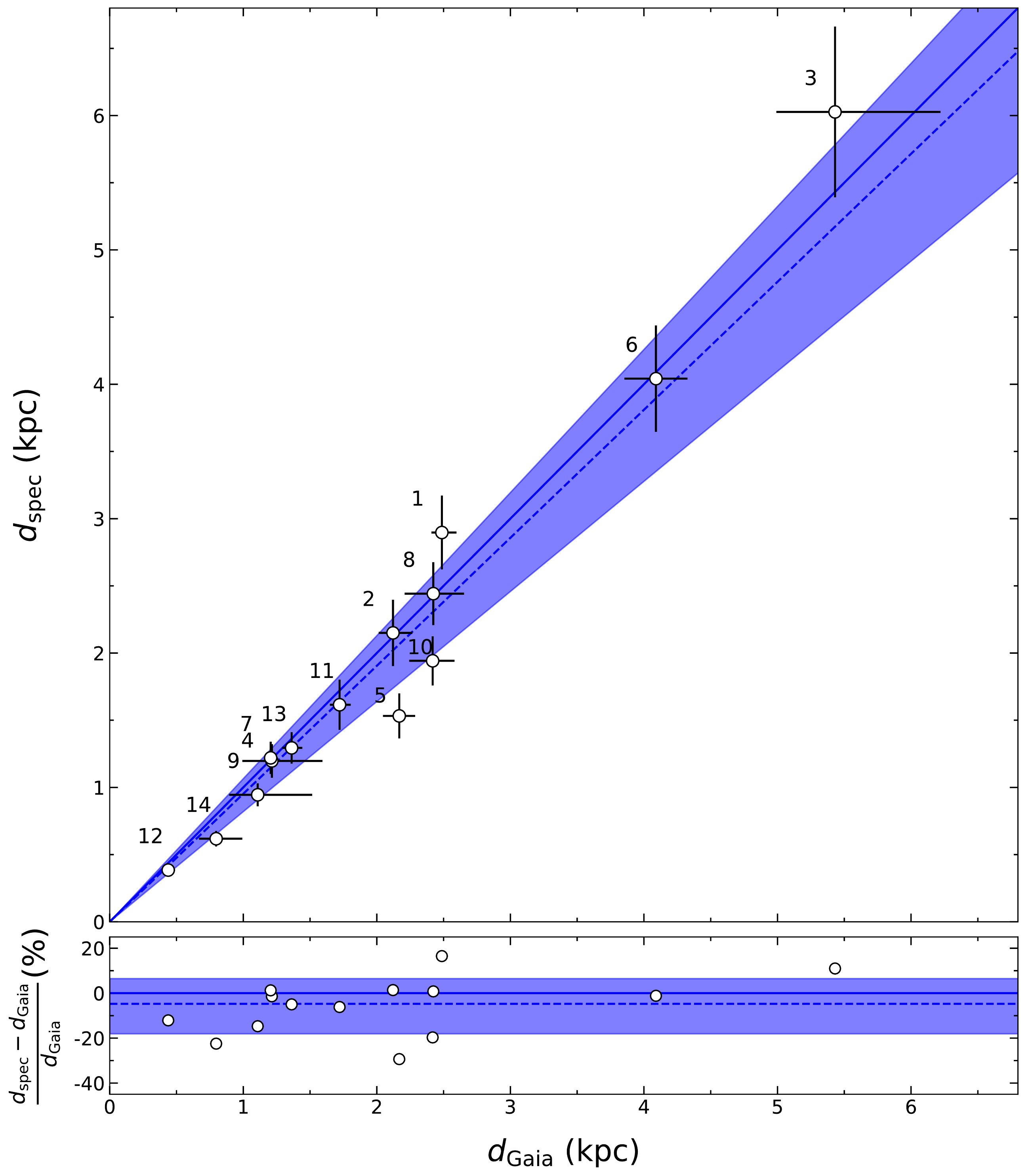}
      \caption{Comparison of spectroscopic distances derived in the present work and distances based on 
      Gaia EDR3 parallaxes (upper) and their relative differences (lower panel). The solid blue lines depict 
      equivalence, while the dashed lines show the best linear fit to the data. The shaded area marks the region of $1\sigma$ standard
      deviation from the mean. }
         \label{fig:distances}
   \end{figure}

As the $N/C$ versus~$N/O$ plot
shows little dependence on the initial stellar masses, rotation velocities, and nature of the mixing processes up to relative 
enrichment of $N/O$ by a factor of about four, it constitutes an ideal quality test for observational results 
\citep{Maederetal14}. The CNO signatures of the present sample supergiants closely follow both the path predicted by models and
the observational data of \citet{Przybillaetal10} and \citet{NiPr12}. This gives confidence that systematic errors 
in the present atmospheric parameters are indeed small.

The star HD~159110 (ID\,\#13) appears at CAS
initial values for CNO abundances, while HD~119646 (ID\,\#11) exhibits CNO abundances consistent with mixing signatures on the the main sequence (i.e. relative enrichment of $(N/O)/(N/O)_\mathrm{ini}$\,$\lesssim$\,3). The majority of the sample stars is noticeably enriched in CN-processed matter, with enhancement almost reaching the high values observed for some of the more evolved BA-type supergiants. %Generally, the trend set forth by the more evolved objects is closely matched by this samples derived abundances.}

For most of the analysed objects, helium abundances are slightly lower than predicted by the models while being consistent with the initial helium 
abundance within the 1$\sigma$ uncertainties. Three of the sample objects (ID\#9, \#11, and in particular \#13) deviate somewhat from this value while they are expected to show no modification. This is true when only statistical uncertainties are considered, which are displayed in Fig.~\ref{fig:cnoy}. However, potential systematic errors also need to be considered and we emphasise that the offset of ID\#13 from the 
initial value corresponds to only 0.02\,dex. Such differences are much smaller than the symbol sizes in the upper panel of Fig.~\ref{fig:cnoy}, stressing the enormous changes in mostly nitrogen (and to a lesser extent carbon and oxygen) abundances versus the enrichment of helium which is difficult to determine.
The highest helium enrichment found in our sample stars is about 15\% above the initial value. This is different to the BA-type supergiants, which show larger enhancement values. We note that different helium lines were
analysed by \citet{Przybillaetal10}, as many of the stars are cooler than the present sample stars. 
However, an investigation of the cause of these differences is beyond the scope of the present paper.

      \begin{figure}
   \centering
   \includegraphics[width=.81\linewidth]{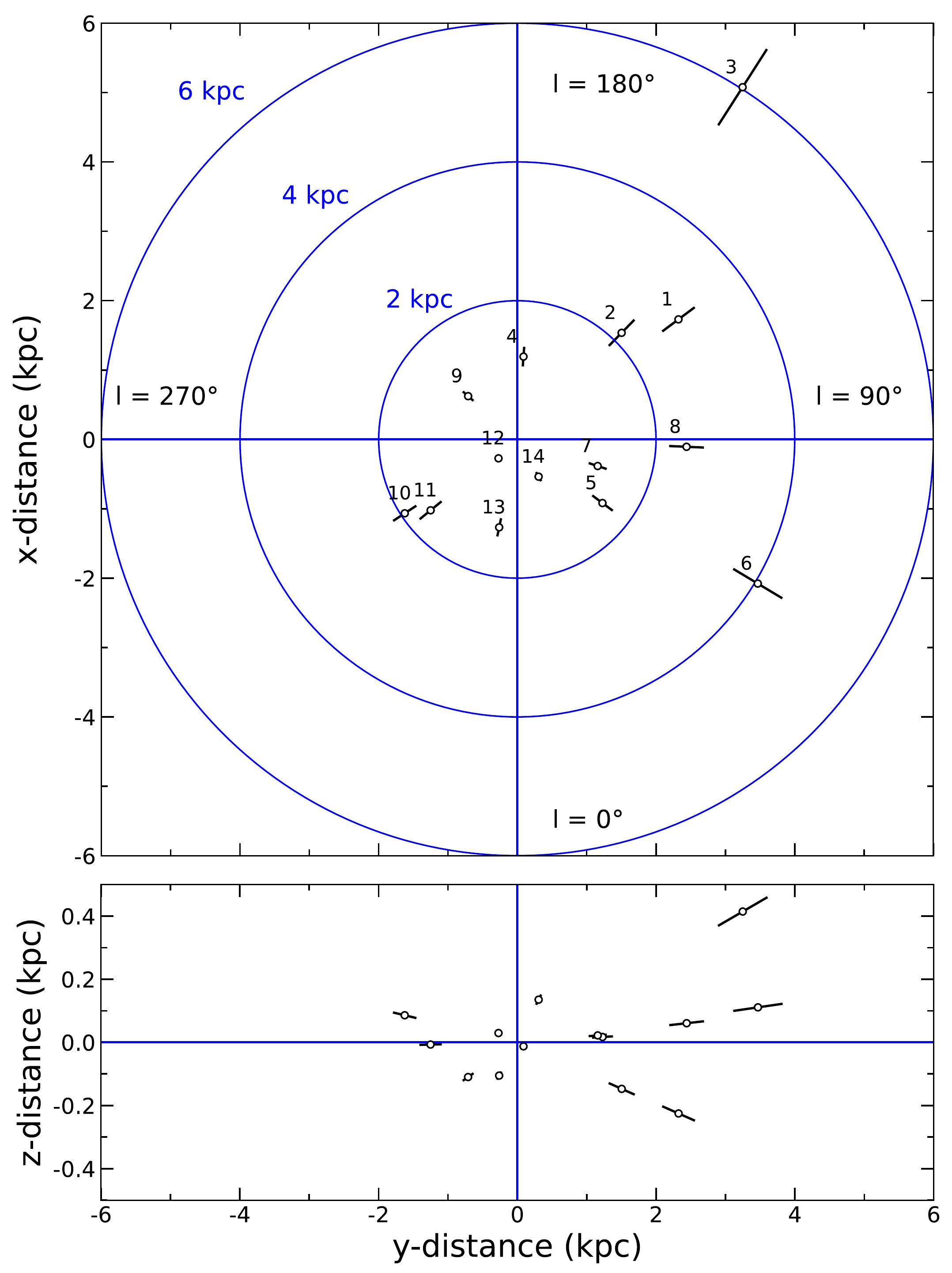}
      \caption{Distribution of sample stars in the Galactic plane (upper) and elevation above the plane (lower 
      panel) based on spectroscopic distances in Cartesian coordinates with the Sun at the origin. The Galactic centre lies
      towards the bottom in the upper panel, at Galactic longitude 0\degr. Blue circles enclose regions of equal distance 
      from the Sun, as indicated.}
         \label{fig:galactic_plane}
   \end{figure}

\subsection{Spectroscopic distances}\label{section:spectroscopic_distances}
The spectroscopic distances derived in this work depend on several parameters deduced both from the quantitative spectral
analysis and inferred from fundamental parameters on the basis of stellar evolution models, spectral synthesis codes, and
photometric data (see Eqn.~\ref{eq:spec_dist}). As already mentioned in brief in Sect.~\ref{sec:spec_dist_method}, the 
comparison with independent distance estimations (e.g. Gaia EDR3) can provide valuable insight into systematic problems in the
derivation of important parameters for the entire sample. It can, however, also highlight individual sample objects that may 
have undergone exceptional evolutionary pathways. Binarity (with and without associated mass transfer) and post-asymptotic 
giant branch evolutionary histories can leave signatures detectable in this approach. Regardless as to whether the star has evolved 
'normally' or not, it may be stated that the primary sources of uncertainty are evolutionary mass $M_{\mathrm{evol}}$ and 
surface gravity $\log g$, so that potential offsets in distances most probably stem from systematically biased parameters.

Figure~\ref{fig:distances} shows a direct comparison (upper panel) and relative difference (lower panel) of our 
spectroscopic distances $d_{\mathrm{spec}}$ and distances $d_{\mathrm{Gaia}}$ derived from Gaia EDR3 parallaxes. Specifically,
$d_{\mathrm{Gaia}}$ is the 'photogeometric' Bayesian estimation of distance by \cite{Bailer-Jones_etal_2021}, which in addition
to Gaia EDR3 parallaxes also takes into account the objects colour and apparent magnitude to achieve yet higher accuracy. In 
the direct comparison, we see a good agreement of the two distance determinations for the individual objects. The 
relative differences display a small mean offset of $\mu_s$\,=\,$-$6\% with a sample standard deviation of $\sigma_s$\,=\,12\%,
showing the excellent agreement between the distances. Even though most objects lie within about 2.5\,kpc from the Sun, the 
relationship does not seem to degrade noticeably at larger distances, as can be seen for the cases of HD~184943 (ID\,\#6,
$d_{\mathrm{spec}}$\,=\,4\,kpc) and HD~25914 (ID\,\#3, $d_{\mathrm{spec}}$\,=\,6\,kpc). Two of our sample stars, HD~7902 (ID\,\#1) and HD~183143 (ID\,\#5), depart somewhat from the mean relationship. While we cannot offer a robust explanation for the discrepancy in distance of either of these objects, we note that both  are evolved stars towards the upper mass limit of our sample. Small scale systematic errors in mass estimates, as discussed in Sect.~\ref{section:masses_radii_method}, are maximised in this region. For ID\#1
a mass reduced by 1\,$M_\odot$ would be sufficient to reach agreement within the mutual 1$\sigma$-uncertainties of the two distances. Maximum systematic effects would be needed for ID\#5 in this picture, requiring an initially non-rotating single star, but at the same time it is one of the two stars with the largest CNO mixing signature in the sample. This could possibly be interpreted in terms of a binary history. However, a further discussion of this is not warranted by the information available. Considering the offset $\mu_s$ of the 
relative differences is of the order of $-0.5\sigma_s$, we may conclude that the line of regression is compatible with an 
offset of zero. It may on the other hand reflect some unaccounted low-scale systematics, which, however, have no significant 
impact on the basic conclusions of the present work.

The distribution of the sample stars in the Galactic disk is depicted in Fig.~\ref{fig:galactic_plane}. 
The sample objects span Galactocentric distances in the range of $R_g$\,=\,7-13\,kpc 
\citep[calculated for a distance of the Sun to the Galactic centre of 8.178\,kpc,][]{GravityCol1aboration19}, 
while the range of elevations above and below the Galactic plane is fairly small, 
typically within 200\,pc. Only our outermost supergiant, HD~25914 (ID\#3), is located $\sim$0.4\,kpc above the Galactic plane. 
While the arrangement of the objects along the spiral arms is not immediately obvious, a closer inspection using the spiral arm delineation by \citet{Xuetal21} shows that stars with IDs\,\#10, 11, and 13 are located in the Carina-Sagittarius Arm, \#6, 7, 8, 9, 12, and 14 in the Local Arm, \#1, 2, and 4 are associated with the Perseus Arm and \#3 is situated~in~the~Outer~Arm. Star \#5 lies in the Sagittarius-Carina arm if $d_\mathrm{Gaia}$ is adopted, and otherwise between that and the Local Arm if $d_\mathrm{spec}$ is considered.   

\subsection{Sight lines -- reddening law}\label{section:sightlines}
\begin{figure}
   \centering
\includegraphics[height=7.2cm,width=.98\linewidth]{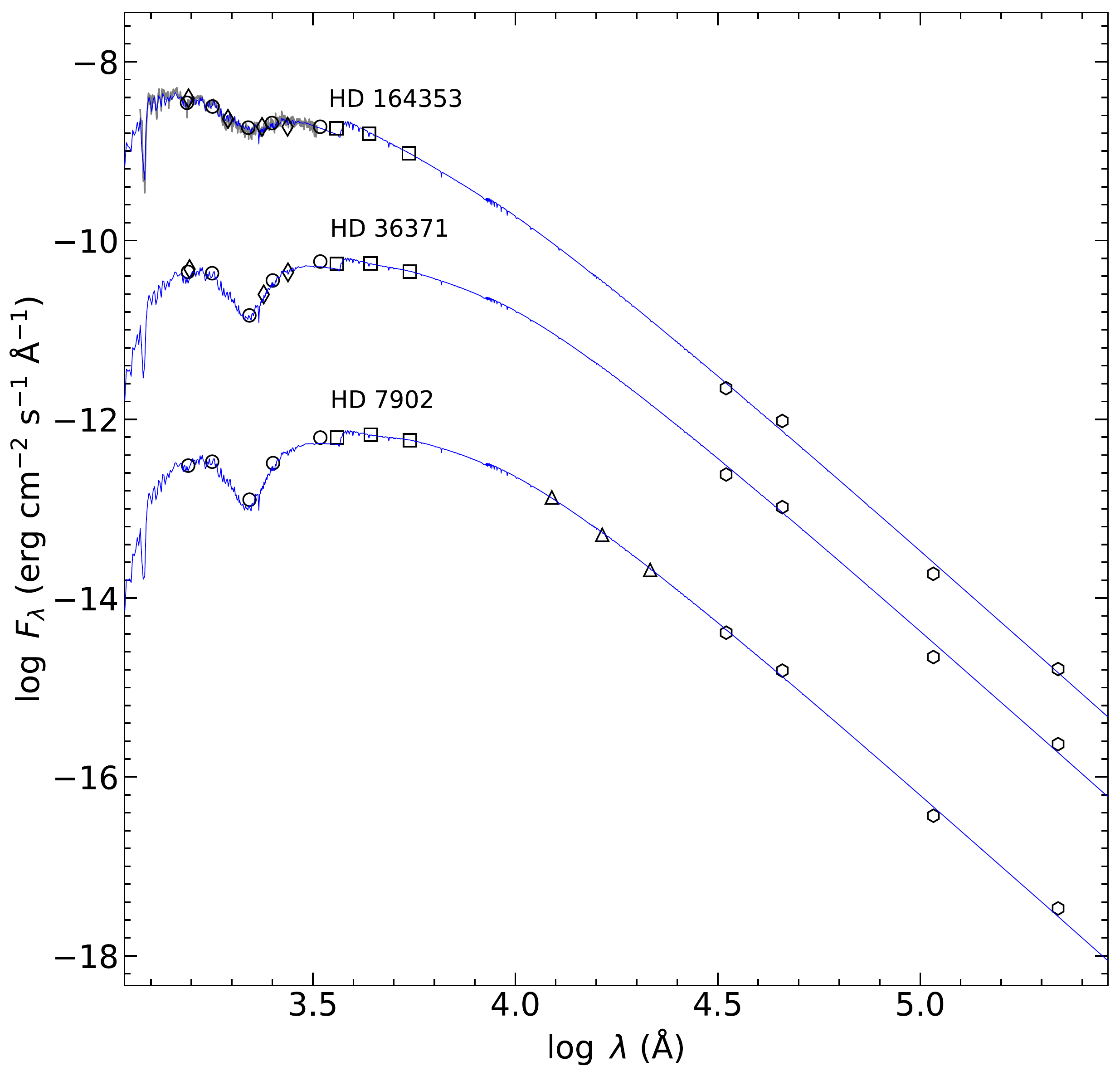}
      \caption{Examples of spectral energy distributions of sample stars. {\sc Atlas9}-SEDs, normalised in $V$ 
      and reddened 
      according to values from Table~\ref{tab:stellar_parameters} (blue lines) are compared  to IUE spectrophotometry 
      (grey lines) and photometric data in various wavelength bands: ANS (circles), TD1 (diamonds),  Johnson (squares), 
      2MASS (triangles), and ALLWISE data (hexagons). Data with bad quality flags were removed. For better visibility, the SEDs and photometry of HD~164353 and HD~7902 were shifted by $+1$ and $-1$\,dex, respectively.}
         \label{fig:sed_fits}
   \end{figure}

For the high-precision determination of the interstellar sight lines to the sample objects, the model \,{\sc Atlas9}-SEDs were 
fitted to photometric observations in various bands as well as to UV spectrophotometry from the IUE-satellite. 
Figure~\ref{fig:sed_fits} exemplarily summarises the result of this fitting process for three out of the 14 sample stars to give an impression 
of the quality of the fits, ordered from top to bottom by increasing values of colour excess $E(B-V)$. For most objects sufficient 
constraining observations were suitable for comparison 
and resulted in very small associated uncertainties in both $R_V$ and $E(B-V)$. 
Values for $R_V$ vary between 2.9 and 3.6, mostly concentrating around the typical ISM value of 3.1, and reddening values vary typically 
between 0.1 and 0.8, see Table~\ref{tab:stellar_parameters} for a summary of the results. The peculiar case of HD~183143 was already briefly
discussed in Sect.~\ref{section:sed_fitting}.

\subsection{Evolutionary status}\label{section:mass_estimates_and_discrepancy}
\begin{figure}
   \centering
   \includegraphics[width=.995\linewidth]{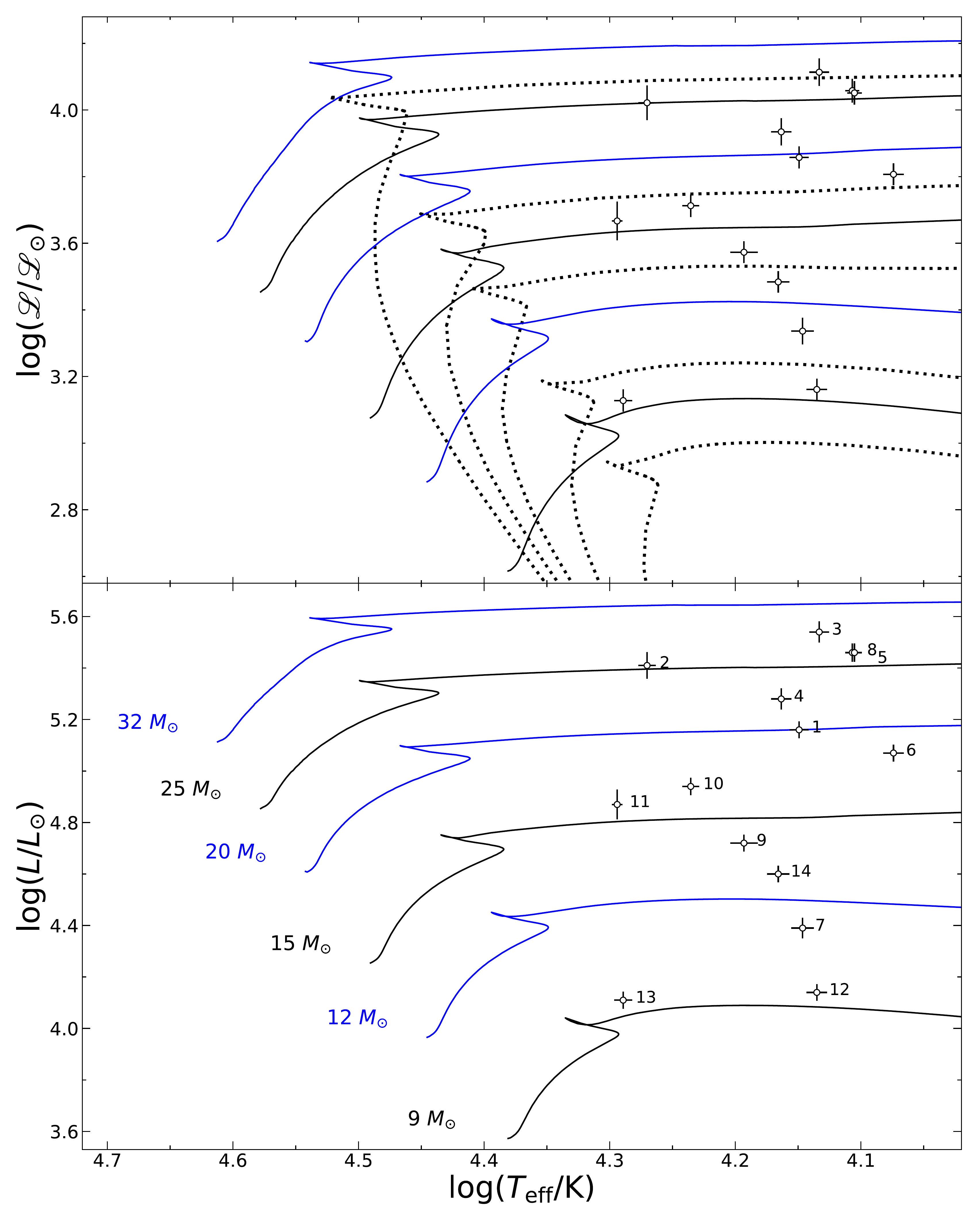}
      \caption{Location of the sample objects in two diagnostic diagrams compared to the loci of evolution tracks for stars 
      rotating with $\Omega_{\mathrm{rot}}$\,=\,0.568\,$\Omega_{\mathrm{crit}}$ by \cite{Ekstroemetal12}, for various
      ZAMS-masses as indicated. Isochrones for the model grid, corresponding to ages of $\log \tau_\mathrm{evol} \in \{6.85, 7.05, 7.20, 7.40, 7.60\}$ are depicted as dotted lines in the upper panel (increasing in age from top to bottom). Upper panel: sHRD; lower panel: HRD. 1$\sigma$ error bars are 
      indicated.}
         \label{fig:hrd_comp}
   \end{figure}

The evolutionary status of the sample stars can be derived by comparison to stellar evolution tracks. Two complementary diagnostic
diagrams may be employed for this, the spectroscopic HRD
\citep[sHRD, $\log(\mathscr{L}/\mathscr{L}_{\odot})$ versus $\log T_\mathrm{eff}$, introduced by][]{LaKu14} and the HRD
($\log L/L_\odot$ versus $\log T_\mathrm{eff}$). The sHRD is based only on observed atmospheric parameters (like the Kiel diagram -- $\log g$ versus $\log T_\mathrm{eff}$ --, not shown here), while the HRD
requires knowledge of the distance and corrections for interstellar extinction to be taken into account. Both were derived in the present
work and we give preference to spectroscopic distances. The positions of the sample stars in both diagrams with respect to evolutionary tracks for
rotating stars by \citet{Ekstroemetal12} are shown in Fig.~\ref{fig:hrd_comp}. We note the very similar positions of the sample
stars relative to the evolution tracks. This is consistent with them likely being post-main-sequence objects with ZAMS masses between about 9 to 30\,$M_\sun$ on the first crossing of the HRD towards the red supergiant phase
(star \#13 is a  potential exception, it may alternatively be in the last stages of core H-burning, depending on its detailed properties). 
They are located on the cool side of the bi-stability jump for stellar winds \citep[e.g.][]{Lamersetal95} and are slowly rotating, with $\varv 
\sin i$ in the range of about 20 to 50\,km\,s$^{-1}$, as expected for such B-type supergiants \citep[see e.g.][]{Vinketal10}. Evolutionary ages vary between about 7\,Myr for the most massive to about 29\,Myr 
for the least massive sample objects, as inferred from isochrones indicated in the upper panel of Fig.~\ref{fig:hrd_comp}. We emphasise again that the masses and ages are derived assuming that the particular rotation rates in evolution models and isochrones are representative on average for the sample. Systematic shifts in mass and age result if the initial rotational velocities had other values, but we expect them to be covered by our uncertainties in most cases.

We note that the sample stars show a variety of metallicities (see Table~\ref{tab:abundances}) because of their different positions in the 
Galactic disk. The most metal-poor star in the present work is HD~25914 at $Z$\,=\,0.010 in the Outer Arm, while several objects reach
super-CAS metallicities in the inner Milky Way, up to $Z$\,=\,0.019, that is the variations reach up to about 30\% below and 40\% above the
CAS value. Moreover, the chemical composition of the sample stars varies from (scaled) solar, as implemented by \citet{Ekstroemetal12}
for the $Z$\,=\,0.014 models, to the bracketing analogous $Z$\,=0.006 \citep{Eggenbergeretal21} and $Z$\,=\,0.020 models 
\citep{Yusofetal22}. The net effects are a more efficient transport of angular momentum and CNO-processed material with decreasing metallicity 
and a higher mass-loss with increasing metallicity. However, we do not see the resulting differences as critical for the present work in
terms of parameters deduced from the comparison such as ZAMS or evolutionary masses. The evolutionary tracks remain similar
throughout the metallicity range \citep[see e.g.][their Fig.~5]{Yusofetal22}, such that the resulting systematics are expected to lie within our uncertainties.

The number of sample stars is too low to investigate the effects responsible for the mixing of the surface layers with
CNO-processed material from the core systematically. Two findings are in line with the general picture of rotational mixing: the two stars with CNO 
signatures closest to the pristine values (\#11, \#13) are closest to the terminal-age main sequence, towards lower masses, and the stars
showing the highest processing (\#5 and \#8) are the most evolved (i.e. showing the coolest temperatures) and tend to be among the 
most massive sample stars. On the other hand, star \#3 -- the most massive and most metal-poor object of the sample -- shows only a milder degree of chemical mixing, which may be the consequence of an initially slower rotation than average. The issue has to be revisited based on a much larger sample of objects.

\section{Test for extragalactic applications at intermediate spectral resolution}\label{section:intermediateR}
\begin{figure*}
\centering 
    \resizebox{0.78\textwidth}{!}{\includegraphics{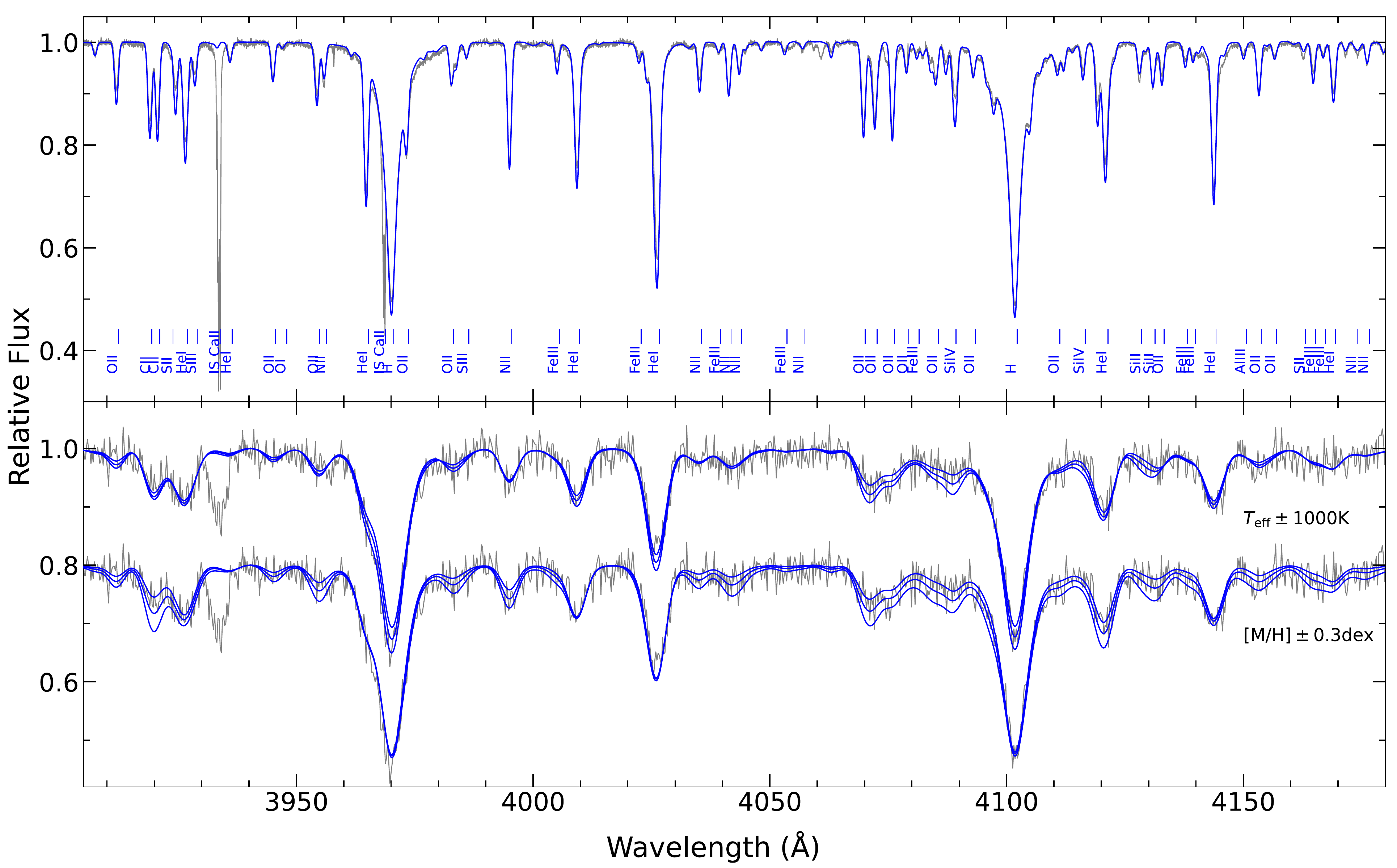}}\\
    \resizebox{0.78\textwidth}{!}{\includegraphics{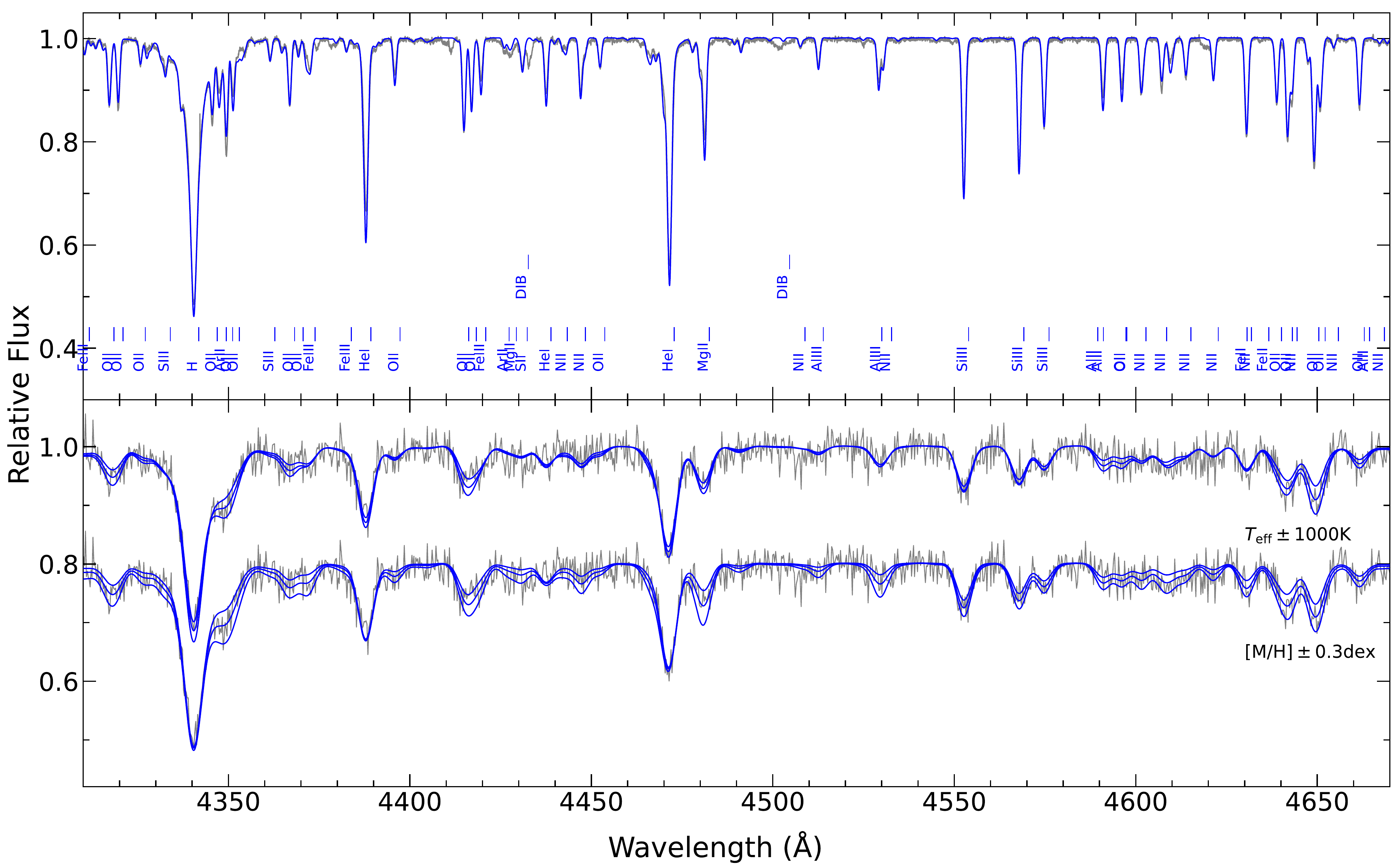}}
      \caption{Testing the suitability of our modelling for analyses at intermediate spectral resolution, e.g. for extragalactic 
      applications, for the blue spectral regions of
      about 3900 to 4180\,{\AA} (upper) and 4310 to 4670\,{\AA} (lower panel set). The upper panel of each set shows the comparison 
      of our best-fit synthetic spectrum for HD~119646 (blue line) with the observed high-resolution spectrum (grey line). The 
      important spectral features are identified, including interstellar features that were not modelled. The lower panels of each set show
      the same spectra, however degraded to $R$\,=\,1000, with the observations degraded to a $S/N$\,=\,50 and showing some degree of 
      oversampling. Two sets of model curves are shown in these cases, exemplifying the impact of a $T_\mathrm{eff}$-variation by 
      $\pm$1000\,K and the effects of a metallicity variation of $\pm$0.3\,dex, as indicated.}  
         \label{fig:extragal_3900_4200}
\end{figure*}

High-resolution spectroscopy of B-type supergiants as presented here can be conducted in galaxies beyond the Magellanic Clouds only at 
the cost of long exposure times of the order of hours on large telescopes \citep[e.g.][]{Urbanejaetal11}. Fortunately, many of the stronger
diagnostic lines are
isolated, so that intermediate-resolution spectroscopy ($R$\,$\simeq$\,1000--5000) suffices to allow quantitative analyses, at 
the loss of only the weaker spectral lines. This also opens up the possibility of employing multi-object spectroscopy, in particular when 
investigating galaxies beyond the Local Group, providing multiplexing of the order a few tens to hundreds of objects to be observed 
simultaneously. This comprises various successful techniques that have been implemented already as multi-slit spectroscopy 
\citep[e.g. with the FOcal Reducer/low dispersion Spectrograph 2, FORS2, on the ESO Very Large Telescope VLT, e.g.][]{Kudritzkietal16},
multi-fibre spectroscopy \citep[with the Large Sky Area Multi-Object Fibre Spectroscopic Telescope, LAMOST, e.g.][]{Liuetal22}
or integral-field spectroscopy \citep[as with the Multi Unit Spectroscopic Explorer, MUSE, on the VLT, e.g.][]{Gonzalez-Toraetal22}.  
   
Isolated spectral lines of H, He, C, N, O, Mg, and Si, or pure blends thereof, are strong enough to allow atmospheric parameters and
individual elemental abundances to be constrained even at intermediate spectral resolution. The blue spectral region from the Balmer jump 
to about 5000\,{\AA} is particularly useful for analyses, preferentially towards the earlier B spectral types, as they show stronger and
a larger number of metal lines, see Fig.~\ref{fig:spect_lum_showcase}. The comparison of the final synthetic spectrum and the observed 
spectrum for the hottest of the sample stars in two extended blue spectral windows is shown in Fig.~\ref{fig:extragal_3900_4200}. The same 
comparison, however for an artificially downgraded $R$\,=\,1000 reachable with FORS2, is shown in the lower subpanels, where a 
$S/N$\,=\,50 is simulated for the observation. Excellent agreement is achieved in both cases, except for some small details. Moreover, 
modified models by $\pm$1000\,K in $T_\mathrm{eff}$ and $\pm$0.3\,dex in metal abundances are also shown in the 
intermediate-resolution case. This shows that a simultaneous evaluation of all the spectral features, both the atmospheric 
parameters ($\log g$ is constrained by the response of the Balmer lines, not shown here) as well as the elemental abundances 
can be performed using $\chi^2$ minimisation techniques in the multi-parameter space, with uncertainties that are only 
slightly larger than in the high-resolution case: $\Delta T_\mathrm{eff}$ in the range of about 300-1000\,K, $\Delta \log g$ of about 
0.10\,dex, and elemental abundances in the range of about 0.10 to 0.15\,dex. We note in particular that the ionisation equilibria 
\ion{Si}{ii/iii(/iv)}, and in the case that red wavelengths are also covered \ion{O}{i/ii}, remain available at intermediate resolution.
The microturbulent velocity can best be constrained from the rather numerous \ion{Si}{ii/iii} and \ion{O}{i/ii} lines 
\citep[in contrast to the minimalistic approach of concentrating only on the \ion{Si}{iii} triplet 4552-4574\,{\AA}, e.g.][]{Hunteretal07}.

We conclude that the present hybrid non-LTE spectrum synthesis technique based on reliable model atoms allows for  comprehensive
quantitative analyses of B-type supergiants on the basis of intermediate-resolution spectra. This opens up the prospect of B-type supergiants as versatile 
tools to address a number of highly-relevant astrophysical topics in the context of extragalactic stellar astronomy. 

Even with available instrumentation on the current generation of 8-10m telescopes, a wide range of detailed studies, in particular concerning
galactic evolution \citep[galactic abundance gradients, the galaxy mass-metallicity relationship, e.g.][]{Urbanejaetal05a,Kudritzkietal12,Kudritzkietal14,Castroetal12} and the cosmic distance scale \citep[via application 
of the FGLR, e.g.][]{Urbanejaetal17}, can be addressed by investigating supergiants in galaxies in the field and in the nearby galaxy groups. With the advent of the
Extremely Large Telescopes (ELTs), the step 
to investigate supergiants in galaxies in the nearby Virgo and Fornax galaxy clusters will become feasible, allowing environmental 
effects to be studied. However, as adaptive optics techniques will be required to reach the full potential of the ELTs in terms of spatial 
resolution, spectroscopic observations will have to concentrate on redder wavelength regions, at least initially. For example, the High Angular
Resolution Monolithic Optical and Near-infrared Integral field spectrograph \citep[HARMONI,][]{Thatteetal21} on the ESO ELT will cover 
wavelengths beyond 4700\,{\AA} and the multi-object spectrograph MOSAIC \citep{Hammeretal21} beyond 4500\,{\AA}. The information content 
will be lower than at bluer wavelengths, but suitable spectral lines for analyses are present, see the figures in Appendix~\ref{section:appendixA}. 
Important for the scientific return will be to achieve wide wavelength coverage.

\section{Summary and conclusions}\label{section:conclusions}
A hybrid non-LTE spectrum synthesis approach for quantitative analyses of luminous B-type supergiants with masses up to about 
30\,$M_{\odot}$ was presented, where most spectral lines are formed in a photosphere that is not significantly affected by the stellar 
wind. It was shown that practically the entire observed optical to near-IR high-resolution spectra can be reliably reproduced, including 
the dozen chemical elements with the highest abundances. The modelling was thoroughly tested for 14 sample objects spanning a 
$T_\mathrm{eff}$-range from about 12\,000 to 20\,000\,K (i.e. spectral types B8 to B1.5) and luminosity classes II, Ib, Iab, and Ia. 
The present work helps to connect the region of late O- and early B-type stars on the main-sequence with luminosity classes V to III \citep{NiPr12,NiPr14} 
and the cooler BA-type supergiants \citep{Przybillaetal06,FiPr12}, which will allow stellar evolution to be tracked observationally throughout the hot 
regime of the HRD in a homogeneous manner. 

Due to the highly interactive and iterative nature of the approach, the time required to carry out the analysis procedure for a comprehensive solution of one sample object amounts to typically 2 weeks for experienced users. For a demonstration of the applicability of a method and a first application, this is an acceptable time investment. But, obviously, a combination of the models with faster, more automatised state-of-the-art analysis techniques \citep[see e.g. Sect. 3.8 of][]{Simon-Diaz20} is required for future larger-scale applications.

It has been shown that the atmospheric parameters of B-type supergiants can be determined with high precision and accuracy using the 
hybrid non-LTE approach. The effects of turbulent pressure were taken into account for the first time for B-type supergiants, and they lead to 
(small) systematic shifts in the atmospheric parameters. Effective temperatures can be constrained to 2-3\% uncertainty, surface gravities 
to better than 0.07\,dex uncertainty, and elemental abundances with uncertainties of 0.05 to 0.10\,dex (statistical 1$\sigma$-scatter) and
about 0.1\,dex (systematic error). Classical LTE analyses that can be partly successful for the analysis of main-sequence stars at 
similar $T_\mathrm{eff}$ cannot be expected to yield any meaningful results for supergiant analyses 
(Fig.~\ref{fig:tlusty_atlas12_HHE_20k250v10} gives an impression of the differences). 

Precise and accurate atmospheric parameters  also allow an improved characterisation of the interstellar reddening and the reddening 
law along the sight lines towards the supergiants to be made. The importance of B-type supergiants in this context lies in their large 
luminosities, so that sight lines to very distant parts of the Milky Way may become traceable in the era of large spectroscopic surveys
\citep[e.g.][]{Xiangetal22}. A comparison with stellar evolution models then also allows the fundamental parameters to be determined. In
particular future Gaia data releases will help to further reduce the uncertainties for Galactic supergiants by providing stronger astro- and photometric constraints to cross-check resulting spectroscopic solutions.

Most of the sample stars show signatures of the surface layers having experienced (rotational) mixing with CNO-processed material from the
core, and the positions of the stars in the HRD are consistent with hydrogen shell-burning being active but core He-burning probably having not yet ignited \citep[which happens for the investigated mass range earliest at $\log T_\mathrm{eff}$\,$\simeq$\,4.1, and cooler, according to the models of][]{Ekstroemetal12}. Unlike main-sequence early B-type stars in the solar neighbourhood \citep{NiPr12}, the B-supergiant sample does not show 
chemical homogeneity for the heavier elements. However, this is not unexpected, as the objects cover a wider range of 
Galactocentric distances, that is they will be subject to Galactic abundance gradients \citep[e.g.][]{Mendez-Delgadoetal22}.

Finally, it was shown that the full spectrum synthesis approach makes applications to intermediate-resolution spectra possible, with only 
slightly increased error margins. Extragalactic samples of B- and A-type supergiants below the $\sim$30\,$M_\odot$ limit can therefore be 
analysed homogeneously in the future. A major step for such applications will be reached once multi-object spectrographs on ELTs become
available. Quantitative spectroscopy of supergiants in the star-forming galaxies of the Virgo and Fornax galaxy clusters can then
commence, allowing galaxy evolution in the different environments to be studied -- in the field, in groups, and in clusters -- in more detail 
than currently feasible on the basis of gaseous nebulae.

\begin{acknowledgements}
D.W. and N.P. gratefully acknowledge support from the Austrian Science Fund FWF project DK-ALM, grant W1259-N27. Based on data obtained from the ESO Science Archive Facility with DOI(s): \url{https://doi.org/10.18727/archive/24}. Based on observations collected at the Centro Astron\'omico Hispano 
Alem\'an at Calar Alto (CAHA), operated jointly by the Max-Planck Institut f\"ur  Astronomie and the Instituto de 
Astrof\'isica de Andaluc\'ia (CSIC), proposals H2001-2.2-011 and H2005-2.2-016. Travel of N.P. to the Calar Alto 
Observatory was supported by the Deutsche Forschungsgemeinschaft (DFG) under grant PR 685/1-1. The latter observational data are available under \url{https://doi.org/10.5281/zenodo.6802567}.
We are grateful to A.~Irrgang for several updates of {\sc Detail} and {\sc Surface}. We thank the referee for useful suggestions to improve on the clarity of the paper.
This work has made use of data from the European Space Agency (ESA) mission
{\it Gaia} (\url{https://www.cosmos.esa.int/gaia}), processed by the {\it Gaia}
Data Processing and Analysis Consortium (DPAC,
\url{https://www.cosmos.esa.int/web/gaia/dpac/consortium}). Funding for the DPAC
has been provided by national institutions, in particular the institutions
participating in the {\it Gaia} Multilateral Agreement.
This publication makes use of data products from the Two Micron All Sky Survey, which is a joint project of the University of Massachusetts
and the Infrared Processing and Analysis Center/California Institute of Technology, funded by the National Aeronautics and Space 
Administration and the National Science Foundation.
This publication makes use of data products from the Wide-field Infrared Survey Explorer, which is a joint project of 
the University of California, Los Angeles, and the Jet Propulsion Laboratory/California Institute of Technology, funded 
by the National Aeronautics and Space Administration. This research has made use of the SVO Filter Profile Service 
(http://svo2.cab.inta-csic.es/theory/fps/) supported from the Spanish MINECO through grant AYA2017-84089.
\end{acknowledgements}


\typeout{}
\bibliographystyle{aa}
\bibliography{biblio.bib}

\begin{appendix} %First appendix
\section{Example of a global model fit}\label{section:appendixA}
\begin{figure*}
\centering 
\includegraphics[angle=90,width=0.8\textwidth]{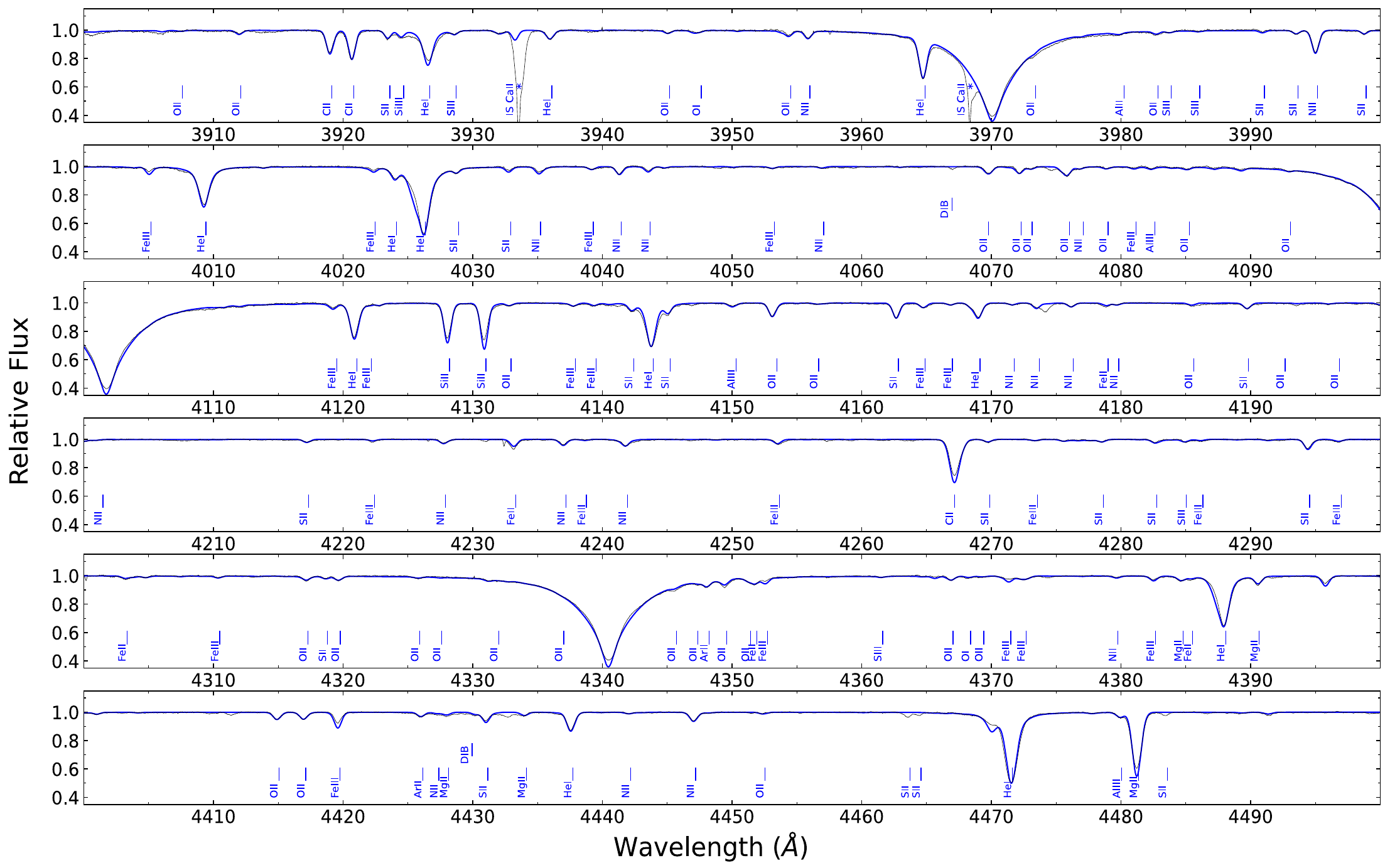}
        \caption{Comparison between the observed spectrum of HD~164353 (black) and the best fitting synthetic spectrum (blue) in the wavelength range of 3900 to 4500\,{\AA}.}
    \label{fig:HD164353_3900_4500}
\end{figure*}

\begin{figure*}
\centering 
\includegraphics[angle=90,width=0.8\textwidth]{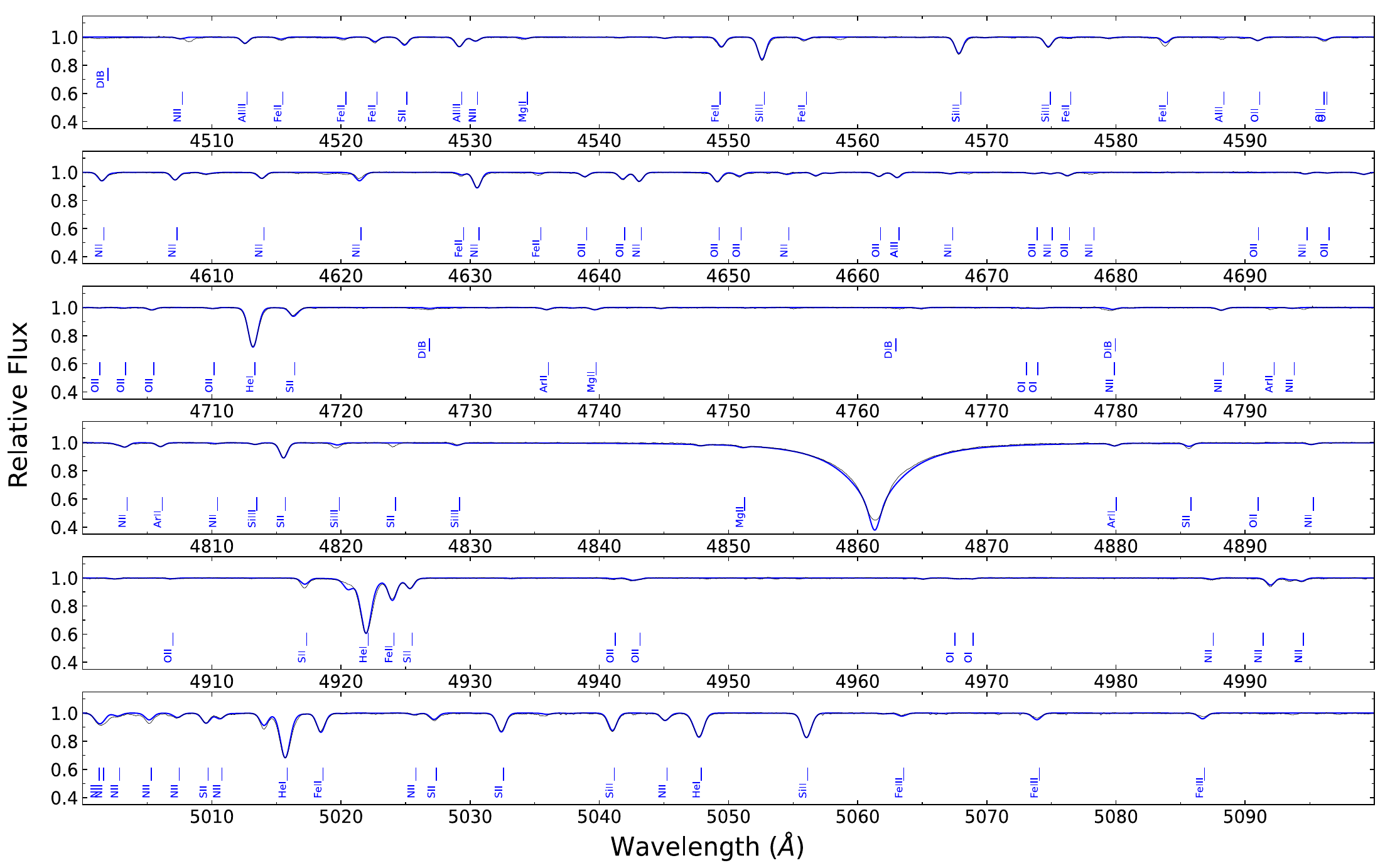}
        \caption{Same as Fig.~\ref{fig:HD164353_3900_4500}, but in the wavelength range $\lambda\lambda$4500--5100\,{\AA}.}
    \label{fig:HD164353_4500_5100}
\end{figure*}

\begin{figure*}
\centering 
\includegraphics[angle=90,width=0.8\textwidth]{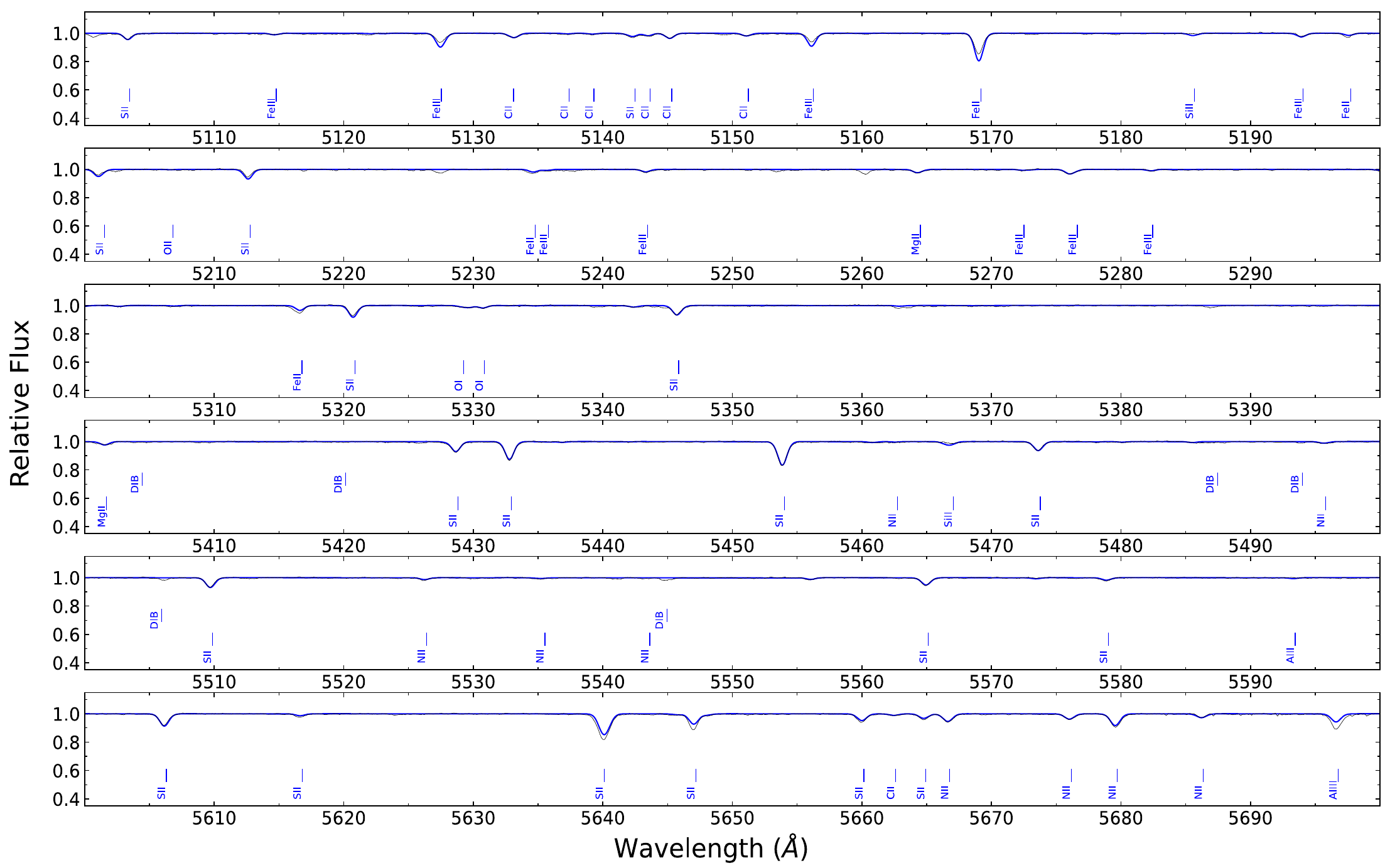}
        \caption{Same as Fig.~\ref{fig:HD164353_3900_4500}, but in the wavelength range $\lambda\lambda$5100--5700\,{\AA}.}
    \label{fig:HD164353_5100_5700}
\end{figure*}

\begin{figure*}
\centering 
\includegraphics[angle=90,width=0.8\textwidth]{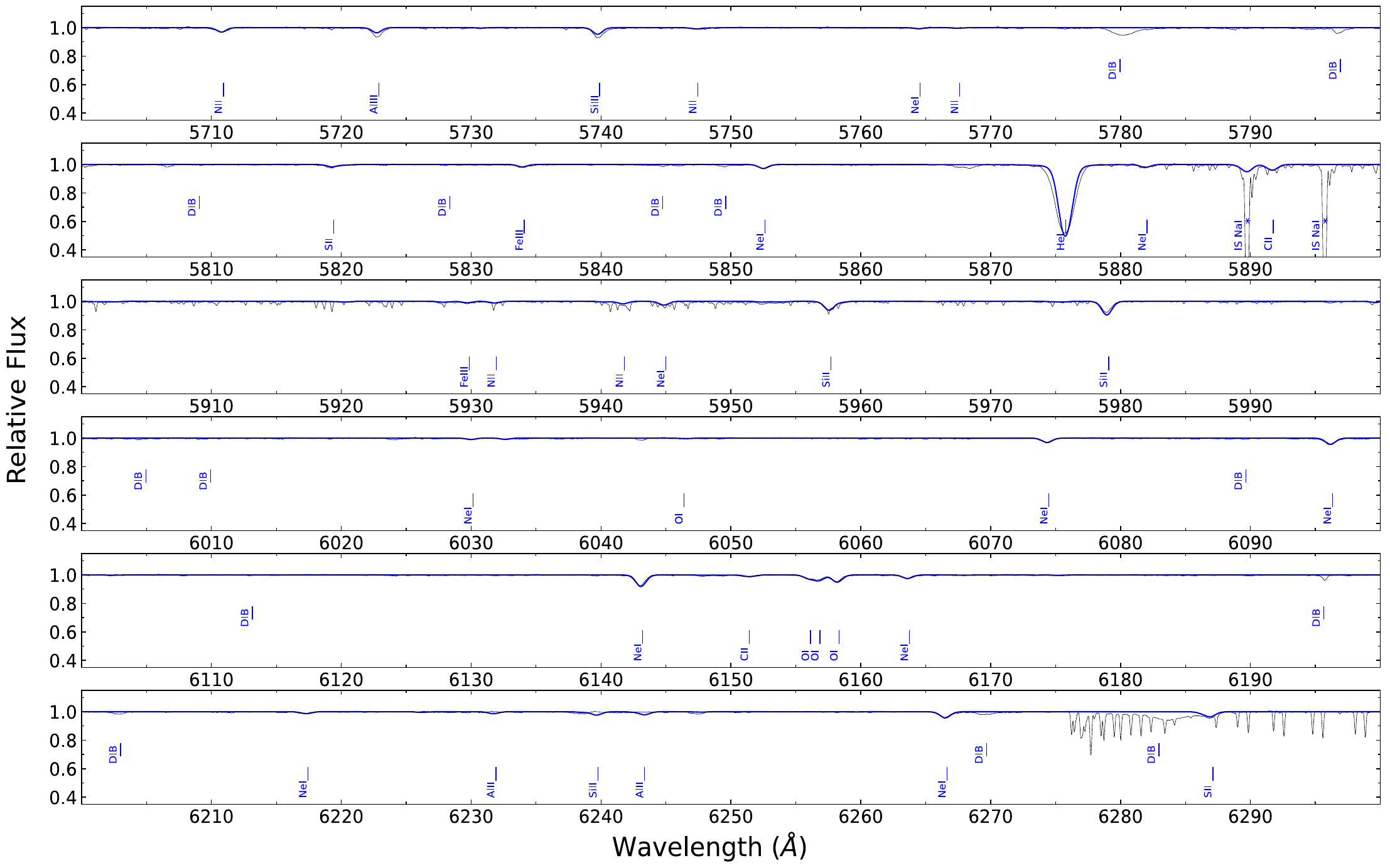}
        \caption{Same as Fig.~\ref{fig:HD164353_3900_4500}, but in the wavelength range $\lambda\lambda$5700--6300\,{\AA}.}
    \label{fig:HD164353_5700_6300}
\end{figure*}

\begin{figure*}
\centering 
\includegraphics[angle=90,width=0.8\textwidth]{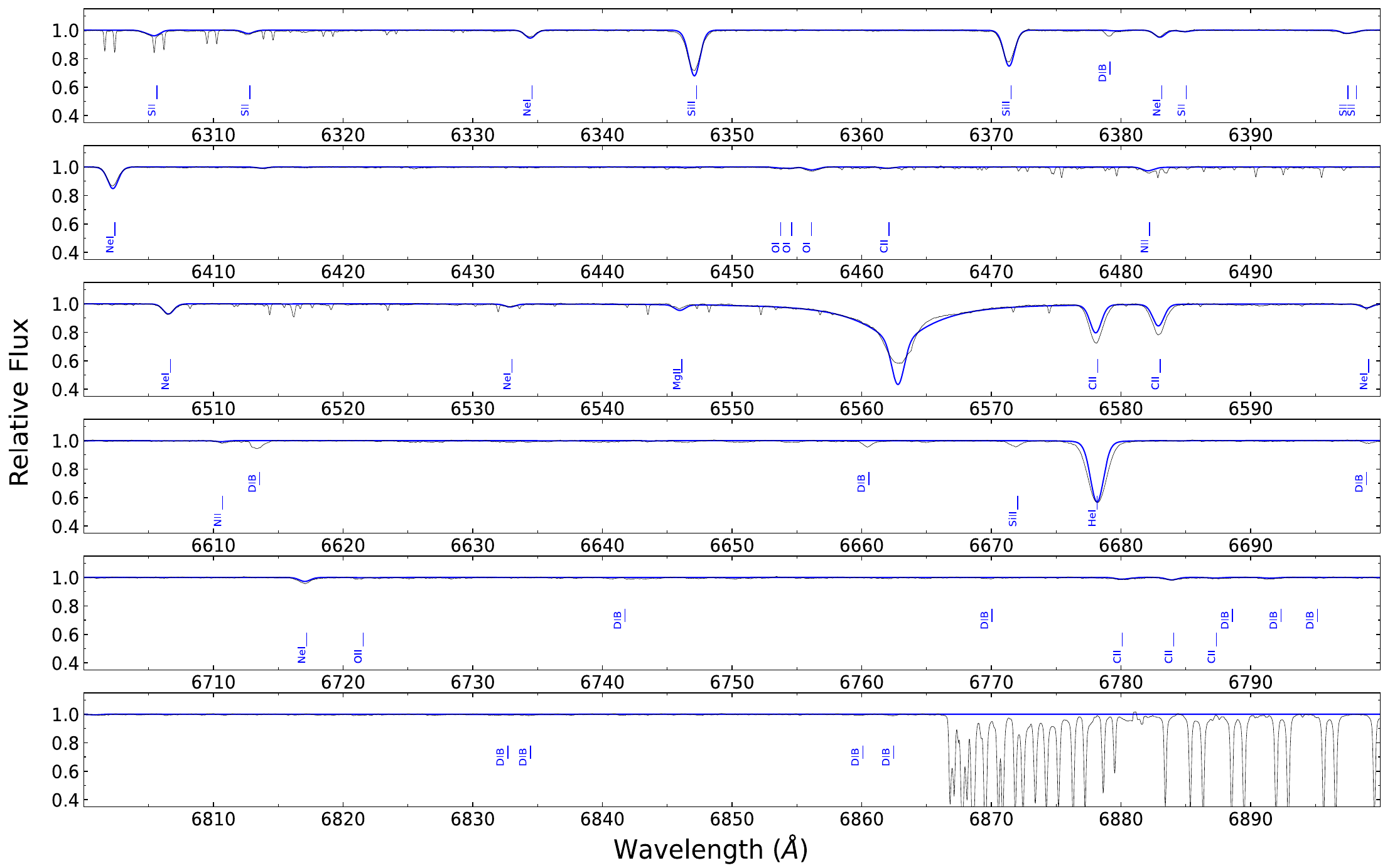}
        \caption{Same as Fig.~\ref{fig:HD164353_3900_4500}, but in the wavelength range $\lambda\lambda$6300--6900\,{\AA}.}
    \label{fig:HD164353_6300_6900}
\end{figure*}

\begin{figure*}
\centering 
\includegraphics[angle=90,width=0.8\textwidth]{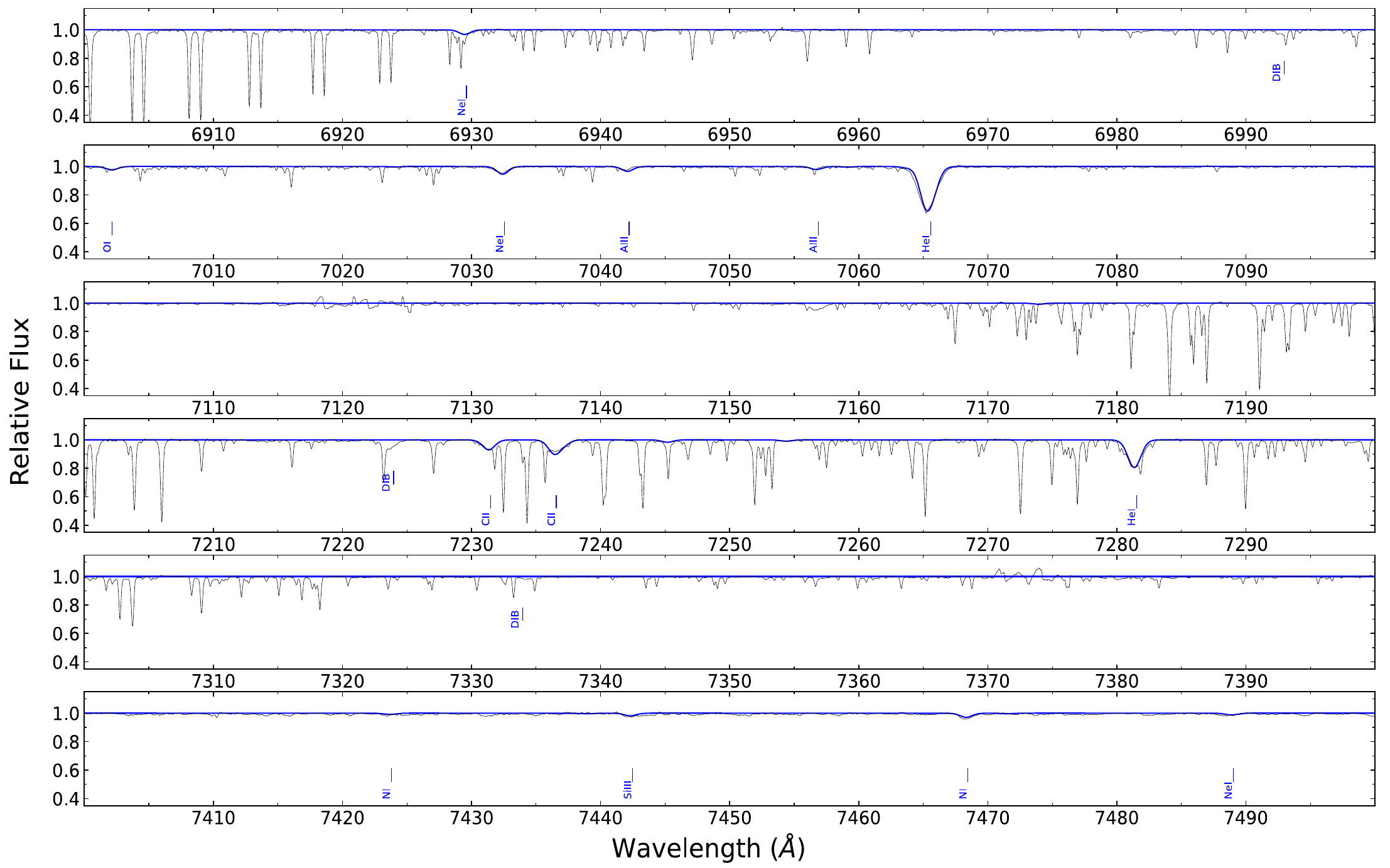}
        \caption{Same as Fig.~\ref{fig:HD164353_3900_4500}, but in the wavelength range $\lambda\lambda$6900--7500\,{\AA}.}
    \label{fig:HD164353_6900_7500}
\end{figure*}

\begin{figure*}
\centering 
\includegraphics[angle=90,width=0.8\textwidth]{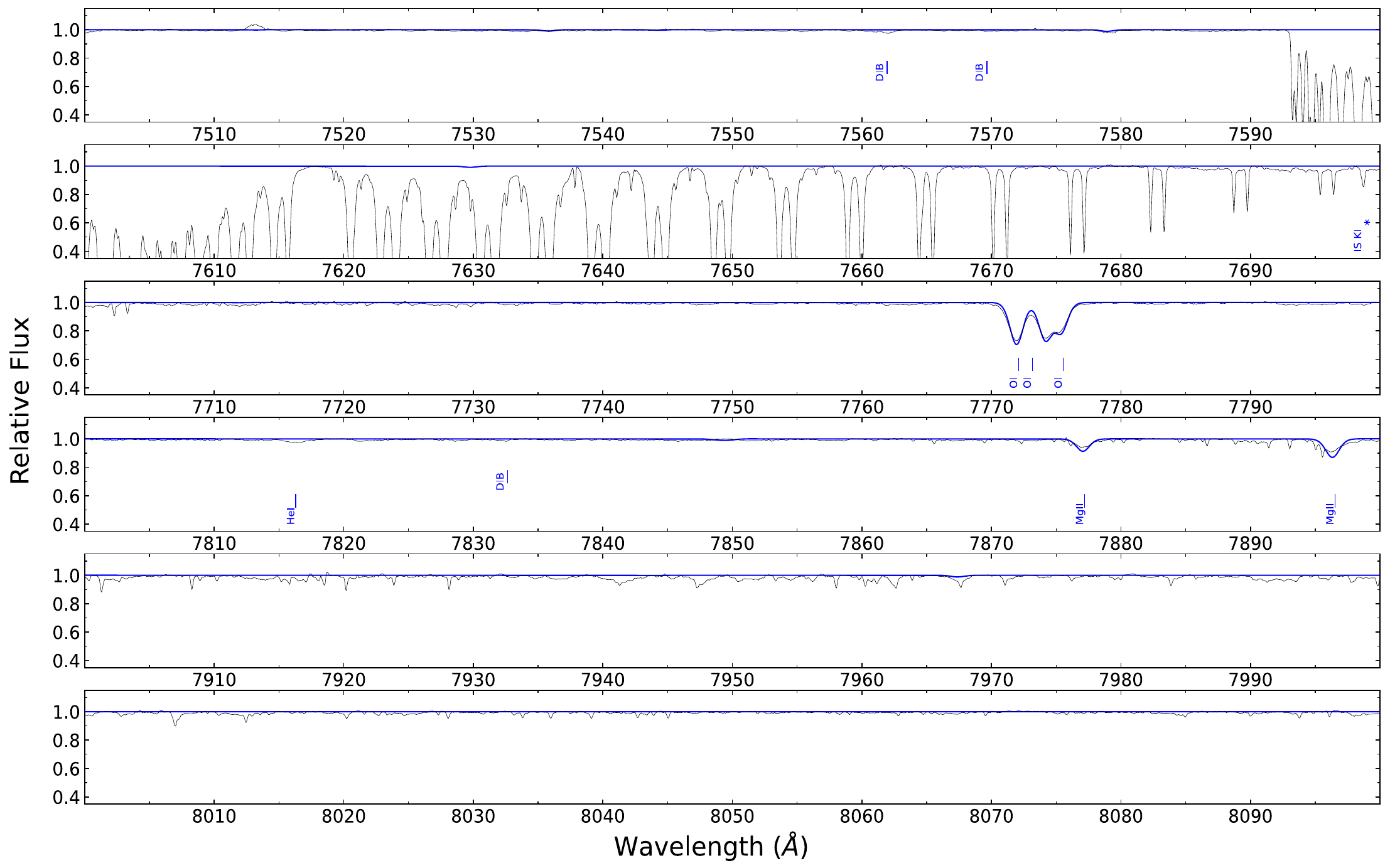}
        \caption{Same as Fig.~\ref{fig:HD164353_3900_4500}, but in the wavelength range $\lambda\lambda$7500--8100\,{\AA}.}
    \label{fig:HD164353_7500_8100}
\end{figure*}

\begin{figure*}
\centering 
\includegraphics[angle=90,width=0.8\textwidth]{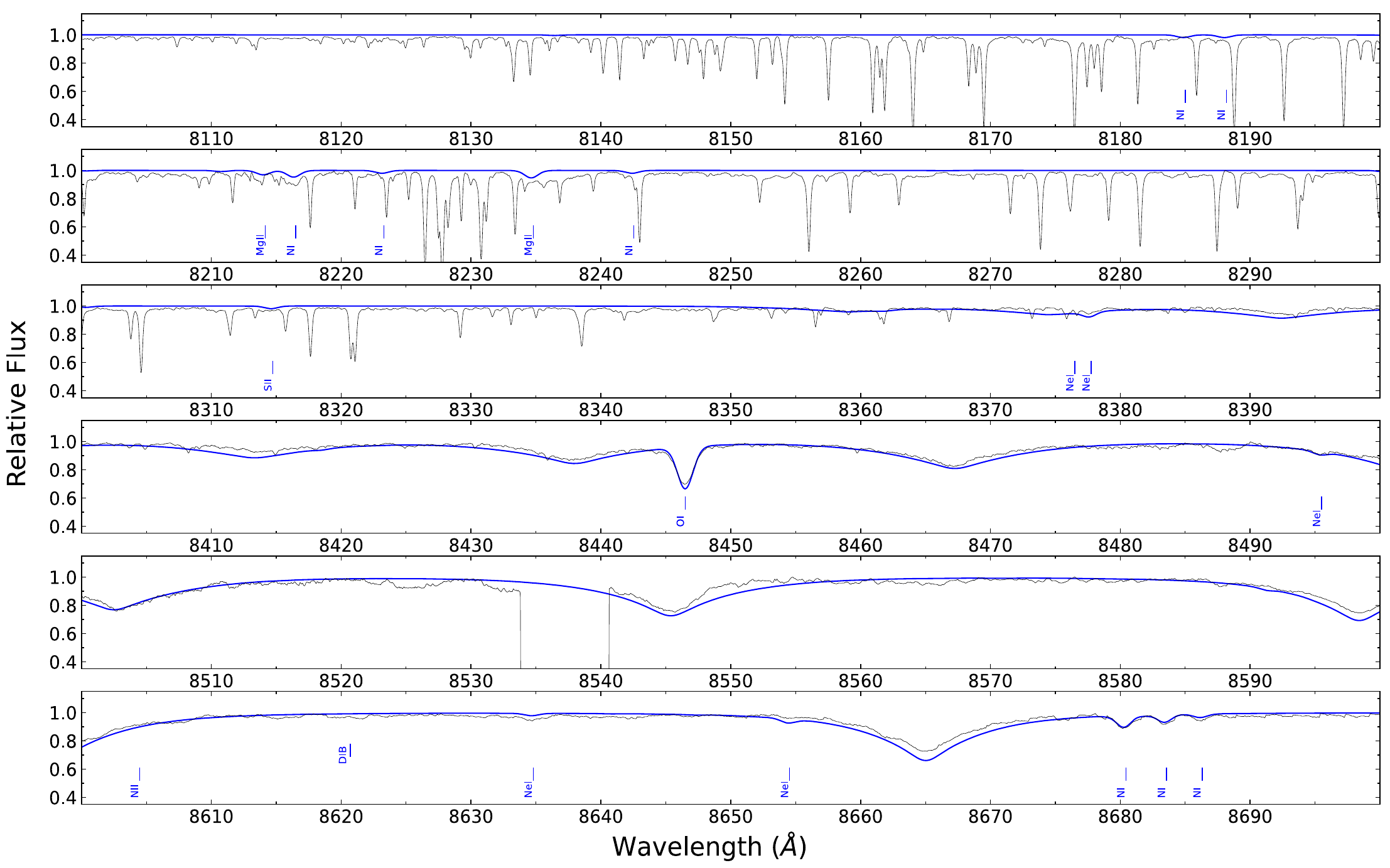}
        \caption{Same as Fig.~\ref{fig:HD164353_3900_4500}, but in the wavelength range $\lambda\lambda$8100--8700\,{\AA}.}
    \label{fig:HD164353_8100_8700}
\end{figure*}

The following figures show a comparison of the spectrum of HD~164353 as observed with FEROS and the 
best fitting global synthetic spectrum. The model was computed with the codes 
{\sc Atlas12/Detail/Surface} on basis of atmospheric parameters and
elemental abundances for the star as summarised in 
Tables~\ref{tab:stellar_parameters} and~\ref{tab:abundances}, respectively. 
The diagnostic stellar lines are identified in Figs.~\ref{fig:HD164353_3900_4500} 
to \ref{fig:HD164353_8100_8700}. A few interstellar ('IS') lines -- the Ca~H+K, 
Na~D and \ion{K}{i} resonance lines -- and several diffuse interstellar bands 
(DIBs) are also identified, but they are missing in the model. Numerous sharp 
unmodelled features redwards of about 5870\,{\AA} are of telluric origin, due to 
H$_2$O or from the O$_2$ A-, B- and $\gamma$-bands.
   
\end{appendix}

\end{document}